\documentclass[12pt,a4paper]{article} 
\usepackage{style}
\usepackage{amsmath,amssymb,epsfig,bm,amsfonts,yfonts}

\relax

\title{Characterization of initial fluctuations for the hydrodynamical description of heavy ion collisions}

\author{Stefan Floerchinger and Urs Achim Wiedemann
\\
\vspace{0.1in}

Physics Department, Theory Unit, CERN, CH-1211 Gen\`eve 23, Switzerland
\vspace{0.1in}

E-mail addresses: {\tt Stefan.Floerchinger@cern.ch, Urs.Wiedemann@cern.ch}
}

\abstract{Event-by-event fluctuations in the initial conditions for a hydrodynamical description of heavy-ion collisions are characterized. We propose a Bessel-Fourier decomposition with respect to the azimuthal angle, the radius in the transverse plane and rapidity. This allows for a complete characterization of fluctuations in all hydrodynamical fields including energy density, pressure, fluid velocity, shear stress and bulk viscous pressure. It has the advantage that fluctuations can be ordered with respect to their 
wave length and that they can be propagated mode-by-mode within the hydrodynamical formalism. Event ensembles
can then be characterized in terms of a functional probability distribution. For the event ensemble of a Monte Carlo Glauber model, we provide evidence that the latter is close to Gaussian form, thus allowing for a particularly simple characterization of the event distribution.}
 
\begin{document}
\maketitle

\section{Introduction}
\label{sec1}
In recent years, data from ultra-relativistic nucleus-nucleus collisions at 
the LHC~\cite{ALICE:2011ab,Aamodt:2011by,Chatrchyan:2012wg,Aad:2013xma} and 
at RHIC~\cite{Adamczyk:2013waa,Adare:2011tg,Alver:2008zza}
have been understood as giving strong support to a dynamical picture according to which the produced soft 
hadronic distributions in transverse momentum, azimuthal orientation, centrality 
and particle species are determined by the fluid dynamic response to fluctuating initial 
conditions~\cite{Mishra:2007tw,Broniowski:2007ft,Sorensen:2008zk,Takahashi:2009na,Alver:2010gr}.  
A detailed dynamical exploration of this picture has the potential of addressing central questions
in the study of hot and dense QCD matter with nucleus-nucleus collisions. In particular, one
expects that fundamental transport properties of dense QCD matter, such as the ratio of shear viscosity
to entropy density~\cite{Kovtun:2004de,Arnold:2000dr,Teaney:2003kp,Meyer:2007ic}, can be constrained 
with unprecedented accuracy from the fluid dynamic propagation 
of fluctuations~\cite{Schenke:2011bn,Qiu:2011iv}. Moreover, to the extent to which the fluid is almost perfect 
and therefore almost transparent 
to the propagation of fluid dynamic perturbations, fluctuation analyses may provide information about the 
initial conditions of ultra-relativistic nucleus-nucleus collisions and their evolution towards equilibrium~\cite{Bhalerao:2011yg,Schenke:2012wb}. 
As we shall shortly recall below, and as summarized in several recent 
reviews~\cite{Muller:2012zq,Heinz:2013th,Gale:2013da}, a large number 
of recent works address this program or parts of it. 

To fully exploit these physics opportunities of fluctuation analyses, one may require that a fluid dynamic formulation 
of ultra-relativistic nucleus-nucleus collisions should be as complete and as differential as possible with respect 
to the characterization of fluctuating initial conditions, their fluid dynamic propagation, and their decoupling at 
freeze-out.
In the present work, we propose to decompose fluctuating initial conditions in a complete, orthonormal basis of fluctuating modes that can be propagated individually, mode-by-mode, as fluid dynamic perturbations on a smooth event-averaged background. To this end, we employ in the following a Bessel-Fourier expansion
that - with the exception of one remarkable work~\cite{ColemanSmith:2012ka} - has not been 
explored for the characterization of initial conditions so far. 

On the level of single events, this can provide, 
for instance, a more differential understanding of how fluctuating modes that differ e.g. with respect to 
wave length are attenuated or enhanced differently during the evolution, thus providing input to the question 
of whether structures arising on some spatial scales in the initial conditions can leave signatures in experimental observables, or whether they will remain experimentally inaccessible since they are washed out in the course of the evolution. 

On the level of event ensembles, the orthonormal basis allows to determine a functional probability distribution that characterizes  weights and event-wise correlations of all fluctuating modes in the initial conditions.
This probability distribution can actually be evolved fluid dynamically by evolving each mode. 
The additional control we gain by this program can help, for example, to relate sub-classes of events defined by cuts on experimental data~\cite{Schukraft:2012ah} to sub-classes of initial conditions, thus opening further possibilities for testing the dynamical relation in between. 

The present paper is devoted to a detailed discussion of the Bessel-Fourier expansion for scalar, vector and tensor fields, the ensuing characterization of event ensembles by probability distributions formulated in this basis, and the relation of this approach to other characterizations of initial conditions for individual events and event samples. 
As emphasized above, one important motivation for the choice of a Bessel-Fourier expansion is that its basis modes
can be propagated individually as fluid dynamic perturbations. A detailed discussion of this fluid dynamic propagation 
will be left to a subsequent publication, but some first results are given already in
a recent letter~\cite{Floerchinger:2013rya}, and we shall comment in the following on properties that make the
Bessel-Fourier expansion particularly suited for such a mode-by-mode fluid dynamic propagation of fluctuations. 

By far the most common characterization of fluctuating initial conditions is in terms of a cumulant
expansion of the initial (entropy) density distribution~\cite{Teaney:2010vd} that underlies the characterization
of spatial azimuthal anisotropies in terms of eccentricities. We shall discuss in section~\ref{sec2} how the 
coefficients in a Bessel-Fourier expansion of initial conditions are related to eccentricities.
Eccentricities have been determined for initial 
conditions from simple model distributions~\cite{Niemi:2012aj,Qian:2013nba,Chaudhuri:2012mr,Alvioli:2011sk,Holopainen:2010gz,Qiu:2011hf,Qiu:2011iv,Gale:2012rq,Schenke:2012wb}, (such as the MC-Glauber, KLN and IP-Glasma conditions), as well as for full dynamical models of 
ultra-relativisitc heavy ion collisions (such as the UrQMD~\cite{Petersen:2012qc}, BAMPS\cite{Deng:2011at} 
and AMTP-codes~\cite{Bhalerao:2011bp}). Eccentricities and closely related cumulant-based formulations
have also been used to characterize angular correlations between different harmonics~\cite{Qiu:2012uy,Jia:2012ju,Jia:2012ma}, 
and they play currently an important role in discussing the specific initial geometry and expected fluid dynamic response of 
collisions between deformed nuclei (e.g. U+U), non-identical nuclei 
(e.g. Cu+Au) and of p-Pb collisions~\cite{Rybczynski:2012av,Qin:2013bha,Petersen:2013vca,Bozek:2012hy,Bzdak:2013zma}.
The fluid dynamic responses that result from initial conditions with characteristic eccentricities 
have been studied in much detail both on the level of single events or event averages \cite{Qiu:2011hf,Qiu:2011iv,Petersen:2012qc,Qian:2013nba}, as well as on the level of 
event ensembles characterized by their probability distributions~\cite{Gardim:2011xv,Bhalerao:2011bp,Deng:2011at,Niemi:2012aj,Chaudhuri:2012mr,Gale:2012rq,Schenke:2011bn}. By demonstrating that data on soft hadronic spectra and
correlations can be reproduced in viscous relativistic fluid dynamic simulations supplemented by realistic freeze-out,
and by constraining the transport properties of matter, these studies have established and are now further exploiting the paradigm that heavy ion collisions produce an almost perfect fluid. 

Despite the obvious use and success of a dynamic framework that relates via fluid dynamic simulations
a cumulant expansion of initial conditions to hadronic observables, there are questions that one may want to address 
within a fluid dynamic treatment of fluctuations and for which a cumulant expansion may not provide an 
optimal parametrization of initial conditions. In particular, any given (positive) transverse density can in principle
be determined fully by the infinite set of its moments or cumulants. But given a finite set of cumulants beyond
the ones that determine a Gaussian, it is not possible to find a positive transverse density corresponding to
them such that higher cumulants vanish. In particular, one cannot find positive transverse density configurations
that correspond to a single cumulant only, as one may want to do if one is interested in studying the propagation
and attenuation of single modes. Ref.~\cite{Teaney:2010vd} had understood this problem and had devised a 
pragmatic approach to work around it by regulating the reconstructed densities to avoid negative values. 
However, introducing a regulator introduces further non-zero cumulants, and therefore, in principle, 
one still cannot formulate initial positive transverse densities to correspond to one cumulant only. Nevertheless,
this approach has been very useful for understanding how specific structures in the initial conditions 
propagate fluid dynamically, in particular when applied to small deviations from a Gaussian transverse density
distribution~\cite{Teaney:2010vd}. But the Bessel-Fourier expansion of initial conditions that we discuss here
(see section~\ref{sec4}) may be better suited for studying the fluid dynamic propagation of fluctuations 
individually mode-by-mode, since it avoids this problem. For a dynamical treatment of individual fluctuations, 
it is also advantageous that this is an expansion in an orthonormal basis, while the cumulant expansion is not. 
Moreover, as we shall discuss in section~\ref{sec5},
the Bessel-Fourier expansion is easily extended to the characterization of initial fluctuations 
in vector and tensor fields and to their fluid dynamic propagation. To the best of our 
knowledge, an extension of the cumulant expansion to vector and tensor fields has not been attempted so far.
While essentially all currently used models of initial conditions neglect fluctuations in the initial fluid velocity and 
shear viscous tensor, they seem to be a natural possibility, and we regard it as an advantage to set up a 
formulation that treats them on an equal footing with fluctuations in the transverse
density.

We have mentioned above that it can be useful to decompose initial fluctuations in an orthonormal basis. 
Such formulations have been explored so far in particular in studies that formulate fluid dynamic perturbations
on top of simple, analytically given background fields~\cite{Gubser:2010ui,Florchinger:2011qf,Staig:2011wj,Staig:2010pn}.
For these special choices of the background field, the orthogonality of the basis modes is then preserved by the
fluid dynamic evolution. However, such a simplification of mode-by-mode fluid dynamics can only be expected 
in the presence of additional symmetries. In particular, for the case of conformal symmetry, the basis functions
used in Ref.~\cite{Gubser:2010ui} do not mix in the fluid dynamic evolution. And for the case of translational
invariance in the transverse plane as it is realized for a background field with Bjorken flow, a
two-dimensional Fourier expansion of modes has this property~\cite{Florchinger:2011qf}. We note that also
the orthonormal modes of the Bessel-Fourier expansion discussed here will not mix during fluid dynamic 
evolution if embedded as fluctuations of a Bjorken background field with transverse translational invariance. 
This feature may be helpful for instance if one plans to check the numerical accuracy of the fluid dynamic 
simulation of fluctuations against simple analytically known limiting cases. 
In general, however, we want to characterize fluctuations in all fluid dynamic fields as event-wise perturbations 
of smooth realistic background fields that do not share such special symmetries. And for this realistic case, 
the modes of fluid dynamic fields will mix under time evolution, and a differential understanding of how this
mixing occurs may provide additional physics insights. 

The present paper is organized as follows: in section~\ref{sec2} we introduce the Bessel-Fourier transform for
scalar fields, we explain how the coefficients of an expansion in this basis can be determined in a 
CPU-inexpensive way via Lemoine's method of discrete Bessel transformation, and we explain how these
coefficients are related to eccentricities. In section~\ref{sec3}, we illustrate first the accuracy and use of 
this expansion by applying it to a simple model of fluctuating initial conditions. We then turn to the question of
how event ensembles can be characterized in terms of probability distributions, and we show that the latter
take a particularly  simple and explicit form if expressed in the expansion coefficients of the Bessel-Fourier
transform. In particular, we emphasize that a Gaussian ansatz for the probability distribution, specified
in terms of event-averaged two-mode correlations only, can account with high accuracy for the event distributions
in a model of initial conditions that is currently used in phenomenological studies. In this sense, realistic event
ensembles of initial conditions are very well approximated
by simple, analytically known expressions depending on a finite number of event-averaged input data. 
While the Bessel-Fourier expansion of initial densities, discussed in section~\ref{sec3}, does not remain positive
everywhere if truncated after a finite number of modes, we show in section~\ref{sec4} that this problem does not 
exist if the expansion is applied to normalized density fluctuations. Since this approach underlies our dynamical
treatment of fluctuations in ~\cite{Floerchinger:2013rya}, we discuss it here in some detail. Section~\ref{sec5} is
finally discussing the extension of the Bessel-Fourier expansion to vector and tensor fields. Some general properties 
of Bessel transformations in continuous and discrete form are given in the appendices \ref{sec:appA} and \ref{sec:appB} while appendix \ref{sec:appC} discusses some properties of functional probability distributions for event samples.

\section{Characterizing fluctuating initial conditions}
\label{sec2}

The hydrodynamical description of heavy ion collisions is normally initialized on some space-time hyper surface
 shortly after the collision, at the end of 
 a regime with early non-equilibrium dynamics. In Bjorken coordinates $\tau$, $r$,
 $\phi$ and $\eta$, related to the laboratory coordinates $t$, $x$, $y$, $z$ by
 $t=\tau\, \cosh \eta$, $x = r\, \cos\phi$, $y=r\, \sin\phi$, $z = \tau \sinh \eta$, the initialization hyper surface is 
 usually taken to correspond to fixed $\tau = \tau_0$.
In this section, we consider the initial transverse enthalpy density $w(r,\phi)$ that
characterize the matter distribution at $\tau=\tau_0$. To keep notation simple, we do not denote 
explicitly the dependence of $w$ on time or its possible dependence on the longitudinal 
position along the beam direction (see section \ref{sec5} for a generalization). In practice, one might want to replace $w$ by the initial transverse energy density $\epsilon$, entropy density $s$, pressure $p$ or some charge
density associated to a single event. Our discussion will focus first on how to characterize 
the fluctuating density $w$ of single arbitrary events in terms of a Bessel-Fourier transformation, 
and we shall turn to the discussion of event averages and event distributions only later.  

\subsection{Radial decomposition of $w$: motivation and Lemoine's method}
\label{sec2.1}
Our starting point is the harmonic Fourier decomposition of the azimuthal dependence of $w(r,\phi)$ in terms of the harmonics
\begin{equation}
	w^{(m)}(r) = \frac{1}{2\pi}\int_0^{2\pi} d\phi\,  e^{i\, m\, \phi}\, w(r,\phi) \, .
	\label{eq2.1}
\end{equation}
We recall that the commonly used event eccentricities $\epsilon_{n,m}$ can be defined~\cite{Luzum:2011mm} as the 
normalized moduli
\begin{eqnarray}
\epsilon_{n,m} &=& \vert \tilde\epsilon_{n,m}\vert / \vert \tilde\epsilon_{n,0}\vert 
\label{eq2.2}
\end{eqnarray}
of the radial moments of $w^{(m)}(r)$ 
\begin{eqnarray}
\tilde \epsilon_{n,m} &=&  2\pi \int dr\,  r^{n+1}\,  w^{(m)}(r) 
= \vert \tilde\epsilon_{n,m}\vert e^{i\, m\, \psi_{n,m}}\, .
\label{eq2.3}
\end{eqnarray}
In recent phenomenological studies, one often focusses on one radial moment
per $m$-th harmonic, selecting for instance the subset of eccentricities 
$\lbrace \epsilon_{2,m}\rbrace$ or $\lbrace \epsilon_{m,m}\rbrace$ that is then
denoted by the shorthand $\lbrace \epsilon_{m}\rbrace$. As we shall see in the 
following, this practice may be justified to some extent by the observation that 
the eccentricities $\epsilon_{n,m}$ for phenomenologically relevant density profiles
tend to change only gradually and smoothly with increasing $n$. However,
while the subset of eccentricities $\lbrace \epsilon_{m}\rbrace$ provides 
an incomplete characterization of $w$, the set of all $\vert\tilde\epsilon_{n,m}\vert$
supplemented by the angular orientations $\psi_{n,m}$ 
is complete: the shape of the transverse density $w(r,\phi)$ of a single
event can in principle be reconstructed unambiguously from the complete set of 
complex-valued $\epsilon_{n,m}$'s.

The azimuthal decomposition (\ref{eq2.1}) of $w(r,\phi)$ provides a natural ordering of the azimuthal dependence  
that characterizes increasingly finer azimuthal structures with increasing azimuthal wave number $m$.
In comparison, the connection between the $n$-th moments $\epsilon_{n,m}$ of $w^{(m)}(r)$ and fluctuating 
modes of particular radial wave length is arguably less direct.  
Here we ask how to write an alternative decomposition of the radial dependence
of $w(r,\phi)$ that orders fluctuating radial modes more explicitly in terms of functions 
of increasingly smaller radial resolution scale. And since we 
expect that dynamics changes the wavelength of a fluctuation only gradually and over sufficiently 
long time scale, we may hope that the modes of such an alternative expansion will mix only weakly 
under dynamical evolution, thus facilitating studies of the relation between modes of characteristic
radial wave length in the initial distribution and measurements that are differential in transverse
momenta.

A Fourier transformation of $w$ provides arguably the simplest decomposition of a function in 
terms of modes of increasing resolution scale. However, in the neighborhood of $r=0$, 
an expansion of $w$ in Fourier modes $e^{i\, k\, r}$ is not possible, since these do not
satisfy the boundary condition $w^{(m)}(r)  \propto r^m$ for small $r$. On the other hand, in radial 
coordinates, a two-dimensional Fourier transformation is an expansion in modes 
$\propto e^{i\, k\, r\, \cos\phi}$ and the $m$-th harmonic moment of this Fourier mode is
a Bessel-function, $\int_0^{2\pi} d\phi\, e^{i\, k\, r\, \cos\phi}\, \cos(m\phi) = 2\pi J_m(kr)$.
The Bessel functions $J_m$ do have the desired limiting behavior $\propto r^m$ for $r \to 0$. 
These considerations prompt us to seek an expansion of the $m$-th moment $w^{(m)}(r)$ in a series of Bessel 
functions $J_m(z)$ \cite{ColemanSmith:2012ka},~\footnote{To minimize notation, we distinguish here the moment $w^{(m)}(r)$ from its Bessel transform $w^{(m)}(k)$ by its argument only. No confusion should arise since most of the following discussion will be in terms of the 
coefficients $w_l^{(m)}$ instead of $w^{(m)}(k)$, see eq.~(\ref{eq2.4a}) below.  }
\begin{equation}
w^{(m)}(r) = \int dk\, k\, w^{(m)}(k)\, J_m(k r)\, .
\label{eq2.3a}
\end{equation}
This continuous expansion becomes discrete if restricted to a finite region 
$r \in \lbrack 0,R\rbrack$ with boundary condition
\begin{equation}
   w^{(m)}(r) = 0\, \quad \hbox{for}\, \quad r>R\quad \, \hbox{and all $m$}\, .
   \label{eq2.4}
\end{equation}
One can write then
\begin{equation}
w^{(m)}(r) = \sum_{l=1}^\infty \; w_l^{(m)}\, J_m\left(k_l^{(m)} r\right)
\label{eq2.4a}
\end{equation}
with the complex coefficients $w_l^{(m)}$ given by the integral expressions
\begin{equation}
w_l^{(m)} = \frac{2}{R^2 \left[ J_{m+1}(k_l^{(m)}R) \right]^2} \int_0^R dr\, r\, w^{(m)}(r)\, J_m\left( k_l^{(m)} r\right).
\label{eq2.4b}
\end{equation}
The discrete set of wave numbers
$k_l^{(m)}$ are defined in terms of the $l$-th zero crossings $z_l^{(m)}$ of the 
Bessel function $J_m(z)$, 
\begin{equation}
   k_l^{(m)} = z_l^{(m)}\, \frac{1}{R}\, .
   \label{eq2.6}
\end{equation}
By construction, the expansion (\ref{eq2.4a}) satisfies the boundary condition 
(\ref{eq2.4}), and the terms with increasing $l$ correspond to modes of 
smaller and smaller radial resolution $1/ k_l^{(m)}$. In this way, the characterization of the azimuthal dependence 
of $w(r,\phi)$ in terms of a discrete set of azimuthal harmonics (labeled by $m$) can be 
paralleled by a characterization of its radial dependence in terms of a discrete set of radial 
modes labeled by $l$.~\footnote{We remark that instead of using a Bessel expansion as in 
(\ref{eq2.1}) directly for the transverse density $w^{(m)}(r)$, it may be advantageous for some questions
to expand a normalized version 
$$ \tilde w^{(m)}(r) = w^{(m)}(r) \Big/ w_\text{BG}(r)\, .$$
Here, $w_\text{BG}(r)$ denotes an appropriately normalized ``background function'' that depends only on the 
radius $r$ and that can be defined e.g. as the event averaged density $\langle w(r,\phi)\rangle$. The
motivations for this formulation and some properties will be discussed in section~\ref{sec4}.}

Lemoine's method~\cite{Lemoine} of discrete Bessel transformation simplifies the determination of the 
weights $w_l^{(m)}$ of the Bessel expansion, since it allows one to replace the integral in (\ref{eq2.4b}) by 
a finite sum, 
\begin{equation}
	w^{(m)}(r) \approx \sum_{l=1}^{N_l}\, w_l^{(m)}\, J_m\left(k_l^{(m)}\, r\right)\, .
	\label{eq2.5}
\end{equation}
Here, $w_l^{(m)}$ are complex-valued expansion coefficients, and the approximation (\ref{eq2.5}) 
can be improved systematically by including a larger number of terms $N_l$. 
Remarkably, according to Lemoine's method \cite{Lemoine}, the determination 
of the coefficients $w_{l}^{(m)}$ does not involve integrations but can be done by matrix 
multiplication of the function $w^{(m)}(r)$ evaluated at a discrete set of radii
\begin{equation}
	r^{(m)}_\alpha = R \frac{z_\alpha^{(m)}}{z_{N_l}^{(m)}}\, .
	\label{eq2.7}
\end{equation}
The coefficients $w_{l}^{(m)}$ take then the form 
\begin{equation}
	w_{l}^{(m)} \approx \sum_{\alpha=1}^{N_l} 
	{\cal M}^{(m)}_{l \alpha} w(r^{(m)}_\alpha,m)\, ,
	\label{eq2.8}
\end{equation}
where the matrix ${\cal M}^{(m)}_{l \alpha}$ is independent of the properties of $w^{(m)}(r)$
and reads
\begin{equation}
	 {\cal M}^{(m)}_{l \alpha} = \frac{4\, J_m\left(k_l^{(m)} r^{(m)}_\alpha\right)}{
	 (z_{N_l}^{(m)})^2\,  
	 J_{m+1}^2(z_l^{(m)})\, J_{m+1}^2(z_\alpha^{(m)})}\, .
	 \label{eq2.9}
\end{equation}
The value of $R$ is a parameter in the analysis that can be choosen freely as long as (\ref{eq2.4})
is satisfied. In practice, it is useful to choose $R$ as small as possible to ensure that the 
expansion (\ref{eq2.5}) does not need to account for regions of $r$ in which the function $w$ vanishes. 
In the numerical studies for Pb-Pb collisions, discussed in later subsections, we choose $R = 8$ fm. 
Once $R$ is fixed, one can tabulate the matrix (\ref{eq2.9}) and determine the complex
expansion coefficients $w_{l}^{(m)}$. If one wants to change the number $N_l$ of terms in
the expansion, both the matrix ${\cal M}$ in (\ref{eq2.9}), and all the coefficients $w_{l}^{(m)}$ need
to be re-evaluated. 

From the expansion coefficients $w_{l}^{(m)}$, the spatial density distribution can then be 
reconstructed,
\begin{eqnarray}
	w_{{\rm reco}(N_m,N_l)}(r,\phi) &=& \sum_{l=1}^{N_l} w_{l}^{(m=0)}\, 
	      J_0\left(z_l^{(0)} r/R \right)\nonumber \\
	      && + 2 \sum_{m=1}^{N_m} \sum_{l=1}^{N_l}  
	       \vert w_{l}^{(m)} \vert \cos\left[ m \left( \phi - \varphi_l^{(m)} \right) \right]
	       \,  J_m\left(z_l^{(m)} r/R \right) \, .
	       \label{eq2.10}
\end{eqnarray}
Here we have made the phase dependence of the complex-valued Bessel coefficients
explicit,
\begin{equation}
	w_l^{(m)} = \vert w_l^{(m)}\vert  \exp\left[ i m\, \varphi_l^{(m)} \right]\, .
	\label{eq2.10b}
\end{equation}
As we shall illustrate with examples in the next subsection, the reconstructed
spatial transverse density becomes an increasingly better approximation of $w(r,\phi)$
if one increases the numbers $N_m$ and $N_l$ of azimuthal and radial modes included in
(\ref{eq2.10}). 

We finally note that by inserting (\ref{eq2.5}) into (\ref{eq2.3}), one can express the
eccentricities $\tilde{\epsilon}_{n,m}$ in terms of a complete set of coefficients $w_{l}^{(m)}$,
\begin{equation}
	\tilde{\epsilon}_{n,m} \approx R^{n+2} \sum_{l=1}^{N_l} {\cal K}_{(m)}^{n\, l}\, 
	w_{l}^{(m)}\, ,
	\label{eq2.11}
\end{equation}
where
\begin{equation}
	{\cal K}_{(m)}^{n\, l} = \frac{2\pi}{\left( z_l^{(m)} \right)^{n+1}}
		\int_0^{z_l^{(m)}}  d\bar{r}\, \bar{r}^{n+1}\, J_m\left(\bar{r}\right)\, .
		\label{eq2.12}
\end{equation}
For finite $N_l$, relation (\ref{eq2.11}) is approximate and can be viewed as including all
contributions to $\tilde{\epsilon}_{n,m}$ that result from fluctuating modes of wavelength
$1/k_{N_l}^{(m)}$ and larger. With increasing $N_l$, this expression becomes 
more and more accurate.
%
\begin{figure}[h]
\vspace{-0.5cm}
\begin{center}
\includegraphics[width=6.cm]{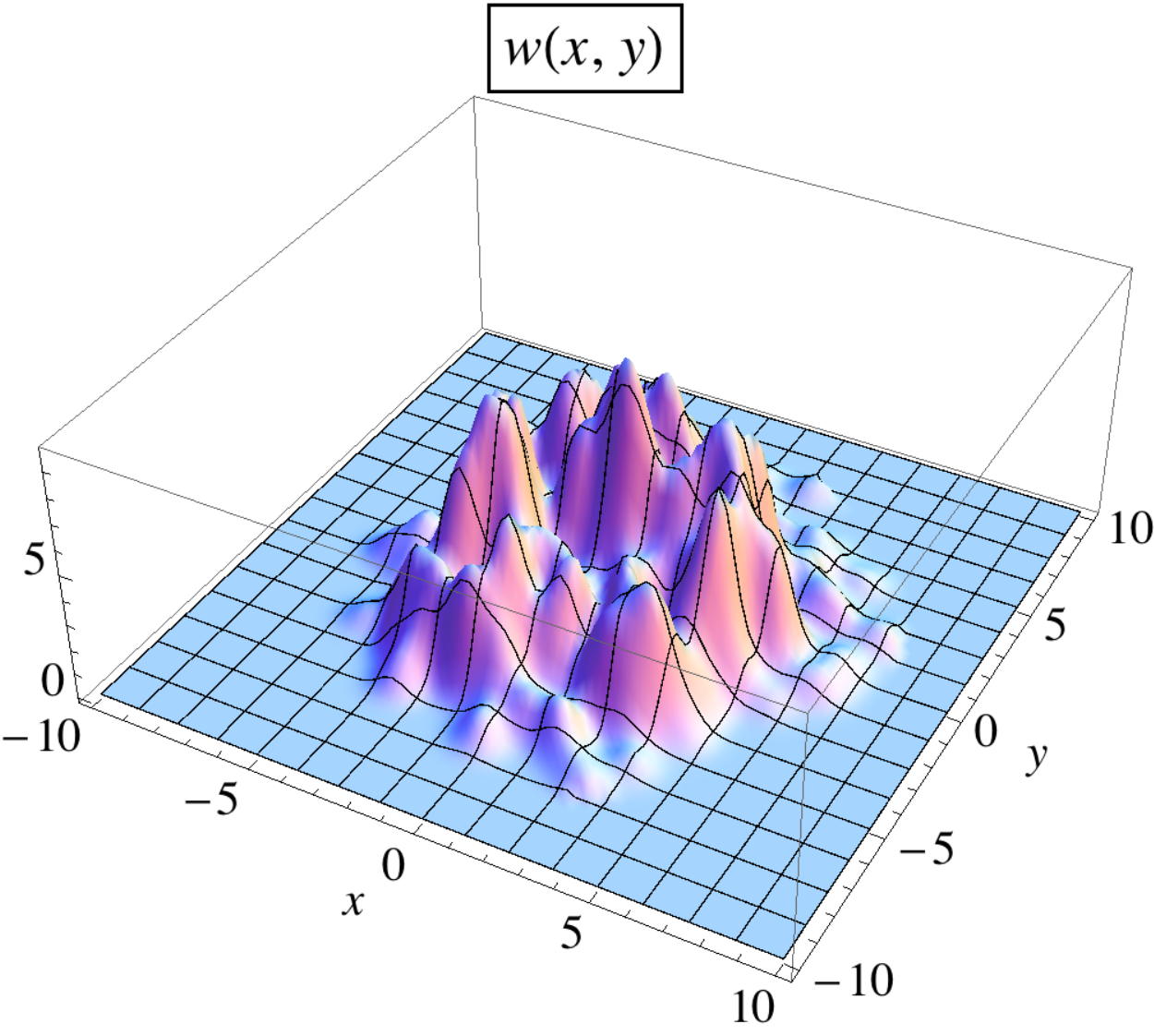}
\includegraphics[width=6.cm]{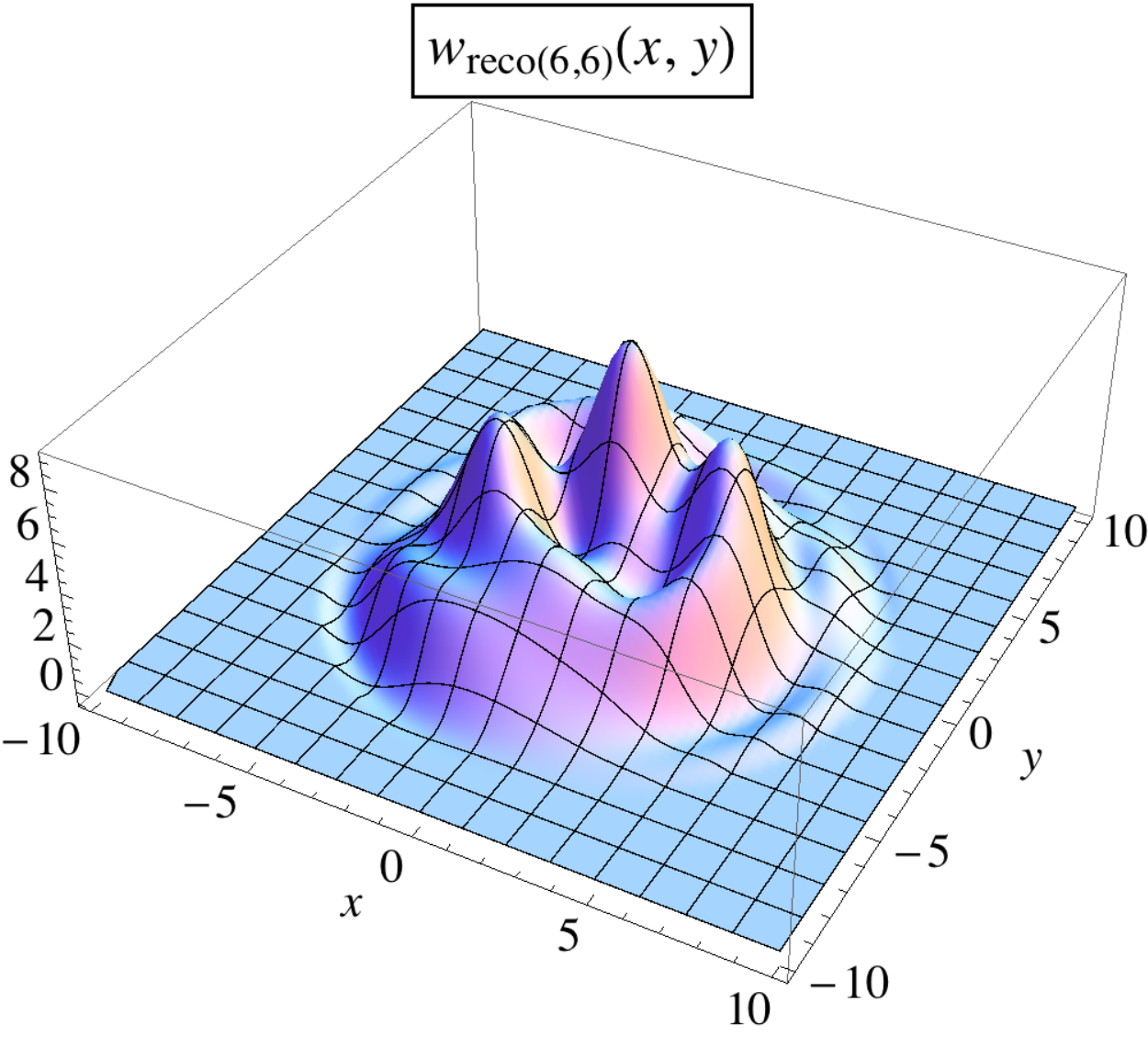}
\includegraphics[width=6.cm]{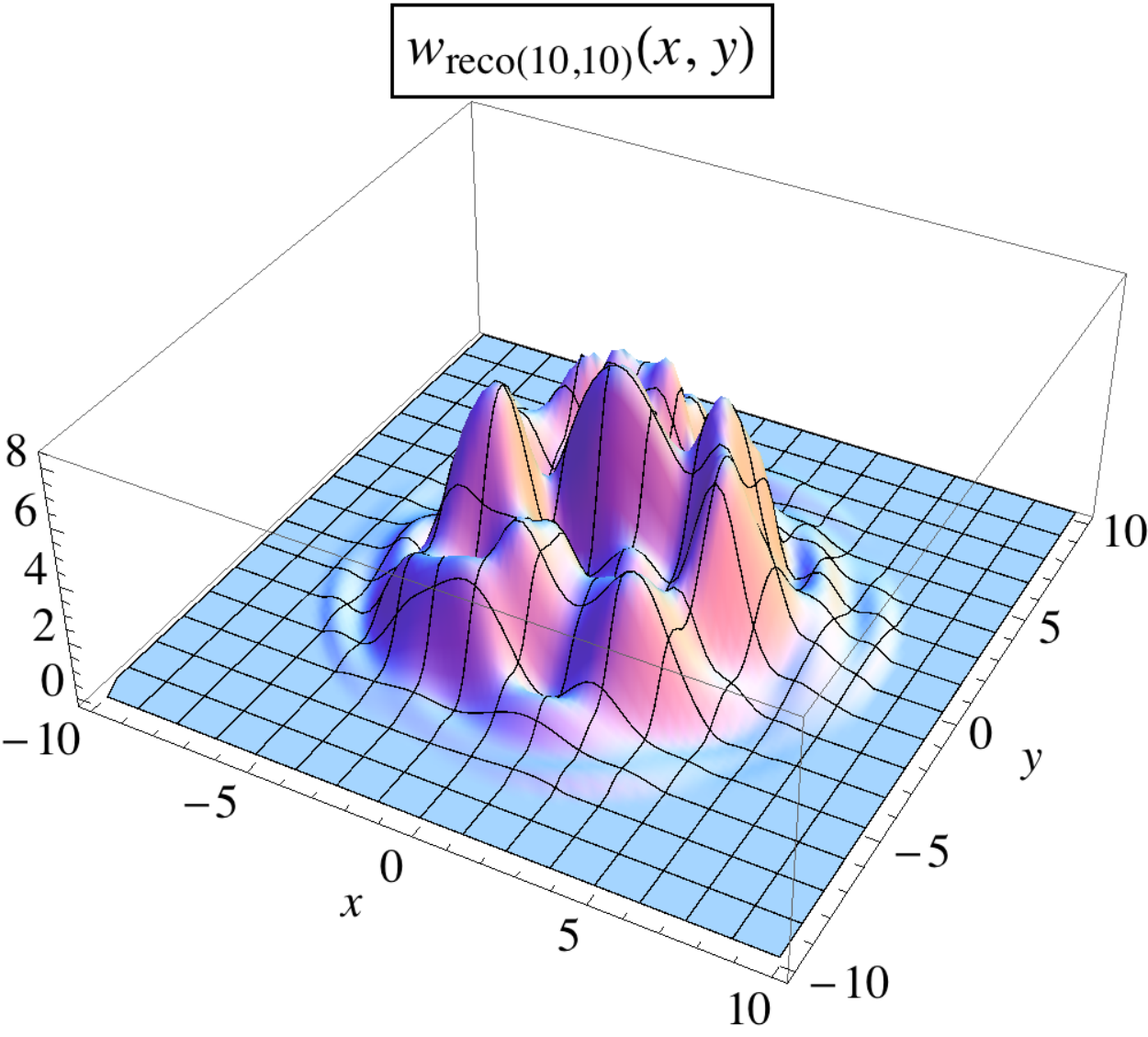}
\includegraphics[width=6.cm]{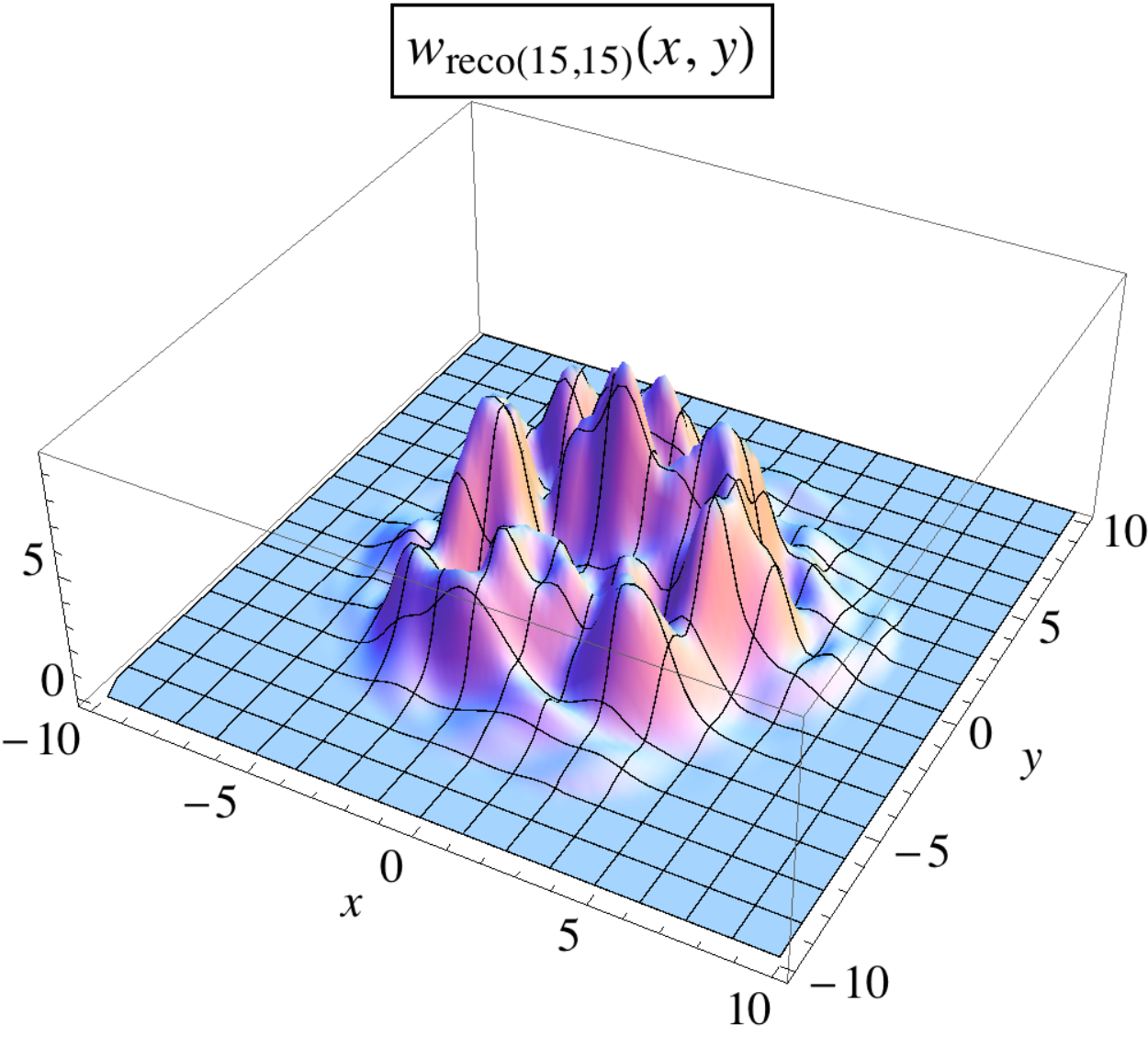}
\includegraphics[width=6.cm]{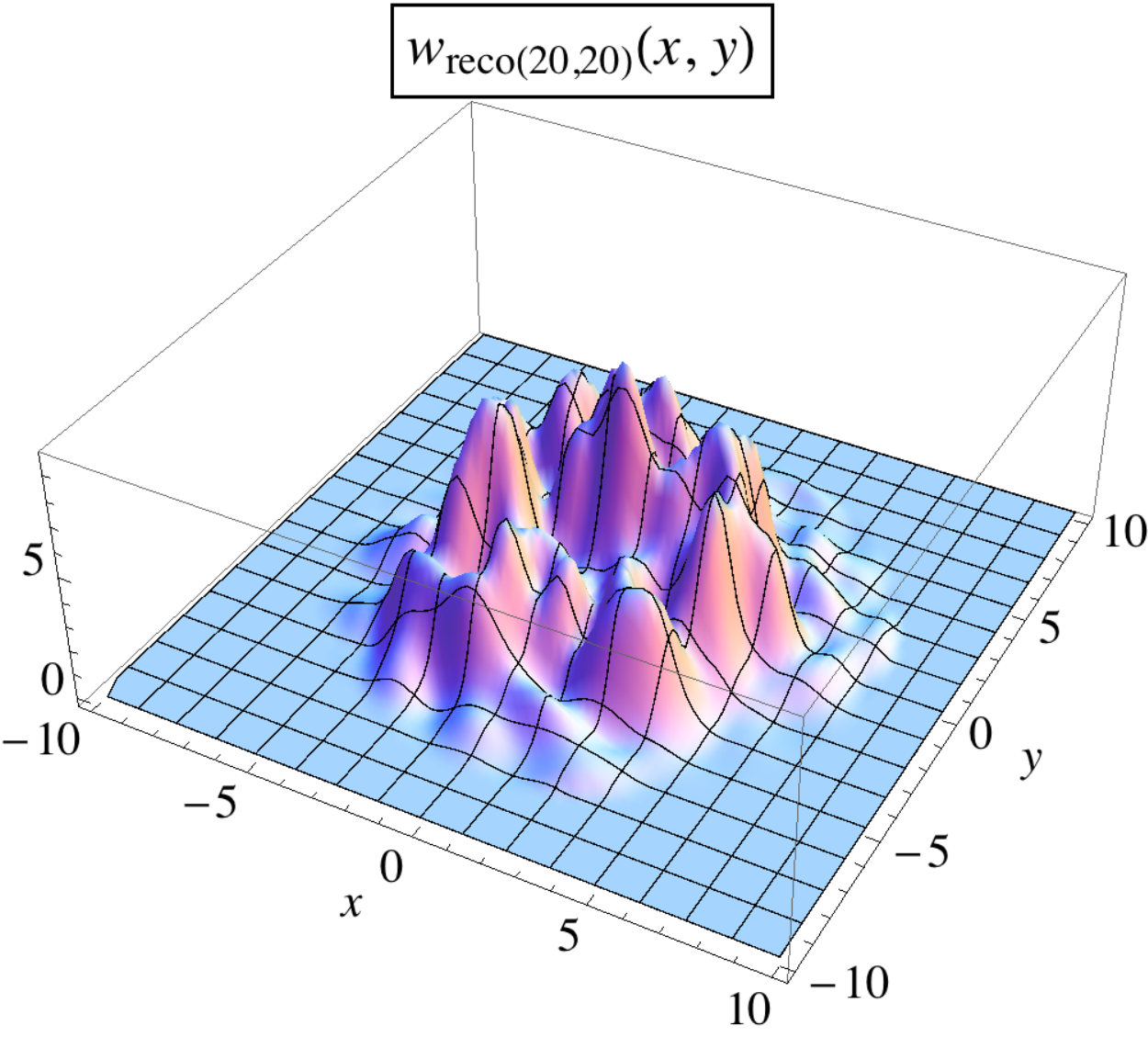}
\includegraphics[width=6.cm]{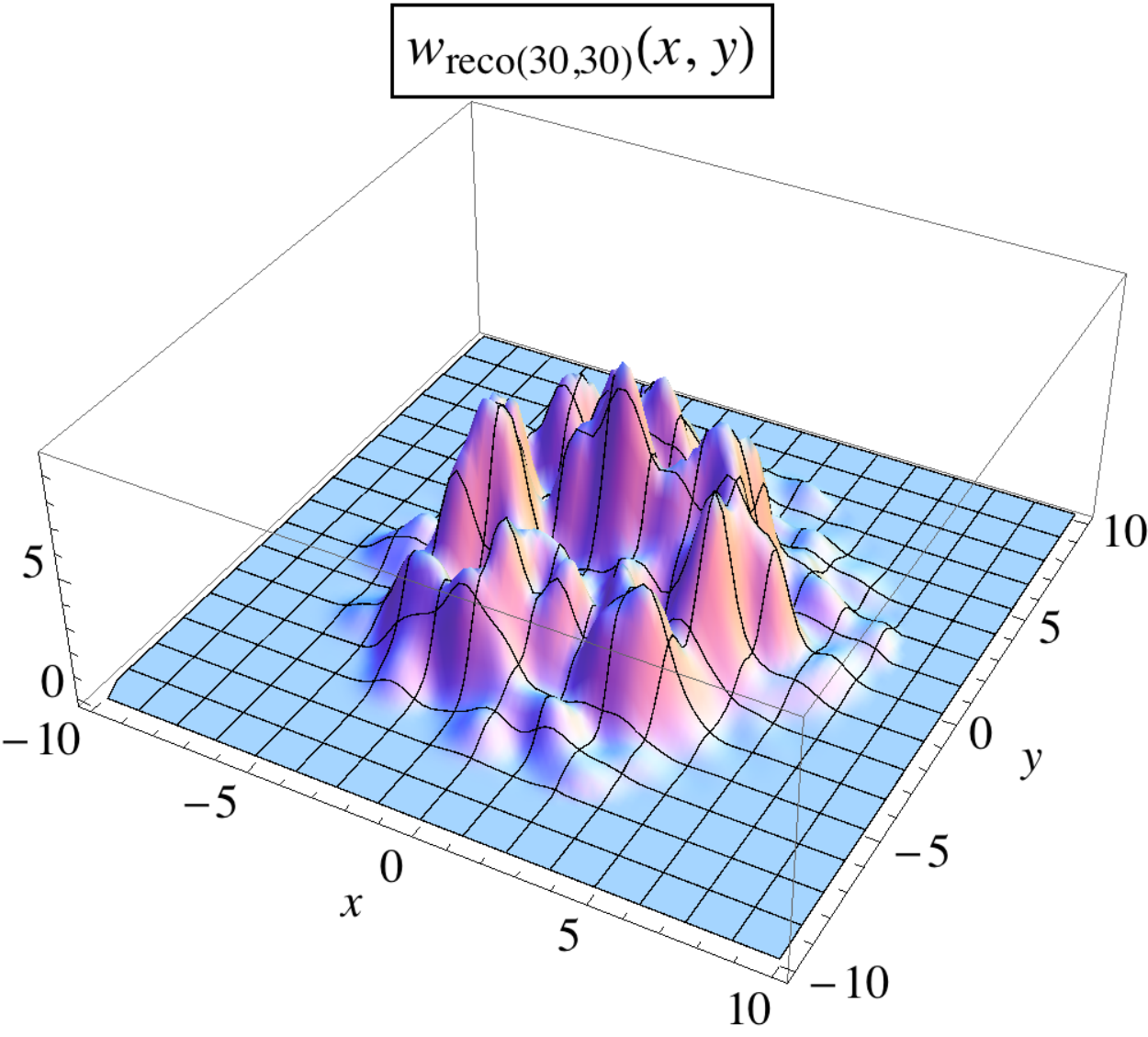}
\end{center}
\vspace{-0.5cm}
\caption{Upper left plot: Transverse enthalpy density distribution $w(x,y)$ for one randomly chosen central
Pb+Pb collision, simulated according to the model described in section~\ref{sec3.1}. 
Remaining five plots: Reconstruction $w_{{\rm reco}(N_m,N_l)}$ of this particular density 
distribution from the data of a discrete Bessel transformation of $w(x,y)$, 
involving an increasing number of modes  in the azimuthal ($N_m$) and in the radial 
($N_l$) direction. The point-by-point differences between the truth $w(x,y)$ and the
reconstruction $w_{{\rm reco}(N_m,N_l)}$ are less than $1 \%$ of the maximal density
for a reconstruction with $N_m =N_l=30$. 
}\label{fig1}
\end{figure}

\section{Characterizing initial conditions with Lemoine's method - a numerical example}
\label{sec3}
The relations (\ref{eq2.11}), (\ref{eq2.12}) illustrate that the information about $w$ contained 
in the Bessel coefficients $w_{l}^{(m)}$ and in the eccentricities $\tilde{\epsilon}_{n,m}$ is
complete and mathematically equivalent. It then depends on the physics problem under
consideration to decide which of these two equivalent characterizations is better suited.
In particular, the relation (\ref{eq2.11}) makes it explicit that for any 
given $m$-th moment, the $n$-th radial moments $\tilde{\epsilon}_{n,m}$ receive in general 
contributions from fluctuating modes of various different radial wave lengths 
$1/k_l^{(m)}$, $l\in \left[1,N_l\right]$. In contrast, the expansion (\ref{eq2.5}) of the fluctuations 
in Bessel functions is explicitly an expansion in modes of increasing radial resolution, and it is
an expansion in an orthonormal basis. This can be helpful. To
illustrate the use of organizing fluctuating modes of $w$ with the help of a discrete Bessel
transformation, we turn now to an explicit numerical example.

\subsection{A simple wounded nucleon model for the initial transverse density}
\label{sec3.1}
%
For illustrative purposes, we consider a simple Monte Carlo Glauber model for the initial conditions in Pb+Pb
collisions, similar to the one described in Ref.\cite{Holopainen:2010gz}. In the simplest version, this model 
determines the enthalpy density $w(r,\phi)$ as proportional to the number of wounded nucleons. 
Nucleons in the incoming projectiles are 
distributed event-by-event randomly in the transverse plane according to the two-dimensional 
projection of a standard spherically symmetric two-parameter Woods-Saxon nuclear density 
profile. Nucleon-nucleon correlations in the incoming projectiles are neglected. The condition
for collision between nucleons $i$ and $j$ of different nuclei is the simple geometric one, namely
that the transverse positions $(x_i,y_i)$ of both nucleons are closer than 
$(x_i-x_j)^2 +  (y_i-y_j)^2 \leq \sigma_{NN}/\pi$. Here, we choose for the inelastic 
nucleon-nucleon cross section a value corresponding to $\sqrt{s_{NN}} = 2.76$ TeV,
namely $\sigma_{NN} = 63$ mb. The transverse enthalpy density in this model is then obtained 
by centering at the transverse position of each participating nucleon an enthalpy 
contribution of Gaussian shape and with width $\sigma_B$,
\begin{equation}
   w(x,y) = {\cal N}\, \sum_{i=1}^{N_{\rm part}} c_i\, \exp \left(- \frac{(x-x_i)^2 +  (y-y_i)^2}{2 \sigma^2_B} \right)\, .
   \label{eq3.1}
\end{equation}
Here, the factors $c_i$ give weights to the contributions from individual participating nucleons $i$. In the
model of Ref.~\cite{Holopainen:2010gz}, $c_i = 1$. Instead, we use a MC Glauber model that determines
for each participating nucleon the number of collisions that this nucleon undergoes. The prefactors $c_i$
are then chosen such that the total entropy of the system scales with 
$\left( (1-x)N_{\rm part}/2 + x\, N_{\rm coll}\right)$ where $x = 0.118$. This model extension is consistent 
with the initial conditions used in recent fluid dynamic simulations of flow~\cite{Qiu:2011iv}. It is unimportant 
for the arguments made in the present paper, but since the present manuscript serves us also to further document 
the input in our recent fluid dynamic study~\cite{Floerchinger:2013rya}, and since this study is based on
state-of-the-art initial conditions, we adhere to it here. The normalization ${\cal N}$ in (\ref{eq3.1}) is then 
fixed by the total enthalpy of the system. For numerical studies, we shall associate to the position of
each participating nucleon a Gaussian of width $\sigma_B = 0.4$ fm, except where states otherwise.
We position the center of mass (the center of enthalpy) of each event at the origin of the coordinate
system.

\subsection{Reconstructing transverse density of a single event from Bessel data}
\label{sec3.2}
We first establish the efficiency of Lemoine's method in reconstructing the transverse enthalpy $w$
of a single event from its Bessel coefficients $w_l^{(m)}$. To this end, we have chosen
one particular Pb+Pb collision simulated with the wounded nucleon model of section~\ref{sec3.1}
for impact parameter $b=0$. The corresponding density distribution is shown in the 
left upper plot in Fig.~\ref{fig1}; it is non-vanishing in a transverse extension of radius 
$\sim 6$ fm, characteristic for a Pb-Pb collision, and it shows significant fluctuations. 
We have determined the Bessel coefficients $w_l^{(m)}$ of this distribution as discussed 
in section~\ref{sec2}, and we have reconstructed the density distribution 
$w_{{\rm reco}(N_m,N_l)}(r,\phi)$, including a varying number of modes $N_m$, $N_l$ in 
the azimuthal and radial direction. As one sees clearly from the various plots in Fig.~\ref{fig1}, 
with increasing number of modes, $w_{{\rm reco}(N_m,N_l)}$
reconstructs increasingly finer details of the transverse density of this single event. 
A larger number of modes is needed to resolve smaller scales. In general, we observe that
for a relatively small number of expansion coefficients, the main features of the transverse
density can be characterized, and that the approximation of the true $w$ in terms of the
reconstructed density (\ref{eq2.10}) improves rapidly in accuracy with increasing number of
modes $N_m$, $N_l$.

\subsection{Characterizing event averages of single fluctuating modes}
\label{sec3.3}

\begin{figure}[h]
\begin{center}
\includegraphics[width=7.cm]{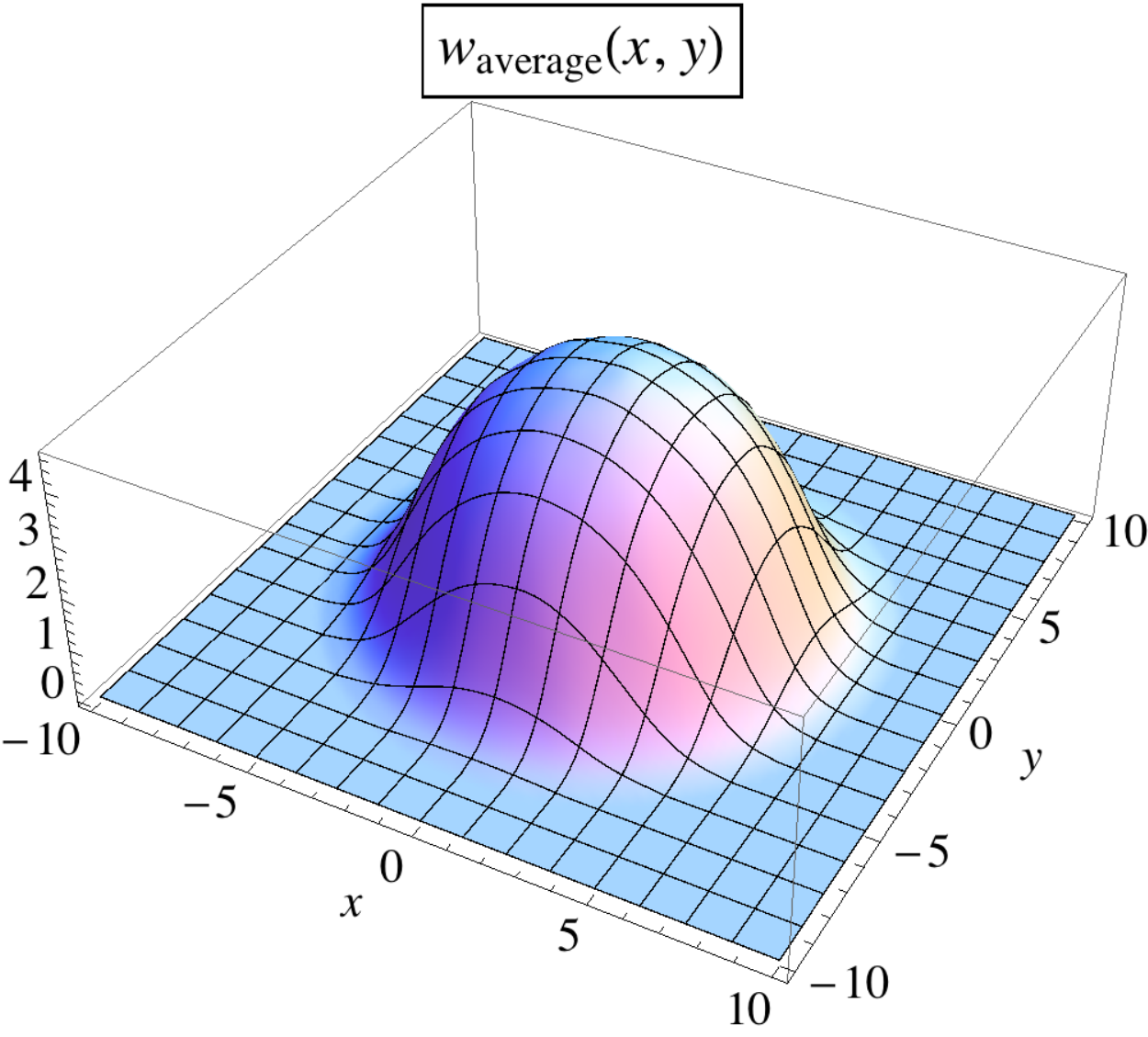}
\includegraphics[width=7.cm]{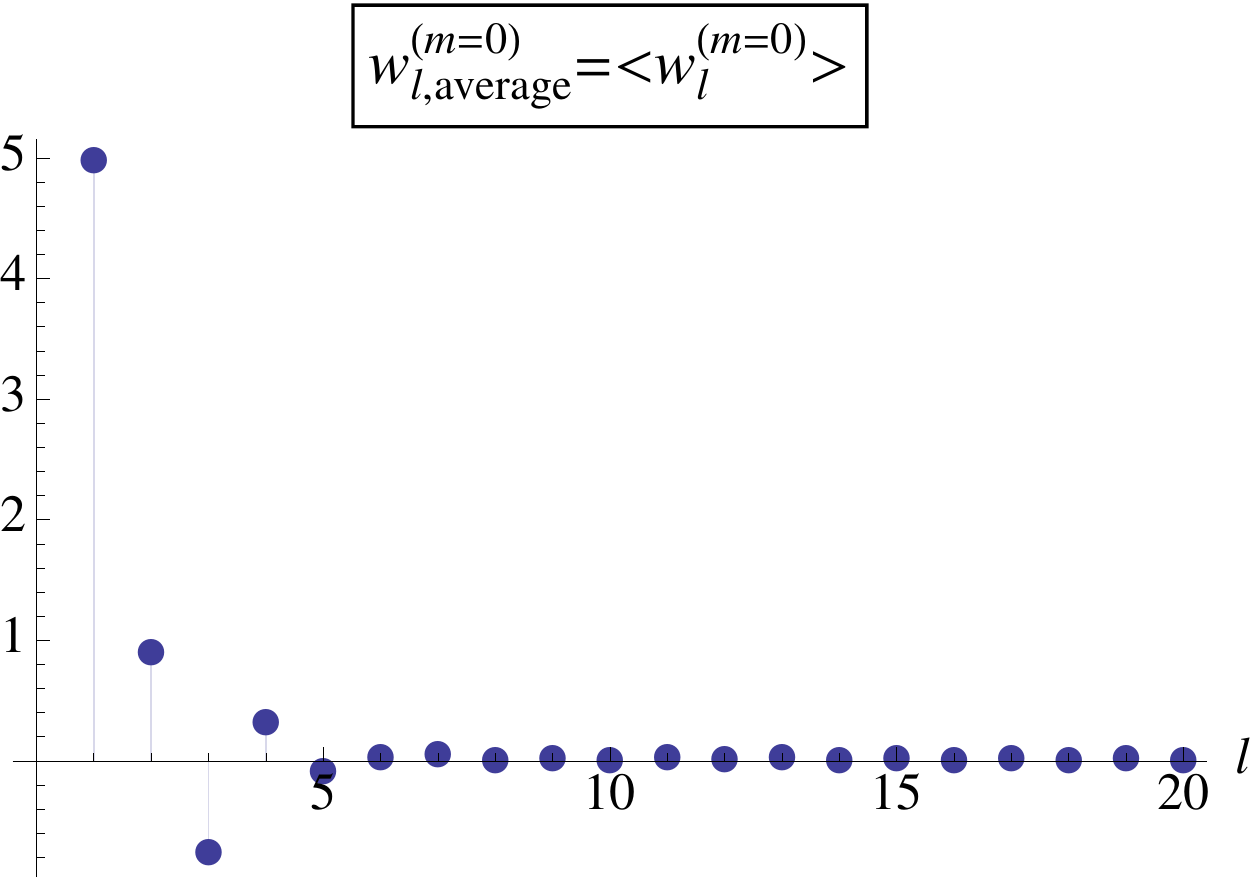}
\includegraphics[width=7.cm]{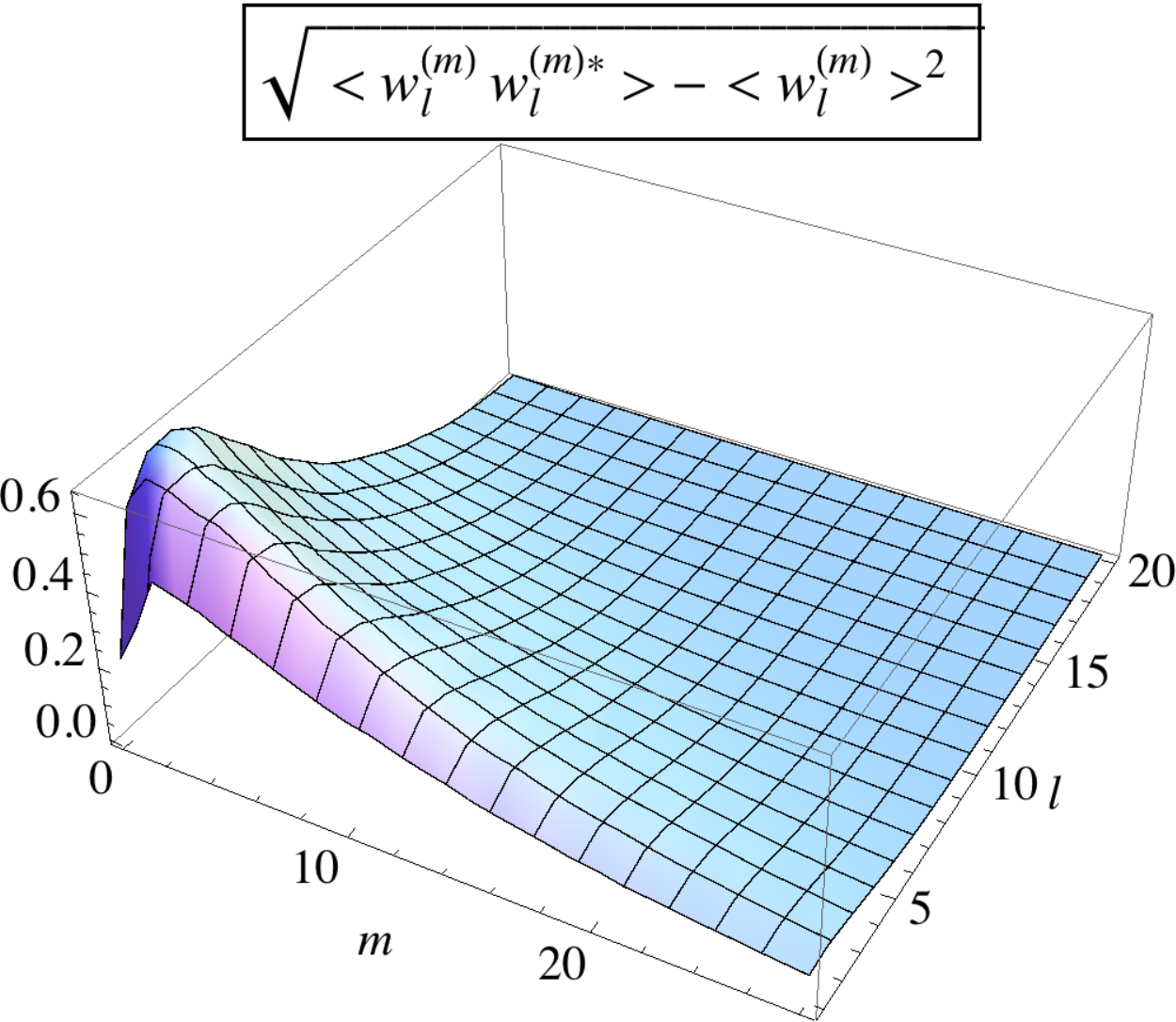}
\includegraphics[width=7.cm]{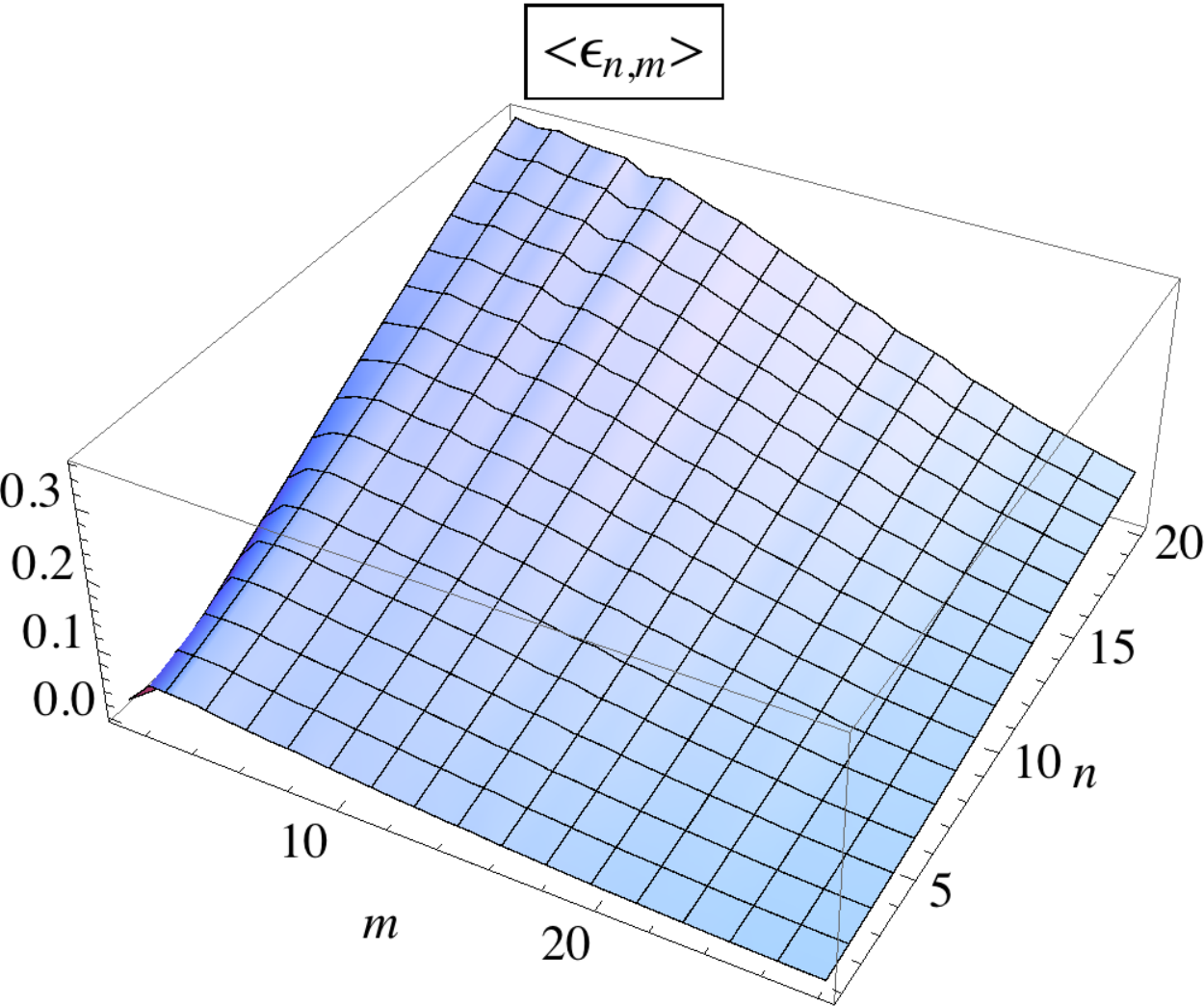}
\end{center}
\vspace{-0.5cm}
\caption{Upper row: the event-averaged density distribution $w_{\rm average}$
for Pb-Pb collision at impact parameter $b=0$ fm and the Bessel coefficients 
$w_{l,{\rm average}}^{(m=0)}$ characterizing it. Lower row: Left hand side:
the dispersion of the
event-wise distribution of $w_l^{(m)}$ around the mean $\langle w_l^{(m)}\rangle$
characterizes event-by-event fluctuations in the initial density.  
Right hand side: characterization of the average event-by-event fluctuations 
by the event-averaged
eccentricities $\langle \epsilon_{n,m}\rangle $. 
Data are simulated with the wounded model of section~\ref{sec3.1}
for a sample of O(1000) Pb-Pb collision at $b=0$ fm. 
}\label{fig2}
\end{figure}

As seen from Fig.~\ref{fig1}, the MC Glauber model for the initial density distribution $w$
generates spatial distributions with significant event-wise variations. We aim at
quantifying these as fluctuations around an event-averaged 
density distribution. To this end, we first determine the event-averaged density distribution by 
averaging over the densities $w_i$ of a large number of events, $i \in \left[1,N_{\rm events}\right]$,
\begin{equation}
 w_{\rm average}(x,y)  \equiv \langle w(x,y)\rangle \equiv
 \lim_{N_{\rm events} \to \infty}
  (1/N_{\rm events}) \sum_{i=1}^{N_{\rm events}} w_i(x,y)
 \, .
 \label{eq3.2}
 \end{equation}
Here and in the following, the brackets $\langle \dots \rangle$ define the event average.
The event-averaged density $w_{\rm average}$ of a Pb+Pb collisions at impact parameter 
$b=0$ is shown in Fig.~\ref{fig2}. At $b=0$, this average is azimuthally symmetric.
As a consequence, all eccentricities of  $w_{\rm average}$ with $m\neq 0$
vanish and similarly the event-averaged Bessel coefficients,
\begin{equation}
	\langle w^{(m)}_{l} \rangle = w^{(0)}_{l,{\rm average}}\, \delta_{m0}\, .
	\label{eq3.3}
\end{equation}
The coefficients $w^{(0)}_{l,{\rm average}}$ quantify the shape of $w_{\rm average}$ completely.
As seen from Fig.~\ref{fig2}, $w^{(0)}_{l,{\rm average}}$ has significant non-zero entries only for
the first few radial modes $l=1,2,3,4$. This reflects the fact that the shape of $w_{\rm average}$ is
smooth and hence only long radial wave-lengths (i.e. modes with small $l$) are needed to characterize
its radial dependence.

As a first step towards quantifying fluctuations on top of the event-averaged 
background distribution $w_{\rm average}$, we display in Fig.~\ref{fig2} the dispersion 
$D(w^{(m)}_{l})$ of the event distribution of $w^{(m)}_{l}$ around its average,
\begin{equation}
	D^2(w^{(m)}_{l}) = \langle w^{(m)}_{l}\, w^{(m)*}_{l}\rangle 
			- \langle w^{(m)}_{l}\rangle\,  \langle w^{(m)}_{l}\rangle \, .
			\label{eq3.4}
\end{equation}
Since fluctuations around  $w_{\rm average}$ break the azimuthal symmetry, 
one finds non-vanishing values for  $\langle w^{(m)}_{l}\, w^{(m)*}_{l}\rangle$, even for
$m \not= 0$, when the event average $\langle w^{(m)}_{l}\rangle $ vanishes.
And since fluctuations vary on smaller scales than the variation of $w_{\rm average}$, 
this dispersion has non-zero entries also for larger mode number $l$.  For a 
physical understanding of the dispersion $D(w^{(m)}_{l})$ shown in Fig.~\ref{fig2},
it is useful to relate the mode number $l$ to the physical scale of the corresponding radial 
wavelenght $1/k^{(m)}_l$. For  $R = 8$ fm, used in Fig.~\ref{fig2}, and for $m=2$, one finds
for instance the following radial wavelengths associated to some modes $l$: 
$1/k^{(2)}_1 = 1.56$ fm, $R/k^{(2)}_5 = 0.45$ fm, 
$R/k^{(2)}_{15} = 0.16$ fm. The fact that for $l > 15$, the dispersions
$D(w^{(m)}_{l})$ around $\langle w^{(2)}_{l}\rangle =0$ 
are very small translates then directly into a statement that within the 
present model, event-by-event fluctuations do not induce significant variations at radial 
scales below $0.16$ fm. In the present case, we know this of course, since the calculation 
of Bessel coefficients in Fig.~\ref{fig2} was done for a MC Glauber model with smearing 
factor $\sigma_B = 0.4$ fm, see eq.~(\ref{eq3.1}).
We further note that in physical units, the radial wavelengths $1/k^{(m)}_l = R/z^{(m)}_l$ 
decrease for increasing $m$ at fixed $l$. This is a consequence of the dependence of the
Bessel zero crossings $z^{(m)}_l$ on $m$ and $l$, and it explains why with 
increasing $m$, less and less modes $l$ give numerically significant contributions
to $w^{(m)}_{l}$.  In this way, the model discussed here illustrates the generic 
fact that higher modes $m$ correspond to increasingly finer azimuthal resolution
and higher Bessel modes $l$ correspond to increasingly finer radial resolution.

Fig.~\ref{fig2} shows also the average eccentricities $\langle \epsilon_{n,m}\rangle$ calculated for
the same sample of 1000 central Pb+Pb events. One sees that $\epsilon_{n,m}$ 
tends to increase smoothly with increasing $n$. In principle, if one would know precisely 
the $n$-dependence of $\epsilon_{n,m}$ for all infinitely many $n$'s, then one could 
reconstruct from this information the radial dependence of $w^{(m)}(r)$ analogously
to the reconstruction given from the Bessel coefficients in Fig.~\ref{fig1}. However,
the radial dependence of $w^{(m)}(r)$ is arguably much less directly characterized
by the $\epsilon_{n,m}$ than by the $w^{(m)}_{l}$. This is so, since a mode characterized by
 $w^{(m)}_{l}$ can be associated with a characteristic radial wave length $1/k^{(m)}_l$,
while a mode $\epsilon_{n,m}$ will in general receive contributions from vastly different 
length scales.
 
\subsection{Characterizing correlations between two fluctuating modes}
\label{sec3.4}
\begin{figure}[h]
\begin{center}
\includegraphics[width=8.cm]{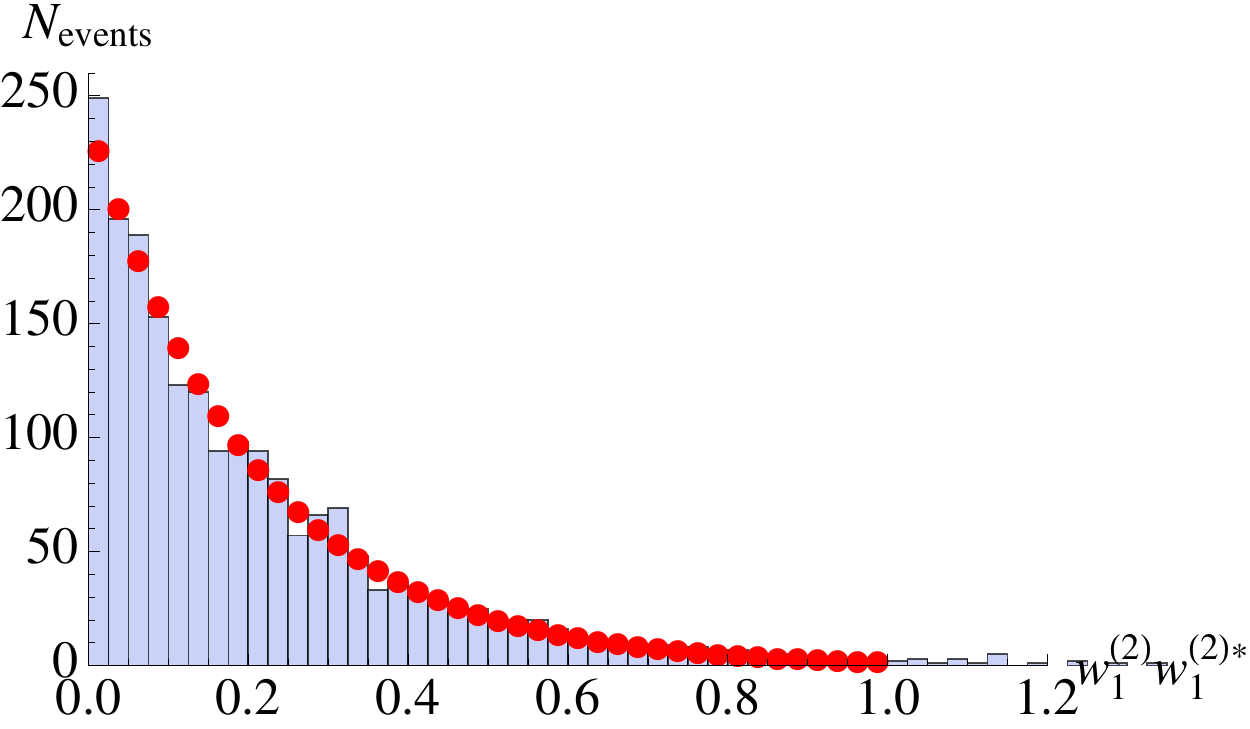}
\includegraphics[width=8.cm]{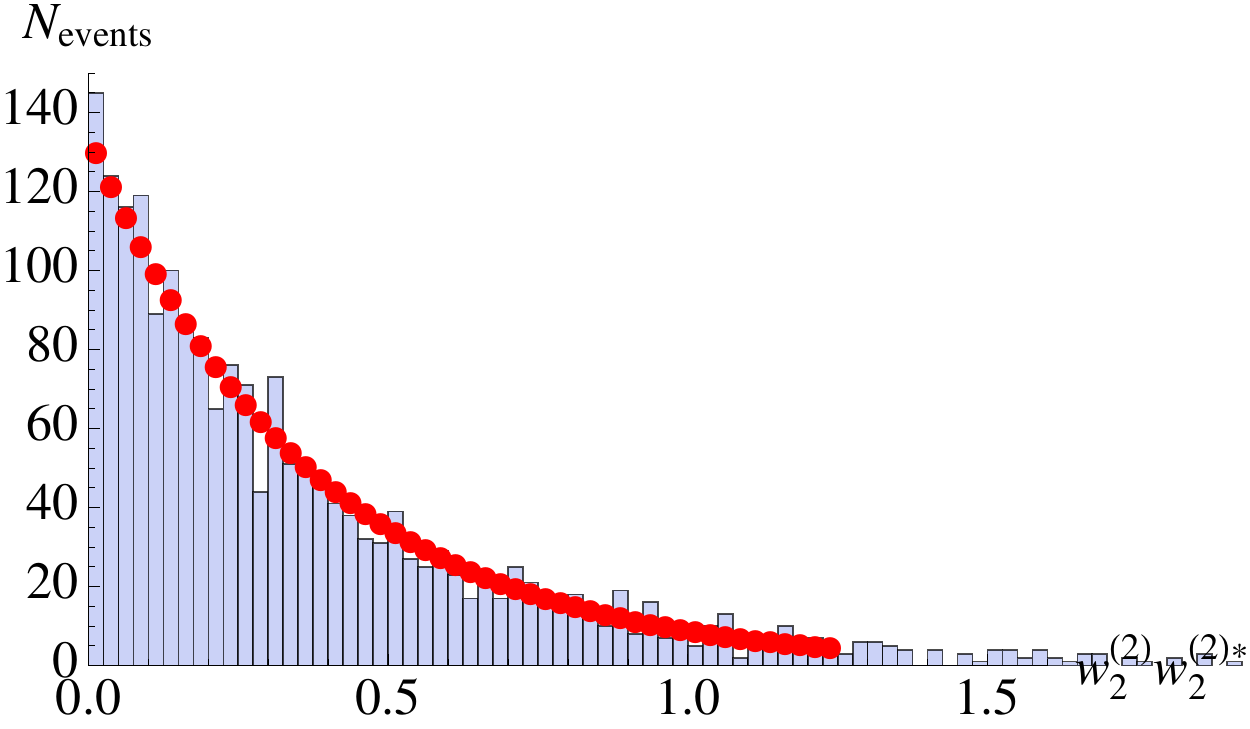}
\includegraphics[width=8.cm]{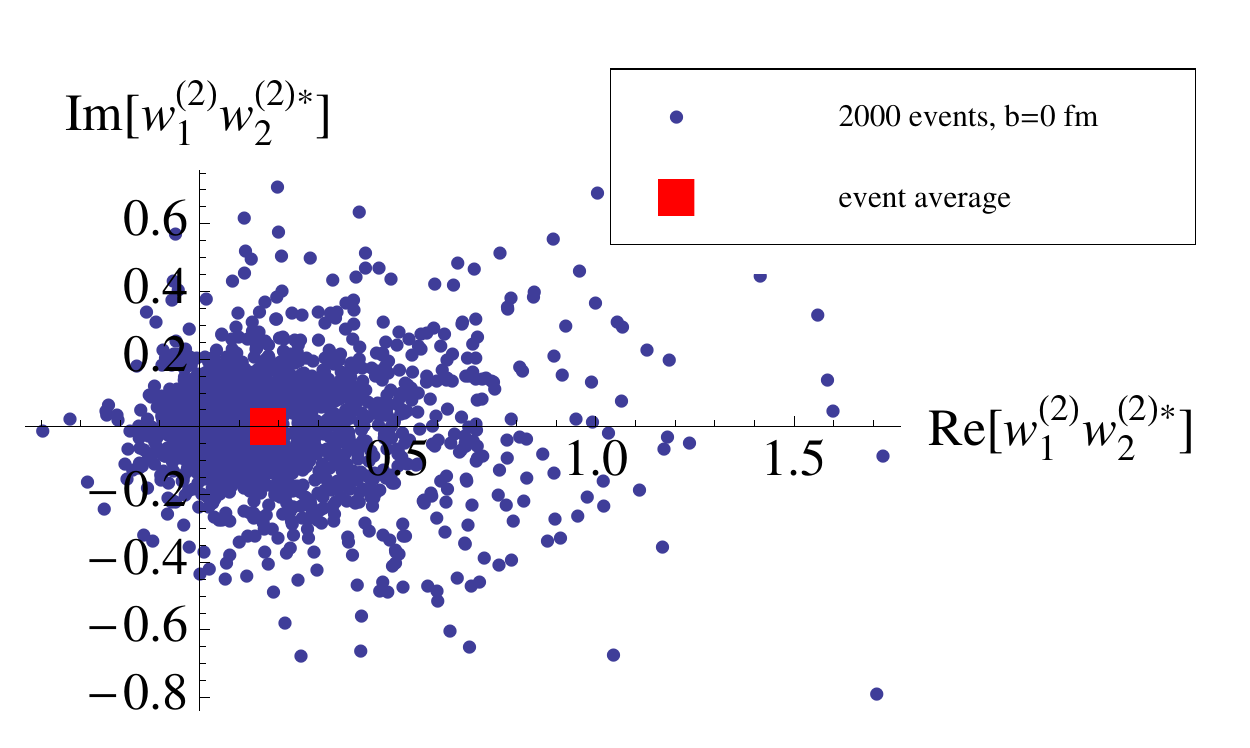}
\includegraphics[width=8.cm]{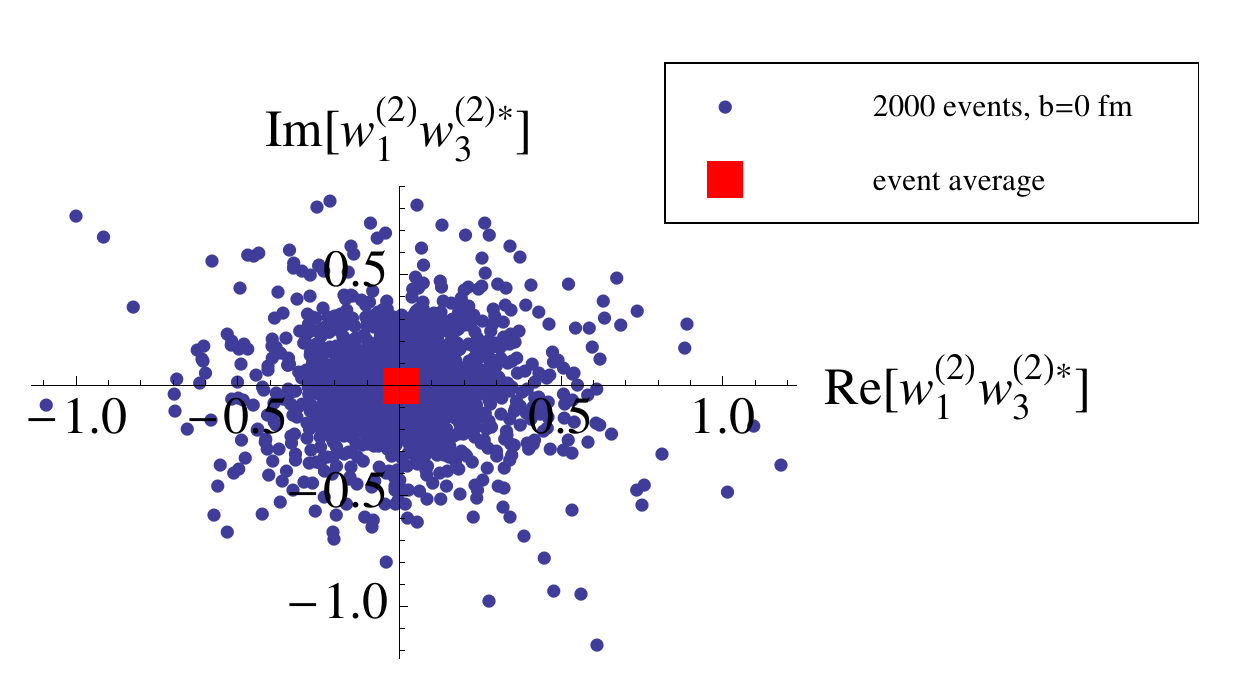}
\end{center}
\vspace{-0.5cm}
\caption{Event distributions of two-mode correlators  $w_{l_1}^{(2)}\, {w_{l_2}^{(2)}}^*$ for
the transverse enthalpy density distribution of 2000 PbPb events simulated with $b=0$. 
Upper row: distribution of diagonal two-mode correlators $w_{l}^{(2)}\, {w_{l}^{(2)}}^*$
for the same radial wave length $l= 1,2$. The simulated event distribution
(blue histogram) is compared to the analytical expectation (\ref{eq3.16}) for a Gaussian 
probability distribution (red dots). 
Lower row: Event distribution of off-diagonal two-mode correlators 
$w_{l_1}^{(2)}\, {w_{l_2}^{(2)}}^*$. The complex phase measures the 
difference $\varphi_{l_1}^{(m)} - \varphi_{l_2}^{(m)}$ in the angular orientation
of two modes in a single event. More details in text. 
}\label{fig3}
\end{figure}

The Bessel coefficients $w_l^{(m)}$ that characterize the transverse density distribution 
$w$ of a single event are complex valued. In principle, a complete characterization of 
event samples amounts to knowing of all $n$-mode correlators
$\langle w_{l_1}^{(m_1)} ...  w_{l_n}^{(m_n)*}\rangle$. As
we shall argue in section~\ref{sec3.5}, knowledge of the two-mode correlators $\langle w_{l_1}^{(m_1)}\, {w_{l_2}^{(m_2)*}}\rangle$ can provide 
a satisfactory characterization for practical purposes. For the model studied here, we
have checked numerically that correlations between different azimuthal
harmonics $m_1$, $m_2$ vanish.\footnote{This feature is actually related to a statistical azimuthal rotation invariance, see appendix \ref{sec:appC}.} Therefore, we focus in the following discussion on two-mode 
correlators  that are diagonal in the azimuthal mode $m$,
\begin{equation}
 	\langle w_{l_1}^{(m)}\, {w_{l_2}^{(m)}}^*\rangle
		= \langle \vert w_{l_1}^{(m)}\vert \, \vert {w_{l_2}^{(m)}}\vert\, 
			\exp\left[i m\, \left(\varphi_{l_1}^{(m)} - \varphi_{l_2}^{(m)}\right) \right]\rangle\, .
			\label{eq3.5}
\end{equation}
 Histograms of the event distribution of $w_{l}^{(m)}\, {w_{l}^{(m)}}^*$ are 
shown for $m=2$ and $l = 1,2$ in the upper row of Fig.~\ref{fig3}. One observes a 
distribution that does not peak at the event average, but that is of approximately exponential 
shape. We shall discuss this shape in the following subsection.  
%
%
For individual events, two fluctuations characterized by the modes $(l_1,m)$
and $(l_2,m)$ can be oriented along different azimuthal directions $\varphi_{l_1}^{(m)}
\not= \varphi_{l_2}^{(m)}$. This shows up in the complex phase
$\exp[i m\, (\varphi_{l_1}^{(m)} - \varphi_{l_2}^{(m)}) ]$ 
of the product $w_{l_1}^{(m)}\, {w_{l_2}^{(m)}}^*$ between
different radial modes $l_1 \not= l_2$, see Fig.~\ref{fig3}. 
If on average two radial modes are completely decorrelated in azimuth, then the
event distribution shows statistical azimuthal symmetry around the origin in the complex
$w_{l_1}^{(m)}\, {w_{l_2}^{(m)}}^*$-plane, and the event average 
$\langle w_{l_1}^{(m)}\, {w_{l_2}^{(m)}}^*\rangle$ vanishes. This case of 
azimuthal decorrelation is (approximately) realized for the distribution 
$\langle w_{1}^{(m)}\, {w_{3}^{(m)}}^*\rangle$ displayed in Fig.~\ref{fig3}. On the other hand,
for second azimuthal harmonics $m=2$, the radial modes $l_1=1$, $l_2=2$ show a
significant positive correlation, characterized by a non-vanishing real value of
$\langle w_{1}^{(m)}\, {w_{2}^{(m)}}^*\rangle$. The plot illustrates also that there is
a significant dispersion in phase and norm around this non-vanishing
event-averaged correlation.

\begin{figure}[h]
\begin{center}
\includegraphics[width=11.cm]{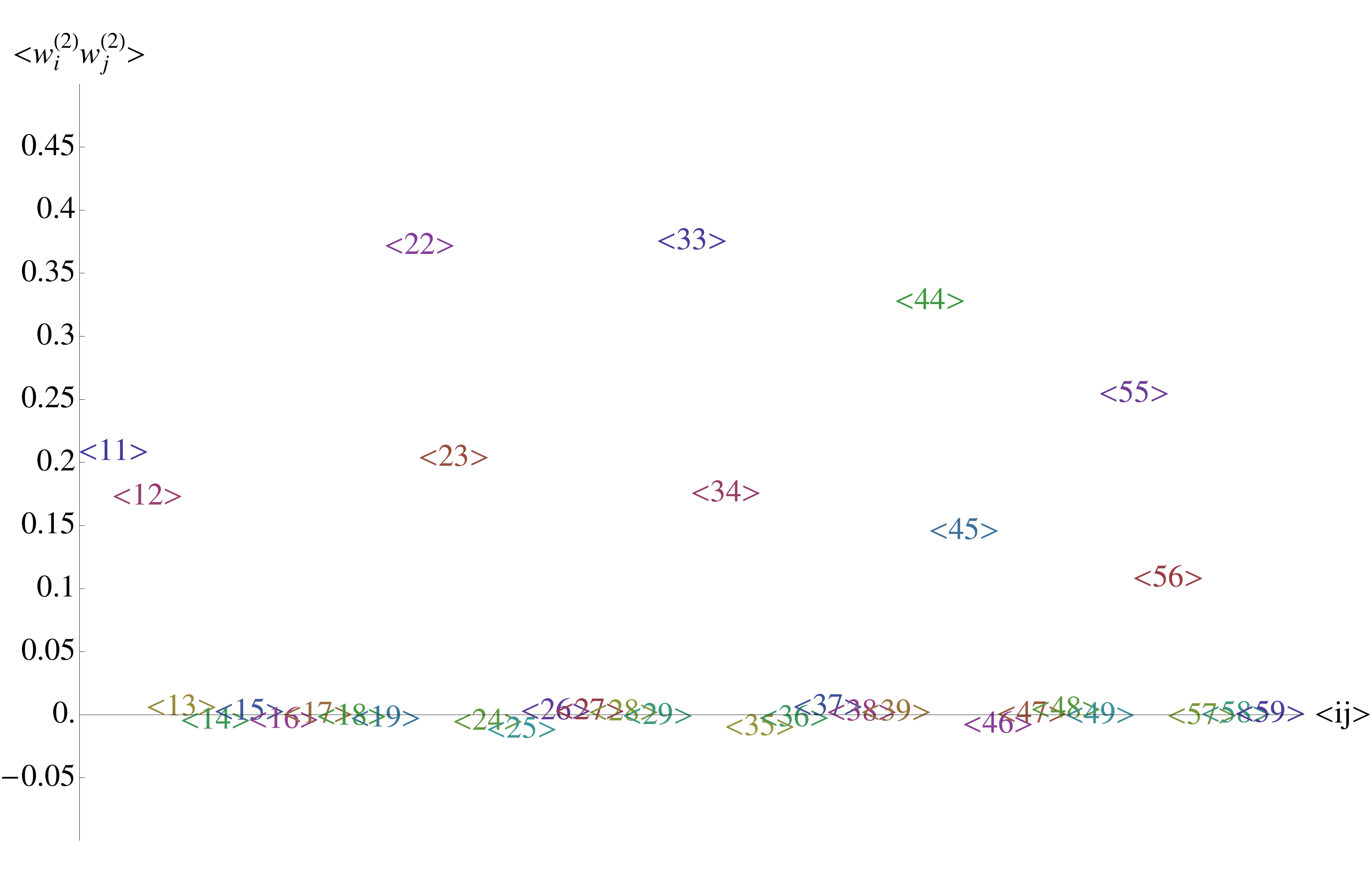}
\includegraphics[width=11.cm]{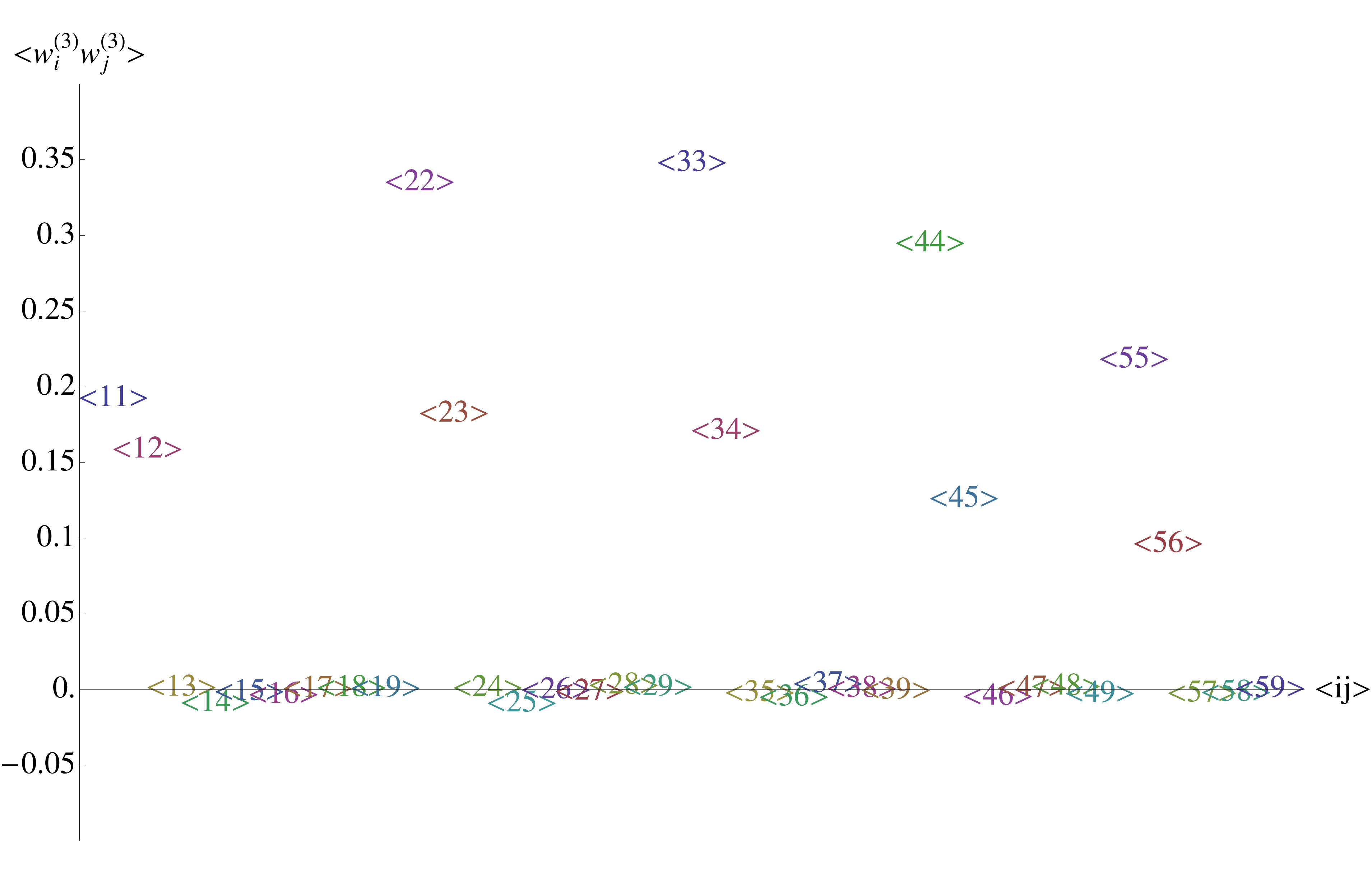}
\end{center}
\vspace{-0.5cm}
\caption{The two-mode correlator 
$\langle w_{l_1}^{(m)}\, {w_{l_2}^{(m)}}^*\rangle$ for $m=2$ (upper plot) and
$m=3$ (lower plot). The position of the brackets $\langle l_1\, l_2\rangle$ 
in these plots indicates the value taken by  $\langle w_{l_1}^{(m)}\, {w_{l_2}^{(m)}}^*\rangle$.
}\label{fig4}
\end{figure}

We have inspected the event-distributions of off-diagonal two-mode products 
$w_{l_1}^{(m)}\, {w_{l_2}^{(m)}}^*$ for azimuthal modes $1 \leq m \leq 5$ and for a
large number of radial modes $1 \leq l_1 \leq l_2 \leq 9$. Some results for the
event-averaged mean $\langle w_{l_1}^{(m)}\, {w_{l_2}^{(m)}}^*\rangle$ are shown
in Fig.~\ref{fig4}. We observe a simple and generic pattern: For fixed $m$, there is
a significant azimuthal correlation between fluctuations of neighboring radial resolution,
$l_1$, $l_2 = l_1\pm 1$. As the difference between radial resolutions increases a bit
(next-to-neighboring modes, $l_2 = l_1\pm 2$), the azimuthal correlation decreases,
and modes with even larger differences in radial resolution $l_2 = l_1\pm n$, $n\geq 3$
show essentially no azimuthal correlation. We have observed the same pattern for
$m=1, 4, 5$ and for higher radial modes $l_1$, $l_2$ (data not shown). The observed
pattern is characteristic of the nature of the fluctuations in the Glauber model of 
section~\ref{sec3.1}. In fact, all fluctuations of this model are built up of elementary uncorrelated 
Gaussian-shaped building blocks of transverse spatial width $\sigma_B = 0.4$ fm.
Event-by-event, this generates fluctuations with a variety of different radial wave lengths,
but the model does not introduce correlations between widely separated radial scales. 
For instance, for $m=2$, radial wavelengths $1/k^{(2)}_2 = R/z^{(2)}_2 = 0.95$ fm and  
$1/k^{(2)}_{2+1} = R/z^{(2)}_{2+1} = 0.68$ fm may be expected to show correlations
since the model will result in some fluctuations of transverse scale around 0.8 fm,
and since fluctuations of this scale will contribute to modes of both radial wavelengths 
$1/k^{(2)}_2  = 0.95$ fm and $1/k^{(2)}_{2+1}  = 0.68$ fm, thus leading to an azimuthal
correlation between them. However, modes of higher wave number $l$ receive contributions 
from structures on smaller scales, and since the model of 
section ~\ref{sec3.1} does not implement correlations amongst fluctuations of different
scale, the pattern observed in Fig.~\ref{fig4} appears to be a natural consequence. 
We expect that a similar pattern emerges also for alternative models of initial state
fluctuations that do not implement correlations amongst fluctuations of very different
scale. 

\subsection{Characterizing event ensembles via functional probability distributions}
\label{sec3.5}
Event samples can be characterized fully via the probability distribution ${\cal P} \left[w \right]$ 
of their initial transverse densities, where ${\cal P} \left[w \right]$ is a functional over the space 
of all functions $w$. Here, we discuss what can be said about the structure of ${\cal P}$.
In the previous subsections, we have seen that the initial transverse density $w$ of each event 
can be characterized completely in terms of the set of complex-valued Bessel coefficients $\lbrace w_l^{(m)}\rbrace$. Therefore, one can characterize an event sample also by a probability distribution  ${\cal P}\left[ \lbrace w_l^{(m)} \rbrace \right]$ that is a function of the set of all $\lbrace w_l^{(m)} \rbrace$. Event-averages of the 
 coefficients $w_l^{(m)}$ and its correlators are then defined by 
 \begin{eqnarray}
    &&\langle w_{l_1}^{(m_1)} \dots w_{l_i}^{(m_i)} 
    w_{l'_1}^{(m'_1)*} \dots w_{l'_j}^{(m'_j)*}\rangle \nonumber \\
    && \quad = \int {\cal D}\left[\lbrace w_l^{(m)} \rbrace\right]
    w_{l_1}^{(m_1)} \dots w_{l_i}^{(m_i)} 
    w_{l'_1}^{(m'_1)*} \dots w_{l'_j}^{(m'_j)*}
     {\cal P}\left[ \lbrace w_l^{(m)} \rbrace \right],\;\;\;\;
     \label{eq3.6}
 \end{eqnarray}
where $ {\cal D}\left[\lbrace w_l^{(m)}\rbrace\right]$ defines the
integration measure 
\begin{equation}
 {\cal D}\left[\lbrace w_l^{(m)}\rbrace\right] \equiv
  \prod_{m=-N_m}^{N_m}  \prod_{l=1}^{N_l} {\rm d}w_l^{(m)}\, .
       \label{eq3.7}
\end{equation}
\subsubsection{Gaussian probability distribution}
For the case of a Gaussian distribution ${\cal P} \left[w \right]$, one has (see also appendix \ref{sec:appC})
\begin{equation}
 	P\left[ \lbrace w_l^{(m)} \rbrace \right] = {\cal N}\, 
	\exp\left[ -\frac{1}{2} \sum_{m_1,m_2=-N_m}^{N_m} \sum_{l_1,l_2=1}^{N_l} \left(w_{l_1}^{(m_1)} 
	- \langle w_{l_1}^{(m_1)} \rangle \right)^*\, 
	T^{(m_1)(m_2)}_{l_1\, l_2}\, 
	\left(w_{l_2}^{(m_2)} - \langle  w_{l_2}^{(m_2)} \rangle \right) \right]\, .
	     \label{eq3.8}
 \end{equation}
 Here, ${\cal N}$ is an appropriate normalization factor.
 The matrix $T$ and the averages $\langle w_l^{(m)} \rangle$
 in (\ref{eq3.8}) are determined by,
 \begin{eqnarray}
  \langle w_{l_1}^{(m_1)} \rangle &=& \int {\cal D}\left[\lbrace w_l^{(m)} \rbrace\right]\, 
    w_{l_1}^{(m_1)} \, {\cal P}\left[ \lbrace w_l^{(m)} \rbrace \right]\, ,
    	     \label{eq3.9}\\
    \langle w_{l_1}^{(m_1)} w_{l_2}^{(m_2)*}\rangle 
	&=& \int {\cal D}\left[\lbrace w_l^{(m)} \rbrace\right]\, 
     w_{l_1}^{(m_1)}\, w_{l_2}^{(m_2)*}\, 
     {\cal P}\left[ \lbrace w_l^{(m)} \rbrace \right] \nonumber\\
     &=& \left( {T}^{-1}\right)_{l_1\, l_2}^{(m_1)(m_2)}
         + \langle w_{l_1}^{(m_1)}\rangle\, 
         \langle w_{l_2}^{(m_2)}\rangle\, .
     	     \label{eq3.10}
 \end{eqnarray}
For collisions at vanishing impact parameter, the azimuthal symmetry of 
event-averages implies that 
 \begin{equation}
  \langle w_{l}^{(m)} \rangle = \delta_{m0}\, w_{l, {\rm average}}^{(0)}\, .
    	     \label{eq3.11}
 \end{equation}
At finite impact parameter, also
non-trivial azimuthal modes can have non vanishing event averages,
$\langle w_{l}^{(m)} \rangle \not= 0$ for $m \not= 0$. However, 
since event-averaged distributions do not display structures at small 
wave-length, one expects generically that $\langle w_{l}^{(m)}\rangle$
is non-negligible only for very small $l$.

In general, since ${\cal P}$ is real, the matrix ${T}$ in (\ref{eq3.8}) is hermitian,
and a non-vanishing complex phase of off-diagonal elements 
$\left( { T}^{-1}\right)_{l_1\, l_2}^{(m_1)(m_2)}$ measures the difference $\varphi_{l_1}^{(m_1)} -\varphi_{l_2}^{(m_2)}$ between the azimuthal orientations of different modes. However, the matrix ${ T}$ is
real and symmetric if the ensemble is symmetric with respect to the parity transformation $\varphi\to-\varphi$. The statistical azimuthal rotation symmetry for $b=0$ implies further that 
the matrix ${T}$ is diagonal in the azimuthal modes $m$,
 \begin{equation}
 	\left( { T}^{-1}\right)_{l_1\, l_2}^{(m_1)(m_2)}
	= \delta_{m_1\, m_2}\, \left( { T}^{-1}\right)_{l_1\, l_2}^{(m_1)}.
		     \label{eq3.12}
 \end{equation}
Within the Monte Carlo Glauber model we observe in Fig.~\ref{fig4} that two-mode correlations decrease quickly 
 with increasing difference in the radial wavelengths of the two modes, that is
 \begin{equation}
  \big\vert\left( { T}^{-1}\right)_{l\, , l}^{(m)}\big\vert >
   \big\vert\left( { T}^{-1}\right)_{l\, ,l+1}^{(m)}\big\vert >
   \big\vert\left( { T}^{-1}\right)_{l\, ,l+2}^{(m)}\big\vert > \dots ,
   	     \label{eq3.13}
 \end{equation}
 and we expect that this feature is shared by other models, as well.

 \subsubsection{Testing the validity of the Gaussian approximation of ${\cal P}$}
 The Gaussian probability distribution (\ref{eq3.8}) is fully specified in terms of the event averages
 $\langle w_{l_1}^{(m_1)}\rangle$, $\langle w_{l_1}^{(m_1)} w_{l_2}^{(m_2)*}\rangle$, and it provides
 a simple ansatz for the  event-wise distribution of arbitrary products  
 $w_{l_1}^{(m_1)} \dots w_{l_i}^{(m_i)}     w_{l'_1}^{(m'_1)*} \dots w_{l'_j}^{(m'_j)*}$. 
 Here, we derive within the Gaussian approximation explicit expressions for some
 of these event-wise distributions, and we establish for the model of section~\ref{sec3.1} that these 
 event-wise distributions are correctly described by the ansatz (\ref{eq3.8}). The practical interest in this 
 statement is that to the extent to which the Gaussian approximation (\ref{eq3.8}) holds, the 
 small set of numbers $\langle w_{l}^{(m)}\rangle$ and 
 $\langle w_{l_1}^{(m_1)}\, w_{l_2}^{(m_2)*}\rangle$ provides then
 complete information not only about all event averages, but also about 
 the functional shapes of all event distributions.
 
 In the recent letter\cite{Floerchinger:2013rya}, we have shown that experimentally measurable flow coefficients
 in nucleus-nucleus collisions can be written as the fluid dynamic response to diagonal and off-diagonal products
 of two modes 
 \begin{eqnarray}
  \xi_a &\equiv& w_{l_a}^{(m)}w_{l_a}^{(m)*}\, ,\nonumber\\
   \chi_{ab} &\equiv& w_{l_a}^{(m)}\, w_{l_b}^{(m)*}\, ,\nonumber
 \end{eqnarray}
 and their event averages.
 We are therefore particularly interested in the event distributions of $\xi_a$ and $\chi_{ab}$
 around the averages $\langle w_{l_a}^{(m)}w_{l_a}^{(m)}\rangle$, $\langle w_{l_a}^{(m)}\, w_{l_b}^{(m)}\rangle$.
 Here, $l_a$ and $l_b$ denote arbitrary but fixed radial wave numbers. For an arbitrary probability distribution
 $P\left[ \lbrace w_l^{(m)} \rbrace \right]$, the distribution in $\xi_a$ and $\chi_{ab}$ can be calculated from 
 \begin{eqnarray}
 	{\cal P}_{\xi}\left( \xi_a\right)
	&=& \int {\cal D}\left[\lbrace w_l^{(m)} \rbrace\right]\, 
         {\cal P}\left[ \lbrace w_l^{(m)} \rbrace \right]\, 
         \delta\left(\xi_a - w_{l_a}^{(m)}\, w_{l_a}^{(m)*} \right)\, ,
     \label{eq3.14}\\
     {\cal P}_{\chi}\left( \chi_{ab} \right)
	&=& \int {\cal D}\left[\lbrace w_l^{(m)} \rbrace\right]\, 
         {\cal P}\left[ \lbrace w_l^{(m)} \rbrace \right]\, 
         \delta^{(2)}\left(\chi_{ab} - w_{l_a}^{(m)}\, w_{l_b}^{(m)*}\right)\, .
     \label{eq3.15}
 \end{eqnarray}
 These are probability distributions in the real variable $\xi_a$ and the complex variable $\chi_{ab}$, respectively. 
For the Gaussian ansatz (\ref{eq3.8}), the integrals in (\ref{eq3.14}), (\ref{eq3.15})  can be done analytically. 
For the diagonal product $\xi_a$, one finds 
 \begin{equation}
 	{\cal P}_{\xi}\left[ \xi_a\right] = \sigma_w e^{-\sigma_w\xi_a}\, 
	I_0\left(2\sqrt{\xi}\, \sigma_w\, \langle w_l^{(m)}\rangle \right)\, 
	e^{-\sigma_w \langle w_{l_a}^{(m)} \rangle^2}\, \Theta(\xi_a)\, , 
	\label{eq3.16}
 \end{equation}
 where the inverse width $\sigma_w$ is given by 
 \begin{equation}
 	\sigma_w = \frac{1}{\langle w_{l_a}^{(m)}\, w_{l_a}^{(m)*}\rangle} = \frac{1}{\langle \xi_{a}\rangle}\, .
	\label{eq3.17}
 \end{equation}
In this subsection, we focus on 
the case of vanishing impact parameter, for which all averages 
$\langle w_l^{(m=2)} \rangle$ vanish, and equation (\ref{eq3.16}) reduces to an exponential 
${\cal P}_{\xi}\left[ \xi_a\right] = \sigma_w \exp(-\sigma_w\xi_a)\, \Theta(\xi_a)$. 
In Fig.~\ref{fig3}, we demonstrate that once the event-average $\langle w_{l_a}^{(m)}\, w_{l_a}^{(m)*}\rangle$
is specified, this provides a parameter-free and accurate desciption 
of the event wise distribution of $\xi_a$ in the model of section~\ref{sec3.1}. 

For the event distribution in the complex variable $\chi_{ab}=\chi_{ab}^r + i\; \chi_{ab}^i$, 
we find for vanishing impact parameter,
(i.e. for $\langle w_l^{(m=2)} \rangle = 0$) 
\begin{equation}
	{\cal P}_{\chi_{ab}}\left(\chi_{ab}^r,\chi_{ab}^i\right) =
	\frac{{\rm det}{\cal T}}{2\pi}\, 
	K_0\left(\sqrt{{\cal T}_{l_al_a}{\cal T}_{l_bl_b}\chi_{ab}\chi_{ab}^*}\right)\, 
	\exp\left[ -{\cal T}_{l_al_b}\, \chi_{ab}^r \right]\, .
	\label{eq3.18}
\end{equation}
Here, ${\cal T}$ denotes the two-dimensional submatrix obtained from ${ T}_{l_1l_2}$, $l_1,l_2 = l_a, l_b$.
In this way, both distributions (\ref{eq3.17}) and (\ref{eq3.18}) are specified completely in terms of 
the one-mode and two-mode correlators eqs.~(\ref{eq3.9}) and (\ref{eq3.10}). For the model studied
here, the finite number of relevant event-averages is shown in Fig.~\ref{fig4}.  

%
\begin{figure}[h]
\begin{center}
\includegraphics[width=8.cm]{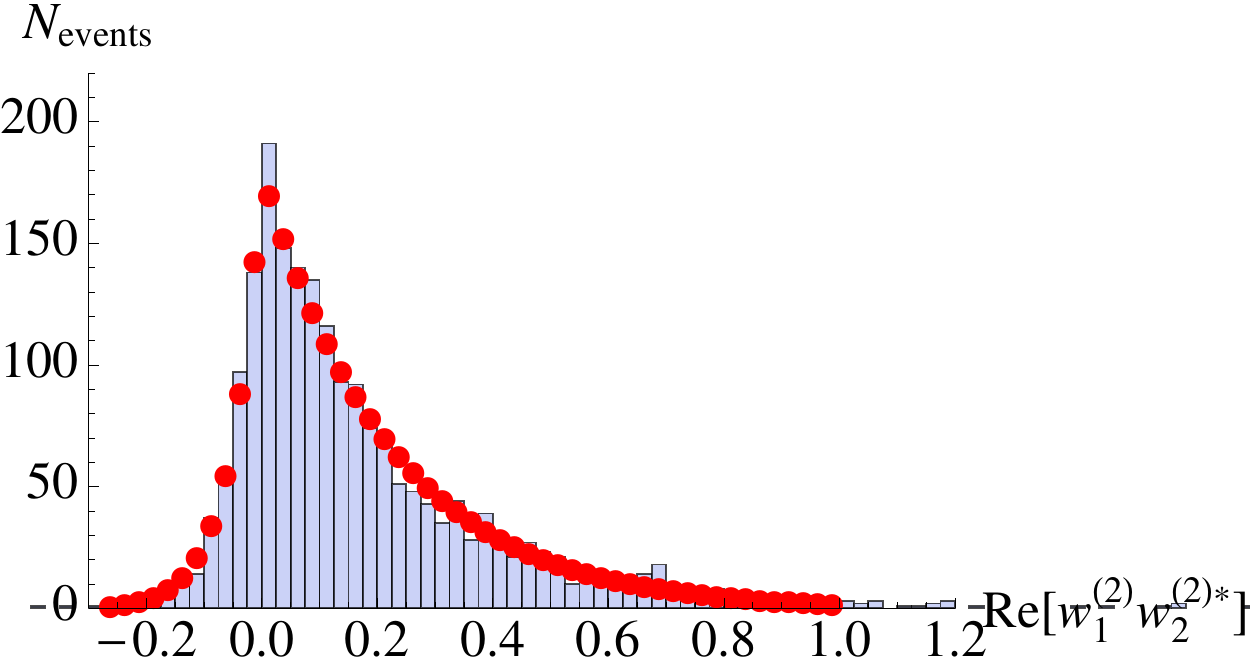}
\includegraphics[width=8.cm]{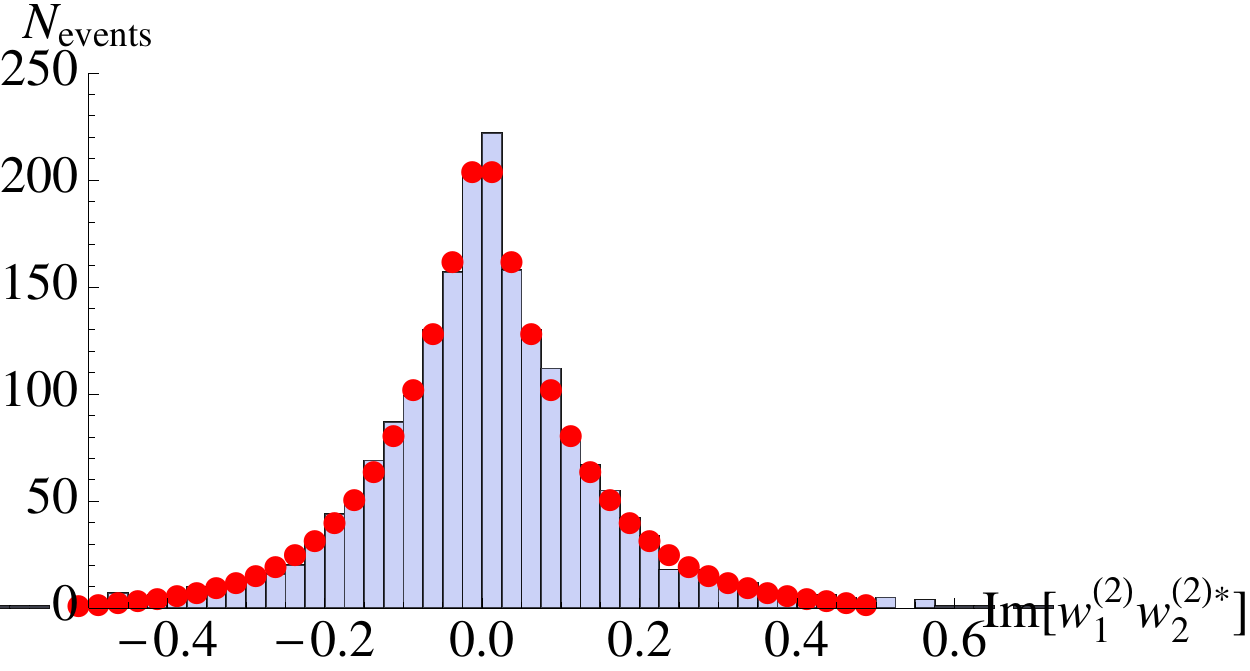}
\includegraphics[width=8.cm]{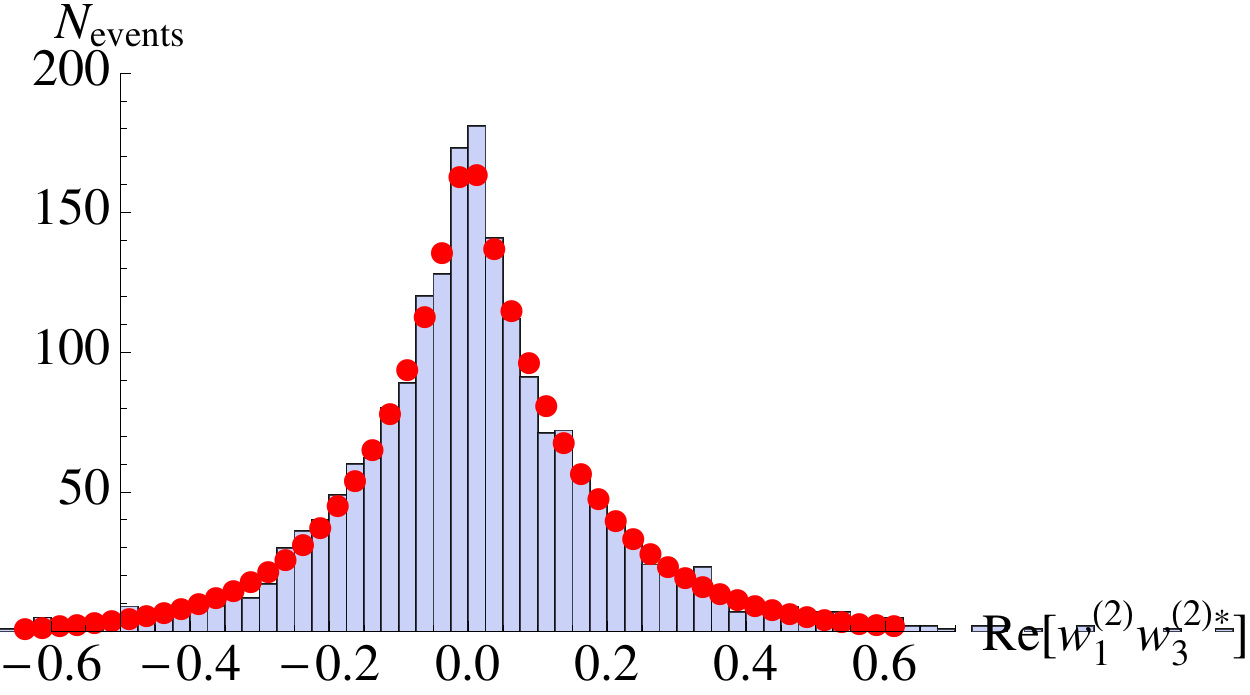}
\includegraphics[width=8.cm]{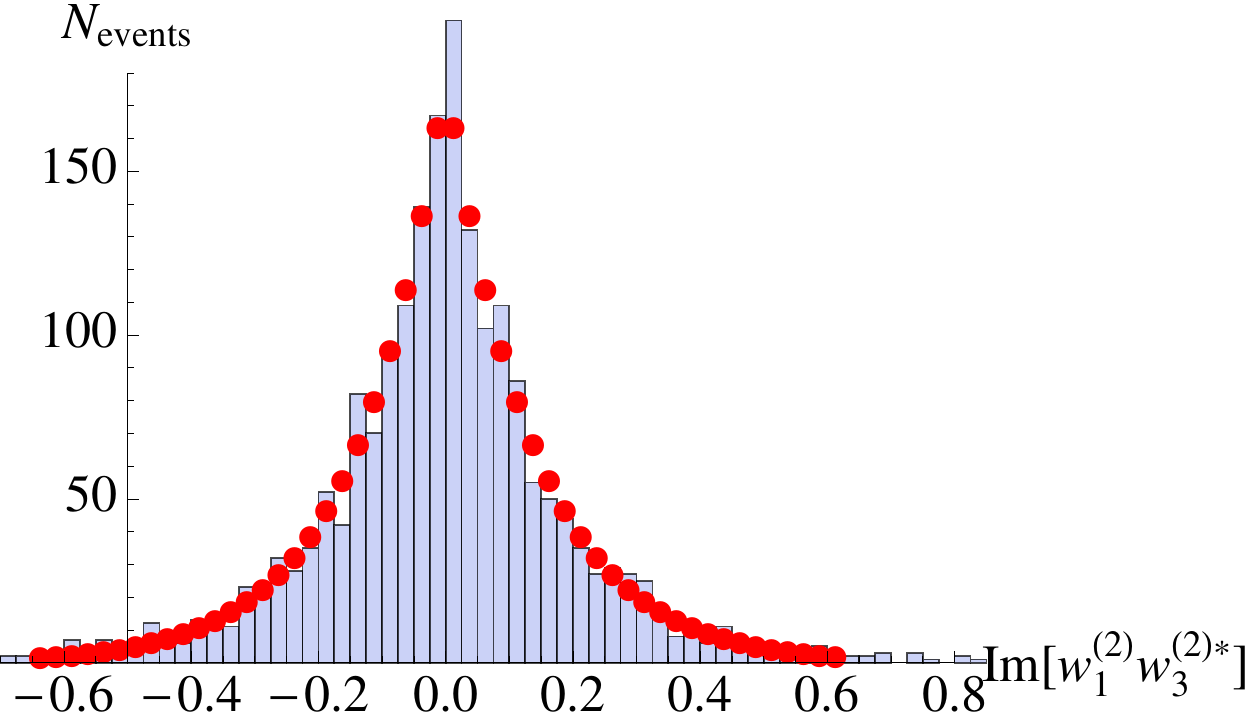}
\end{center}
\vspace{-0.5cm}
\caption{The event distributions of off-diagonal products of two modes
$\chi_{ab} = w_{l_a}^{(m)}\, w_{l_b}^{(m)}$,
shown in the lower column of Fig.~\ref{fig3}, projected on the real and imaginary
axis, respectively.  Results from the simulation of 2000 events (blue histogram) are
compared to analytical results (red dots) obtained from integrating (\ref{eq3.18}). 
}\label{fig5}
\end{figure}
%
To visualize the comparison of (\ref{eq3.18}) to event distributions simulated in the model of section~\ref{sec3.1}, 
we show in Fig.~\ref{fig5} histograms of one-dimensional projections of the off-diagonal two-mode event
distributions plotted in
Fig.~\ref{fig3}. These are compared to the corresponding projection of (\ref{eq3.18}).
We find that the Gaussian approximation
(\ref{eq3.8}) accounts very satisfactorily for the shape of event distributions of off-diagonal 
two-point correlators $\chi_{ab}$, too. For the case of vanishing impact
parameter, these studies indicate that  in practice a small number of two-mode correlations is sufficient to specify fully the shape of event distributions of all products 
of two modes around 
these averages.
 \begin{figure}[h]
\begin{center}
\includegraphics[width=6.cm]{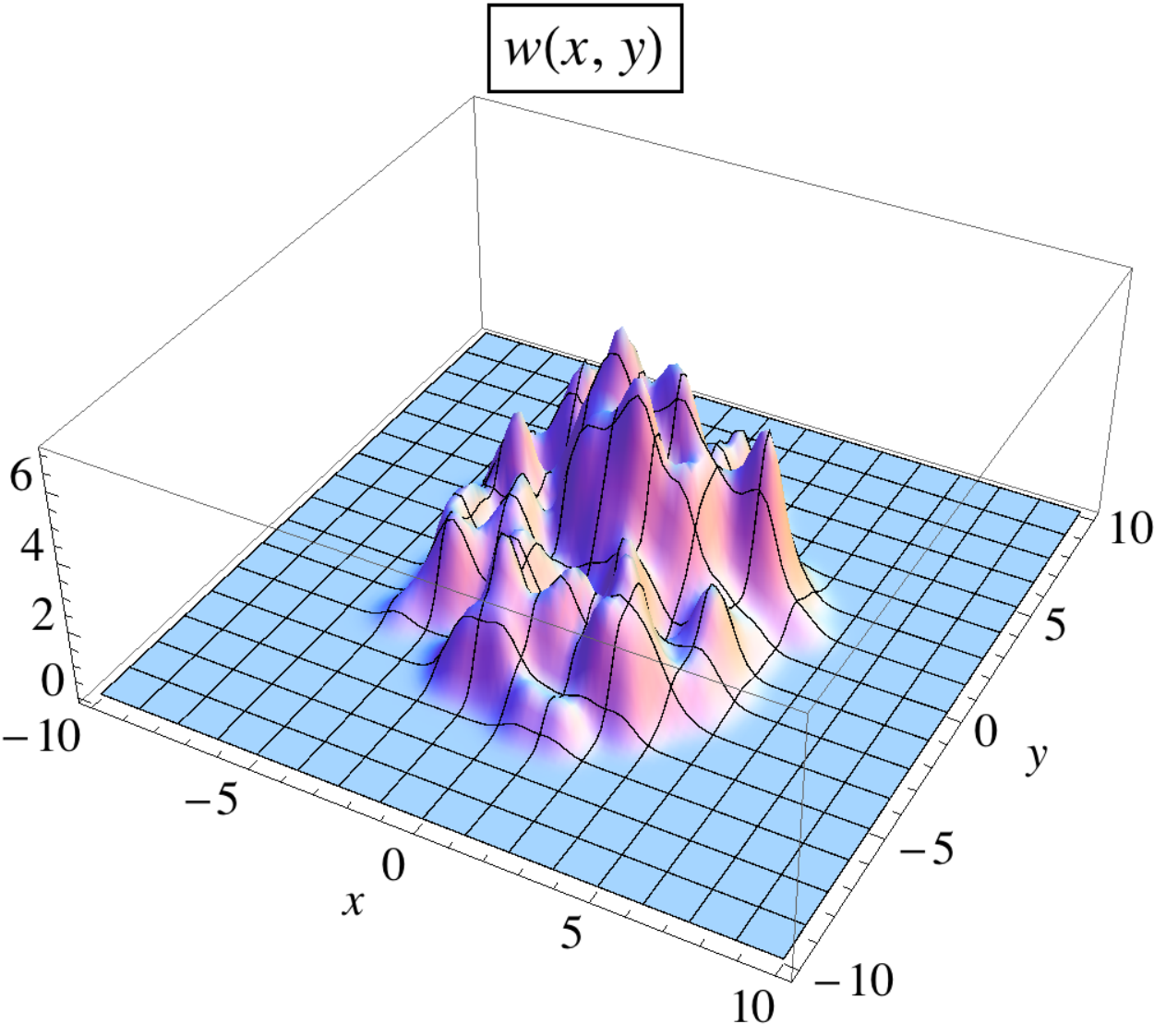}
\includegraphics[width=6.cm]{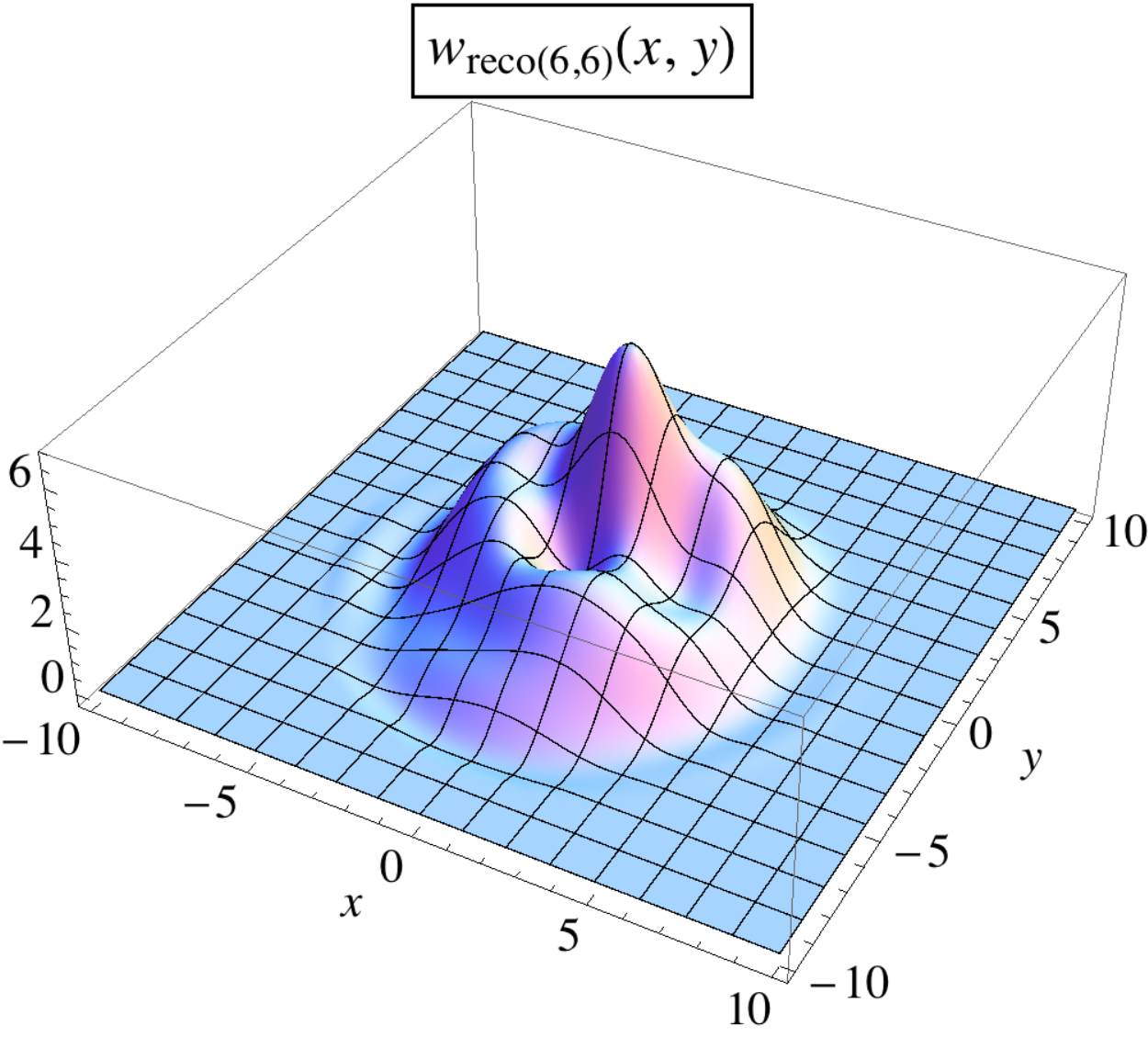}
\includegraphics[width=6.cm]{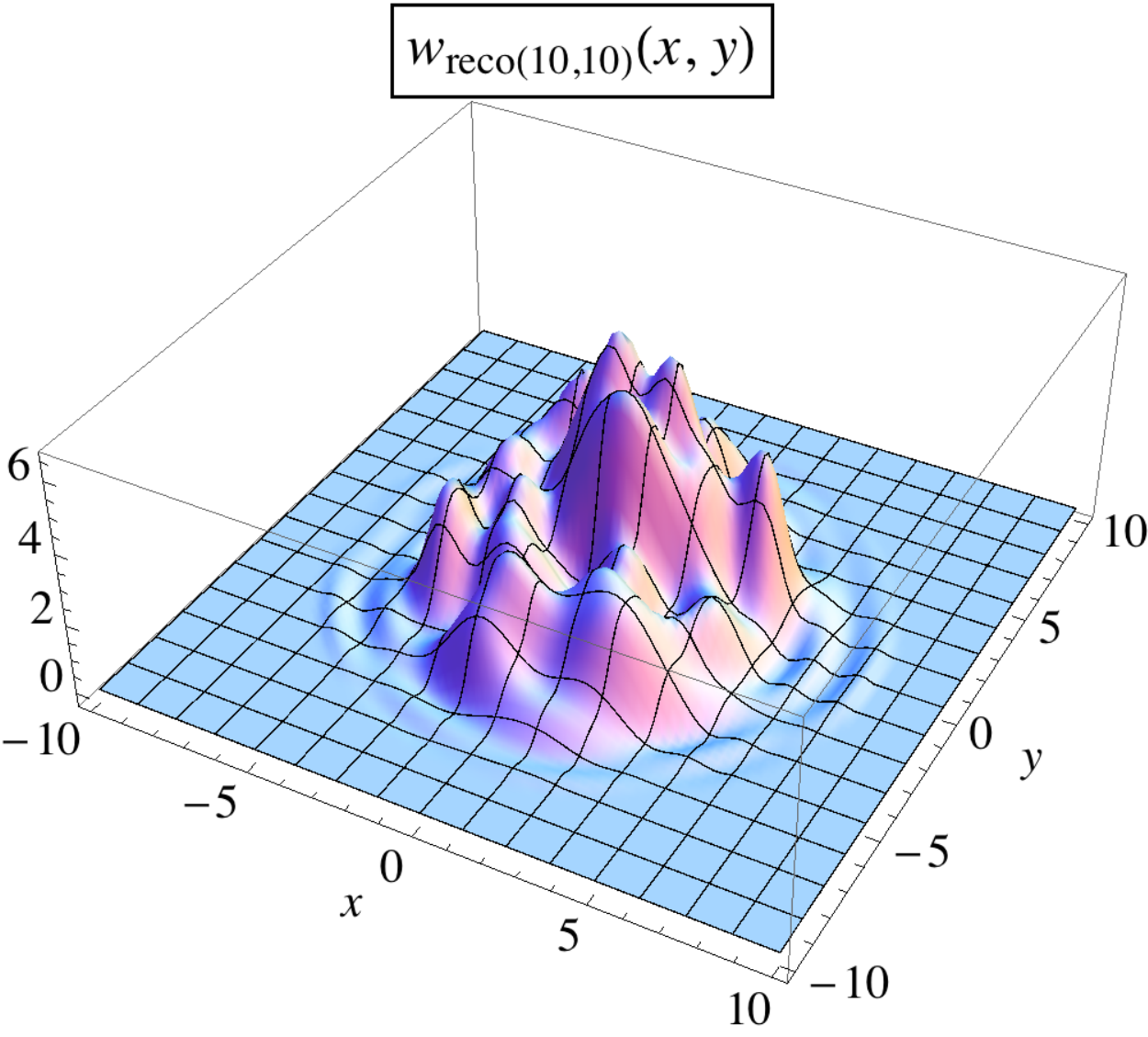}
\includegraphics[width=6.cm]{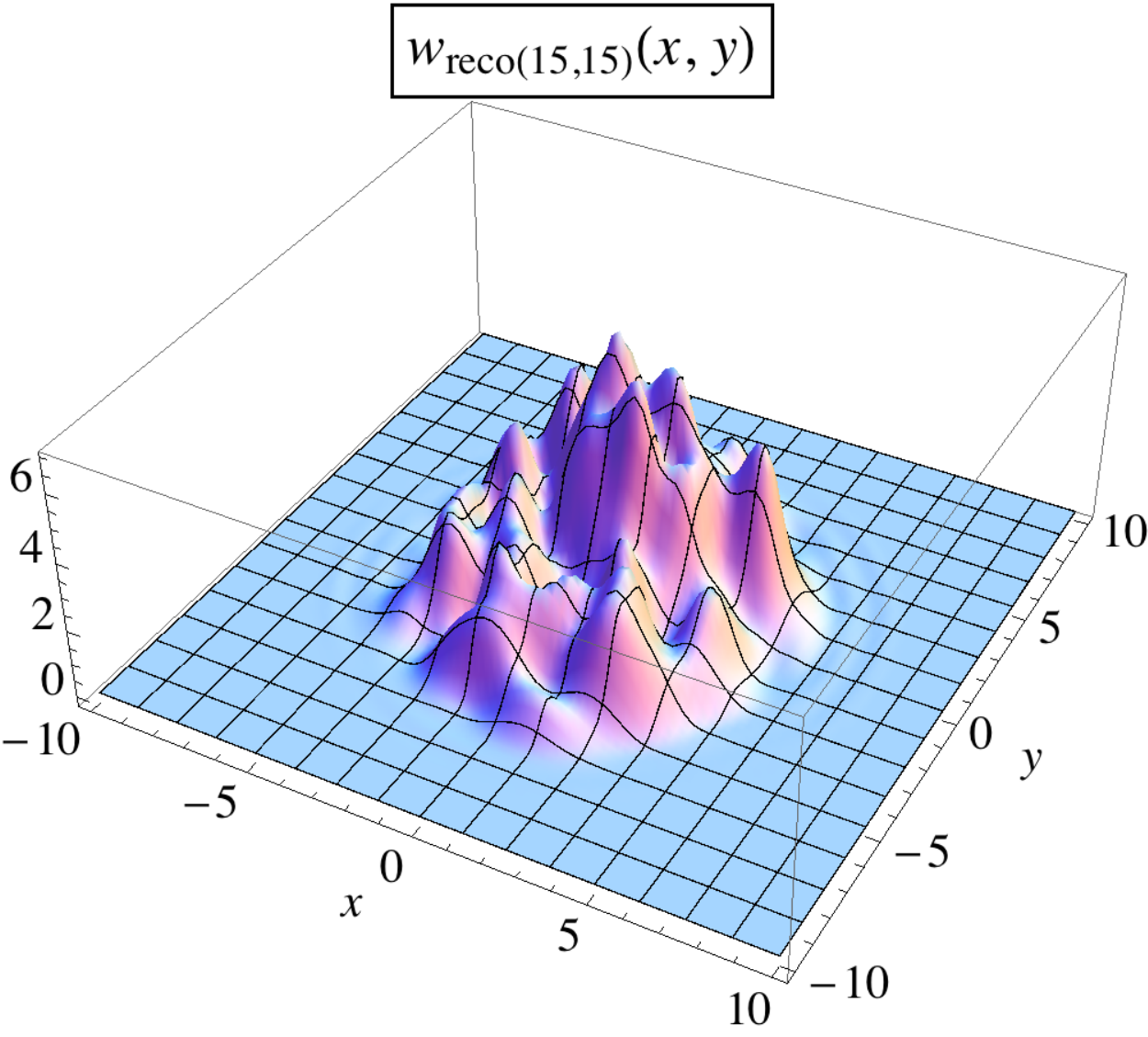}
\includegraphics[width=6.cm]{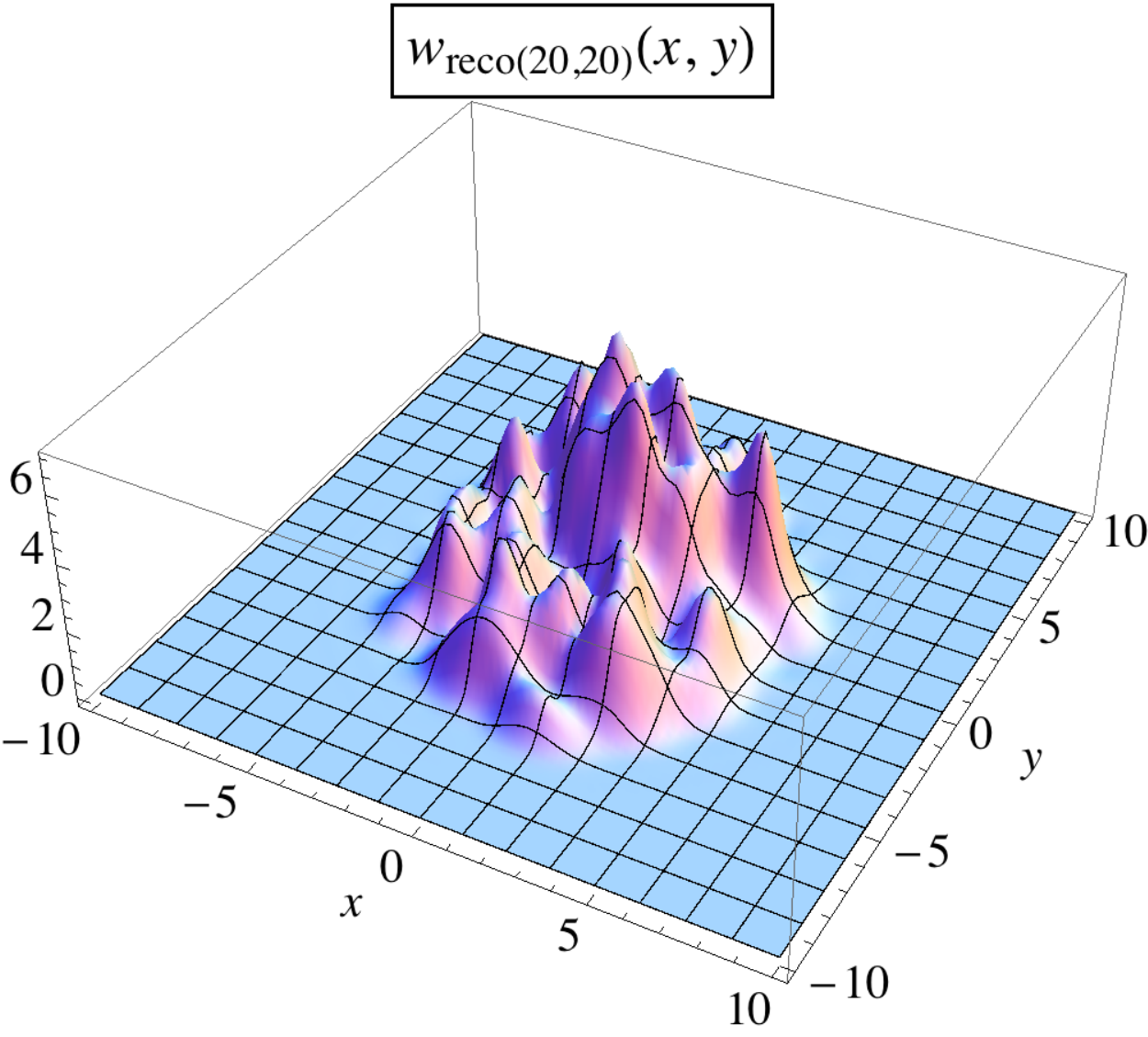}
\includegraphics[width=6.cm]{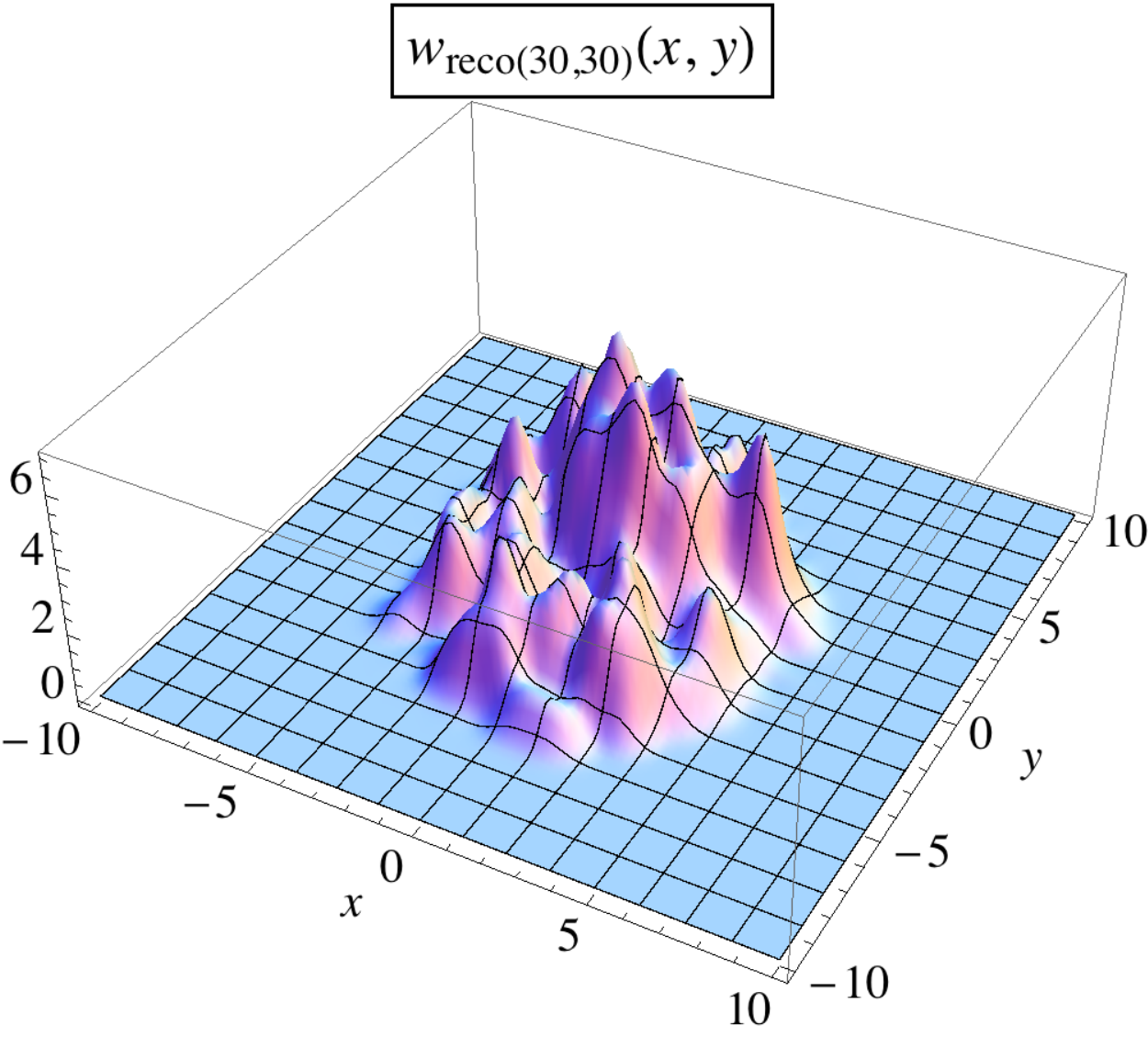}
\end{center}
\vspace{-0.5cm}
\caption{Same as Fig.~\ref{fig1} but  for one randomly chosen Pb+Pb collision
at finite impact parameter $b=6$ fm.
Upper left plot: Transverse density distribution $w(x,y)$ simulated according to the 
Glauber model as described in the text. 
Remaining five plots: Reconstruction $w_{{\rm reco}(N_m,N_l)}$ of this particular density 
distribution from the data of a discrete Bessel transformation of $\rho(x,y)$, 
involving an increasing number of modes  in the azimuthal ($N_m$) and in the radial 
($N_l$) direction. 
}\label{fig6}
\end{figure}

 \subsection{Lemoine's mode decomposition at finite impact parameter: a numerical study for $b=6$ fm}
So far, we have focussed on heavy ion collisions at vanishing impact parameter,
for which event averages are azimuthally symmetric. In this section, we show that
Lemoine's method applies equally well to characterizing initial conditions at finite
impact parameter. To demonstrate this, we repeat in the following the study of 
subsections~\ref{sec3.3}-~\ref{sec3.5} for Pb+Pb collisions at impact parameter $b=6$ fm.
Our discussion will be brief, and we focus only on those points that arise anew at 
finite impact parameter.

In Fig.~\ref{fig6}, we show the density distribution of a randomly chosen Pb+Pb collision
at $b= 6$ fm. In comparison to the collision at vanishing impact parameter, seen in 
Fig.~\ref{fig1}, the active transverse area is clearly smaller. We have checked that 
Lemoine's method characterizes the simulated densities with comparable accuracy 
irrespective of impact parameter. In particular, the point-by-point differences between
the true enthalpy density $w(x,y)$ and the reconstruction $w_{{\rm reco}(N_m,N_l)}$ are less than
$1 \%$ of the maximal density for a reconstruction with $N_m = N_l = 30$. 
%
 \begin{figure}[h]
\begin{center}
\includegraphics[width=7.cm]{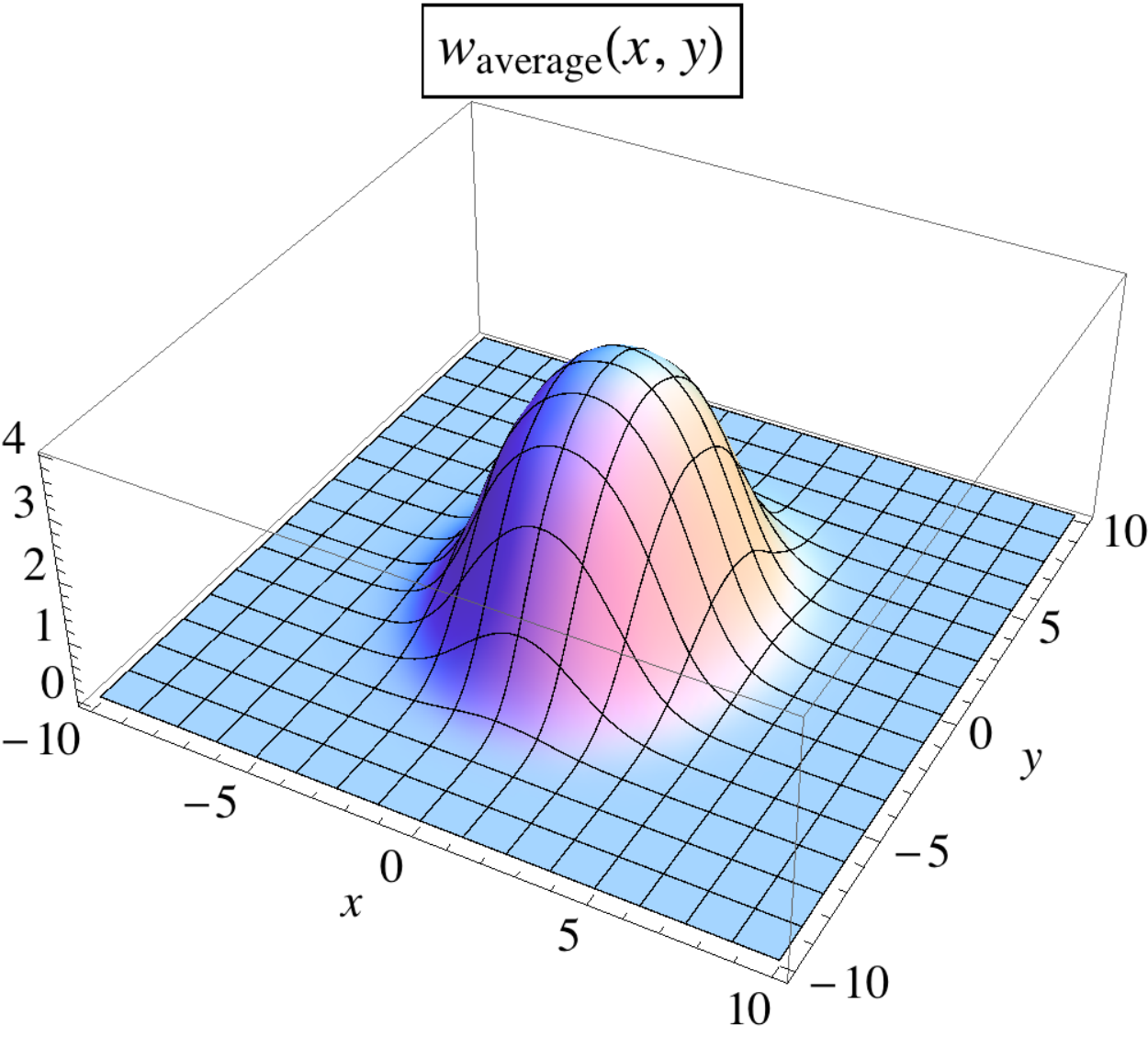}
\includegraphics[width=7.cm]{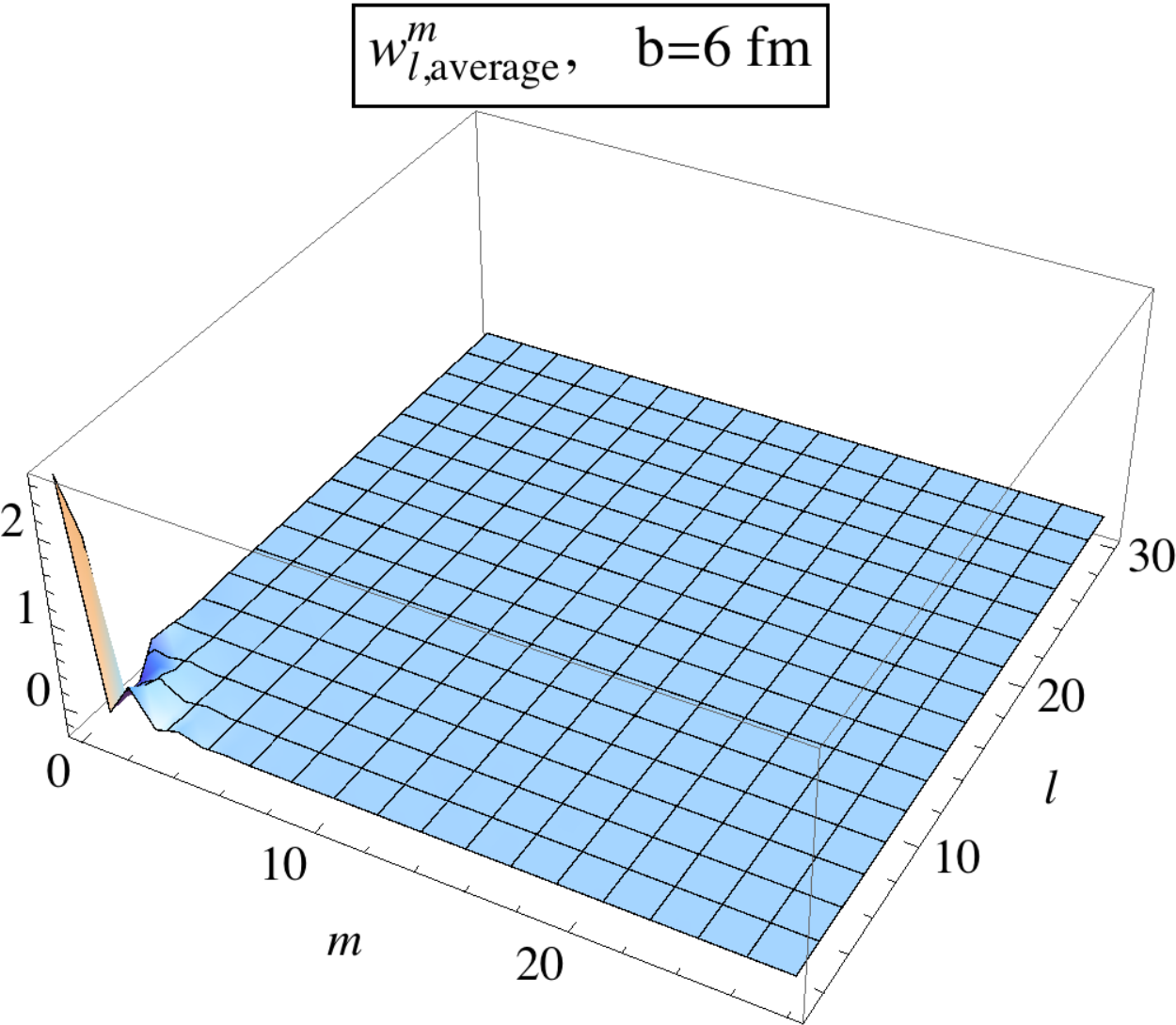}
\includegraphics[width=7.cm]{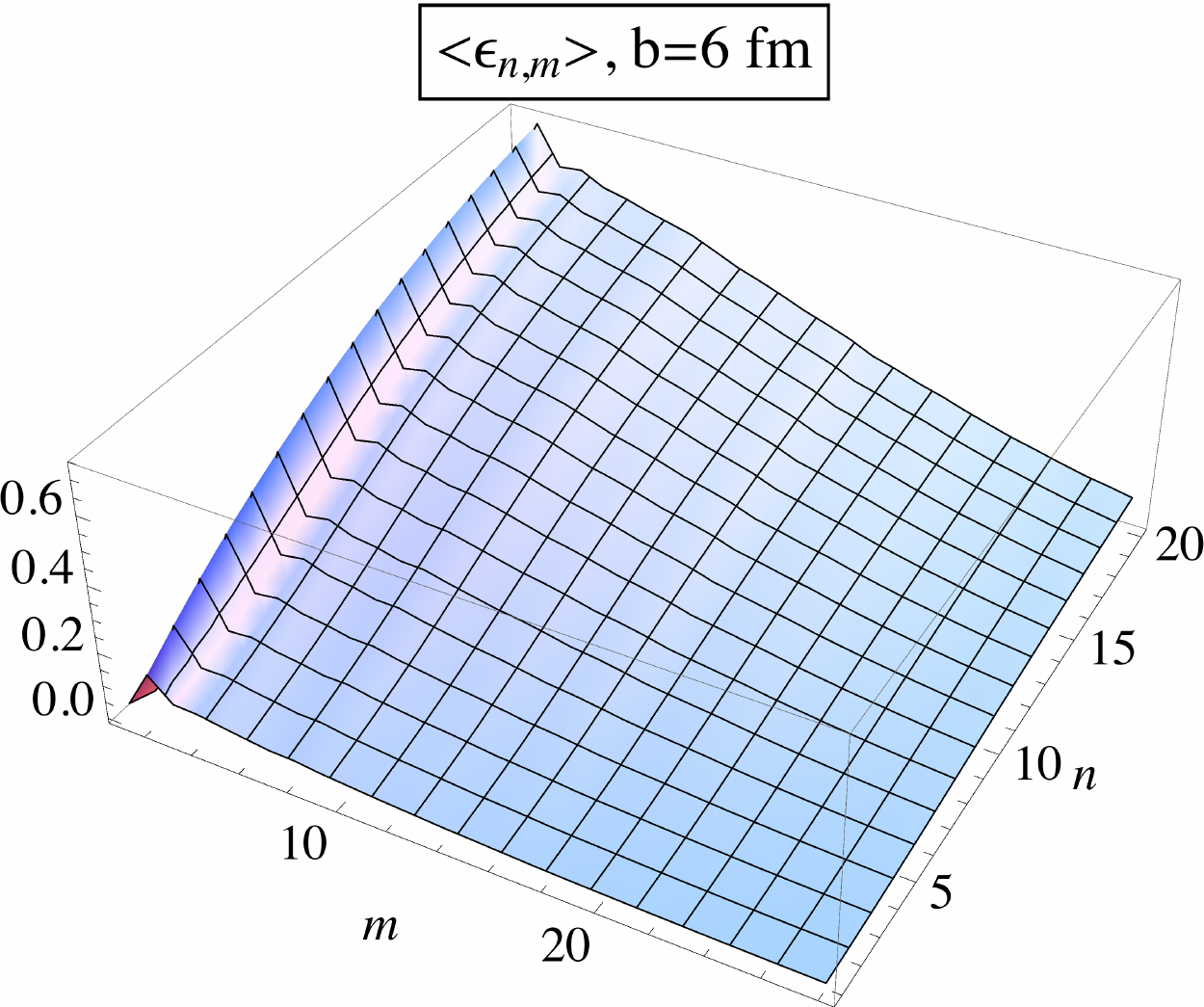}
\includegraphics[width=7.cm]{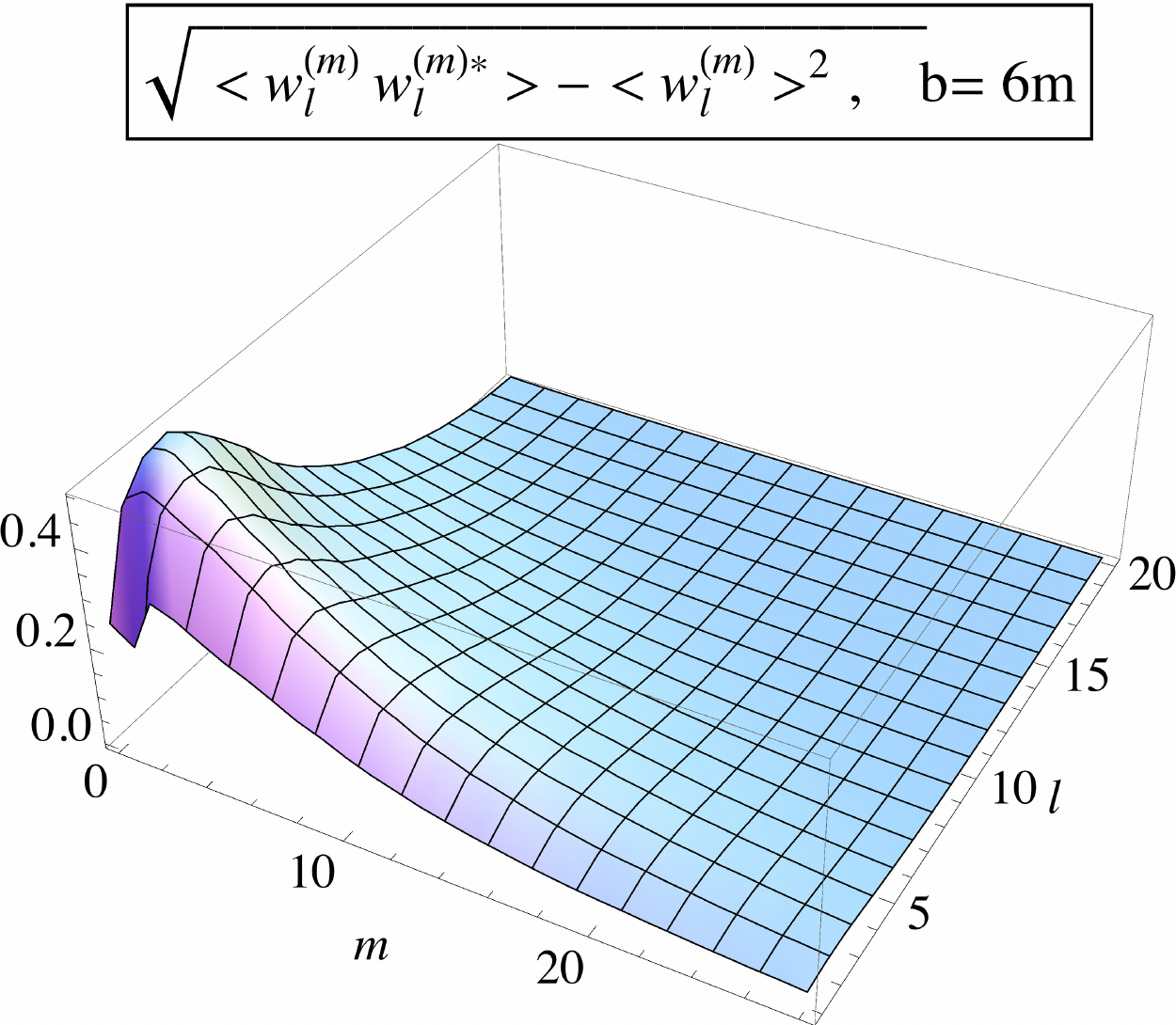}
\end{center}
\vspace{-.5cm}
\caption{Same information as Fig.~\ref{fig2} but for event averages of Pb+Pb collisions 
at impact parameter $b=6$ fm. Upper row: event averaged density distribution
$w_{\rm average}$ and corresponding one-mode averages $\langle w_l^{(m)}\rangle$.
Lower column: event-averaged eccentricities and dispersions around 
$\langle w_l^{(m)}\rangle$.
}\label{fig7}
\end{figure}

Fig.~\ref{fig7} shows some elementary characterizations of event averages at finite
impact parameter when the event-averaged density $w_{\rm average}$ has an approximately ellipsoidal shape that breaks azimuthal symmetry. As a consequence, there are non-vanishing one-mode event averages
 $\langle w_l^{(m)}\rangle$ also for even integers $m\not= 0$. The elliptic variation of 
the average involves only long wavelengths, and therefore $\langle w_l^{(m)}\rangle$ 
takes non-vanishing values only for small $l$. The event-averaged eccentricities 
$\langle \epsilon_{n,m}\rangle$, can be understood as being composed of a 
fluctuating component that is comparable to the one shown for $b=0$ in Fig.~\ref{fig2},
and an  event-averaged non-fluctuating component that contributes to $m = 2$ 
(and slightly to $m=4$) for all values of $n$ and that increases these
coefficients significantly. In the dispersion $\sqrt{\langle w_l^{(m)}\, w_l^{(m)*}\rangle \, -
\langle w_l^{(m)}\rangle^2}$, the non-fluctuating contribution is subtracted by construction.
Comparing this plot to the corresponding one in Fig.~\ref{fig2}, one see that the fluctuations 
around $\langle w_l^{(m)}\rangle$ are similar at vanishing and non-vanishing impact
parameter, although the event averages $\langle w_l^{(m)}\rangle$ are characteristically
different. 

\begin{figure}[h]
\begin{center}
\includegraphics[width=10.cm]{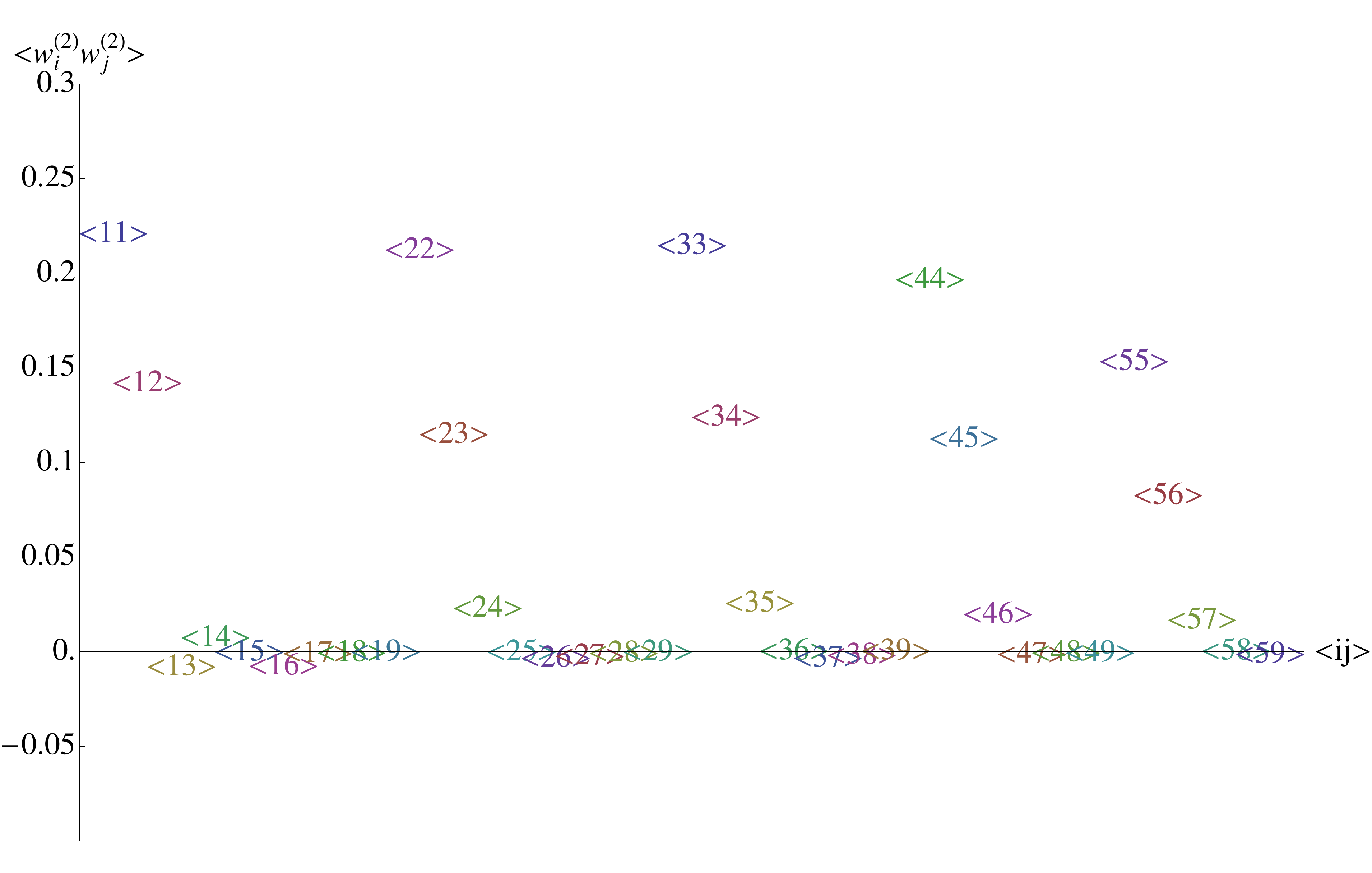}
\includegraphics[width=10.cm]{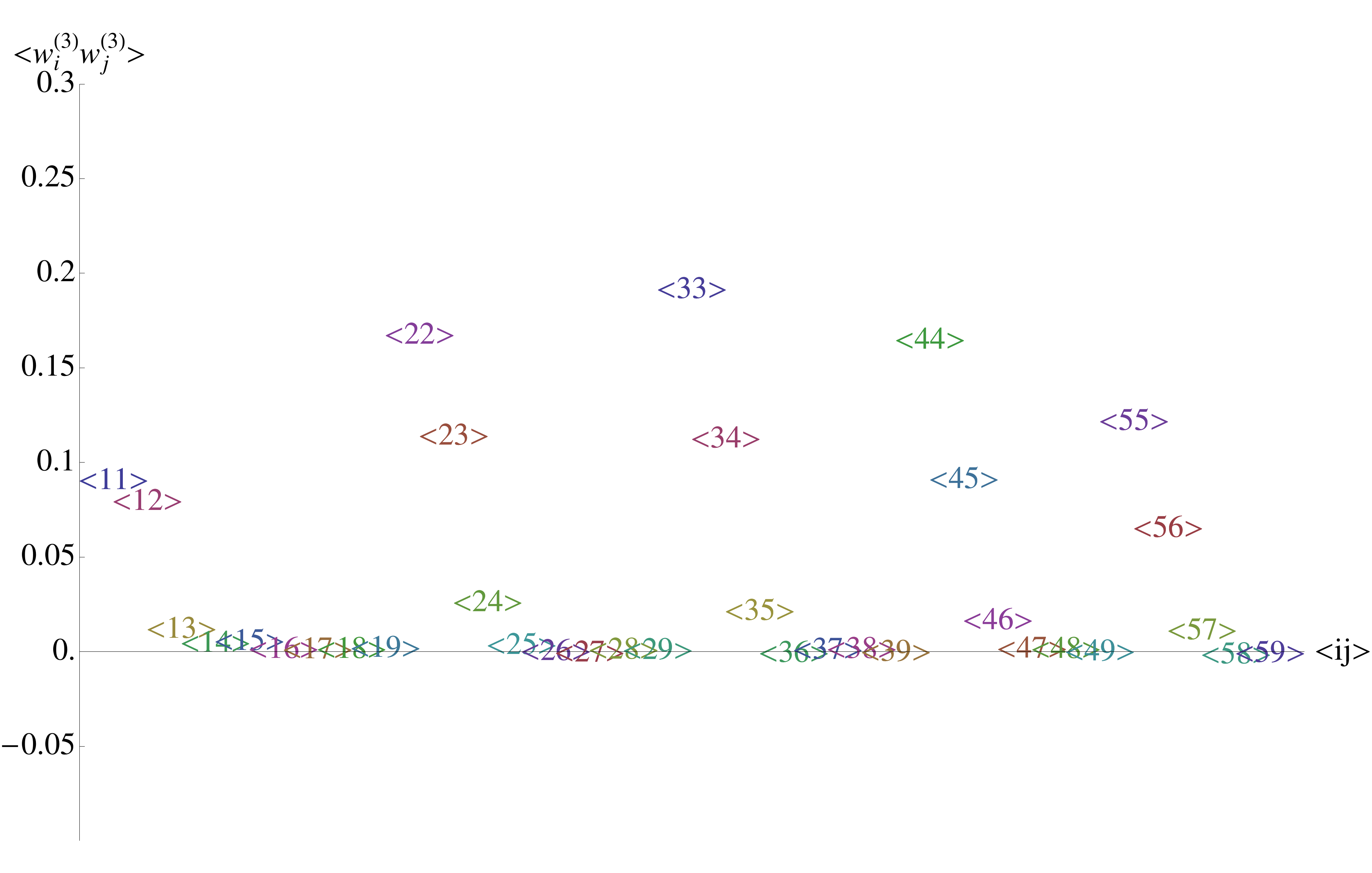}
\end{center}
\vspace{-0.5cm}
\caption{Same as Fig.~\ref{fig3} but for finite impact parameter $b= 6$ fm.
}\label{fig8}
\end{figure}

Also the two-mode correlators $\langle w_{l_1}^{(m_1)}\, w_{l_2}^{(m_2)*}\rangle$
show remarkable similarities between collisions at finite impact parameter (see Fig.~\ref{fig8})
and collisions at vanishing impact parameter (see Fig.~\ref{fig3}). The overall normalization
of all $\langle w_{l_1}^{(m_1)}\, w_{l_2}^{(m_2)*}\rangle$ increases with the total enthalpy
of the distribution, and it is thus larger  for the case $b=0$. 
Compared to the case for $b=0$, however, the relative weight of 
$\langle w_{l_1}^{(m=2)}\, w_{l_2}^{(m=2)*}\rangle$ is significantly increased in the
longest wavelength modes $l_1,l_2 = 1,2$ that characterize the event-averaged
azimuthal anisotropy of $w_{\rm average}$. 
%
\begin{figure}[h]
\begin{center}
\includegraphics[width=10.cm]{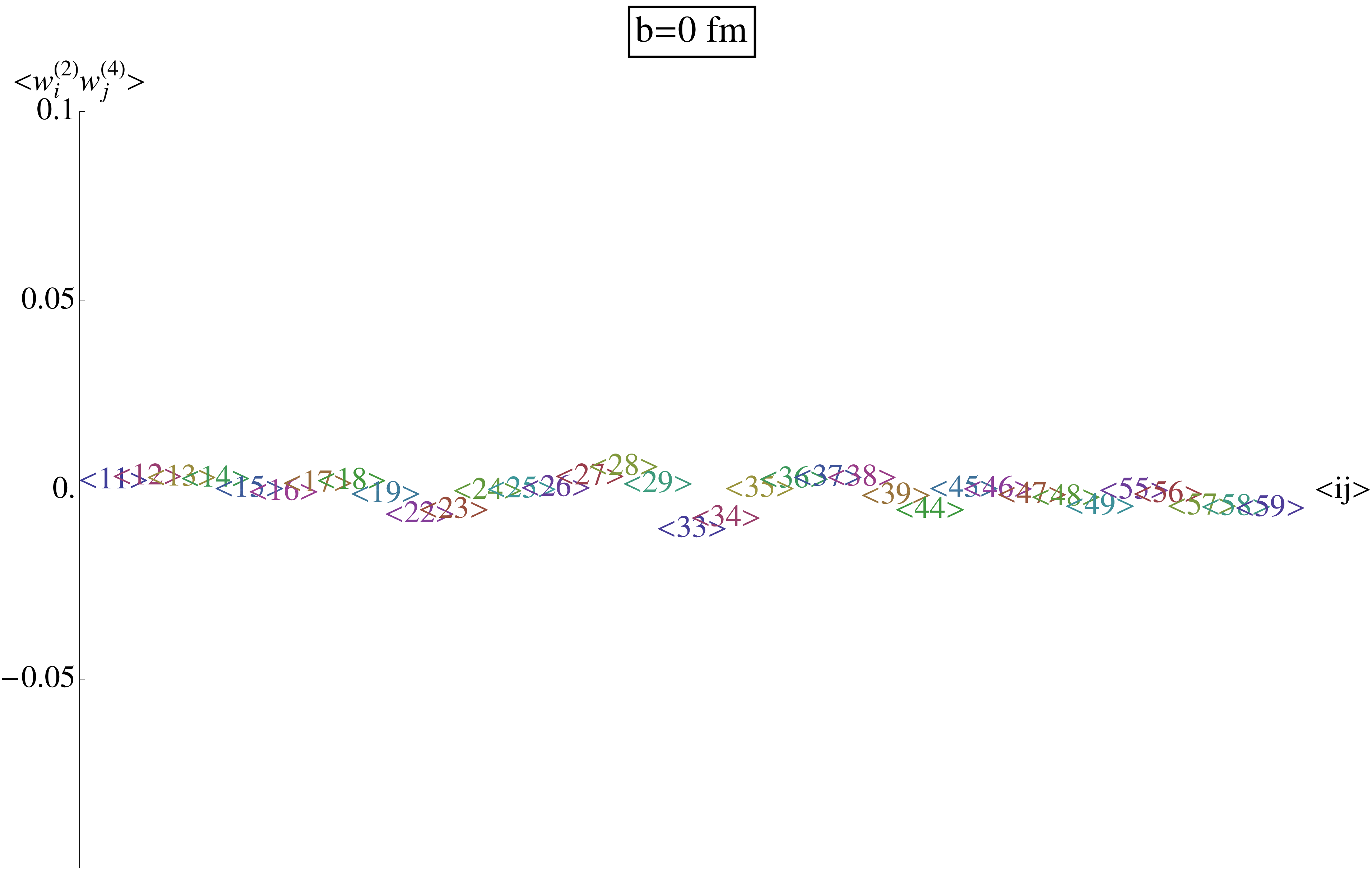}
\includegraphics[width=10.cm]{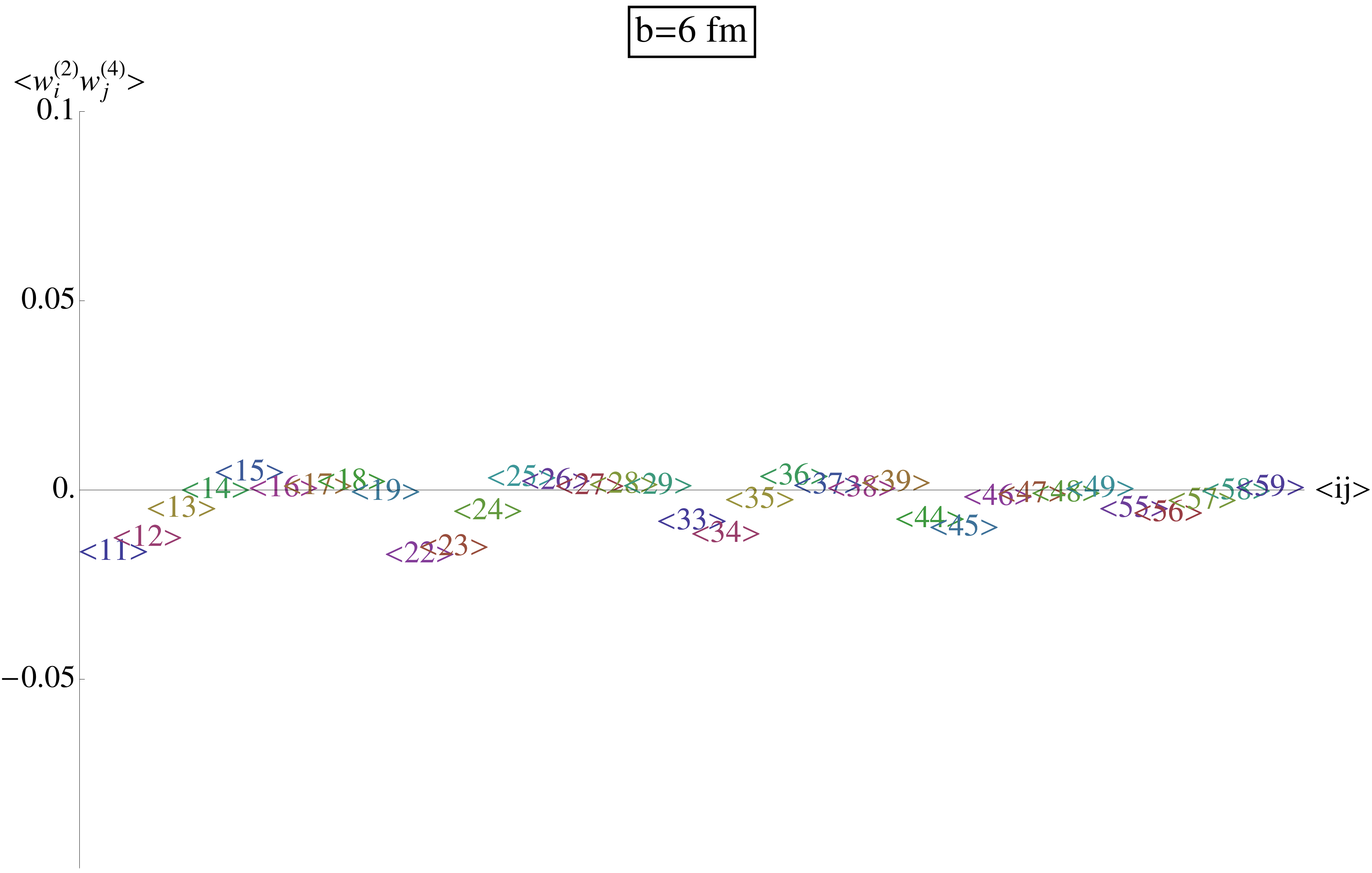}
\end{center}
\vspace{-0.5cm}
\caption{The two-mode correlators $\langle w_{l_1}^{(m_1)}\, w_{l_2}^{(m_2)*}\rangle$
for different harmonics $m_1=2$,
$m_2 = 4$ vanish for vanishing impact parameter (upper plot)  but can take
non-zero values at finite impact parameter (lower plot).
}\label{fig9}
\end{figure}

At finite impact parameter, there can be non-vanishing event-averaged
two-mode correlators $\langle w_{l_1}^{(m_1)}\, w_{l_2}^{(m_2)*}\rangle$ also for modes 
corresponding to different azimuthal harmonics $m_1 \not= m_2$. In particular,
the event-averaged shape of $w_\text{average}$ at finite impact parameter contributes
not only to the second but also to the fourth azimuthal harmonics and this leads to
non-vanishing correlations $\langle w_{l_1}^{(2)}\, w_{l_2}^{(4)*}\rangle$. As seen
in Fig.~\ref{fig9}, such correlations vanish for $b=0$ within statistical uncertainties,
but they are found at finite impact parameter in the model studied here. 
The strength of these correlations is weak if compared to correlations for modes 
at the same azimuthal harmonics. 
%
\begin{figure}[h]
\begin{center}
\includegraphics[width=7.cm]{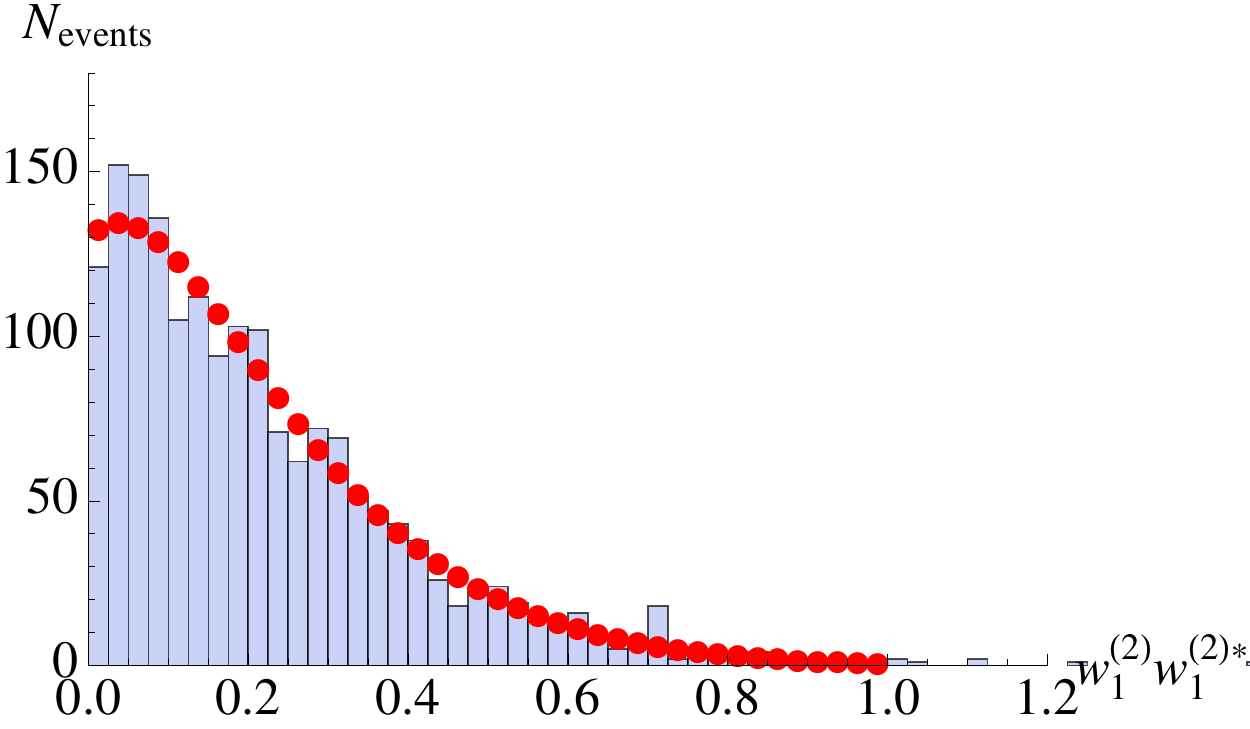}
\includegraphics[width=7.cm]{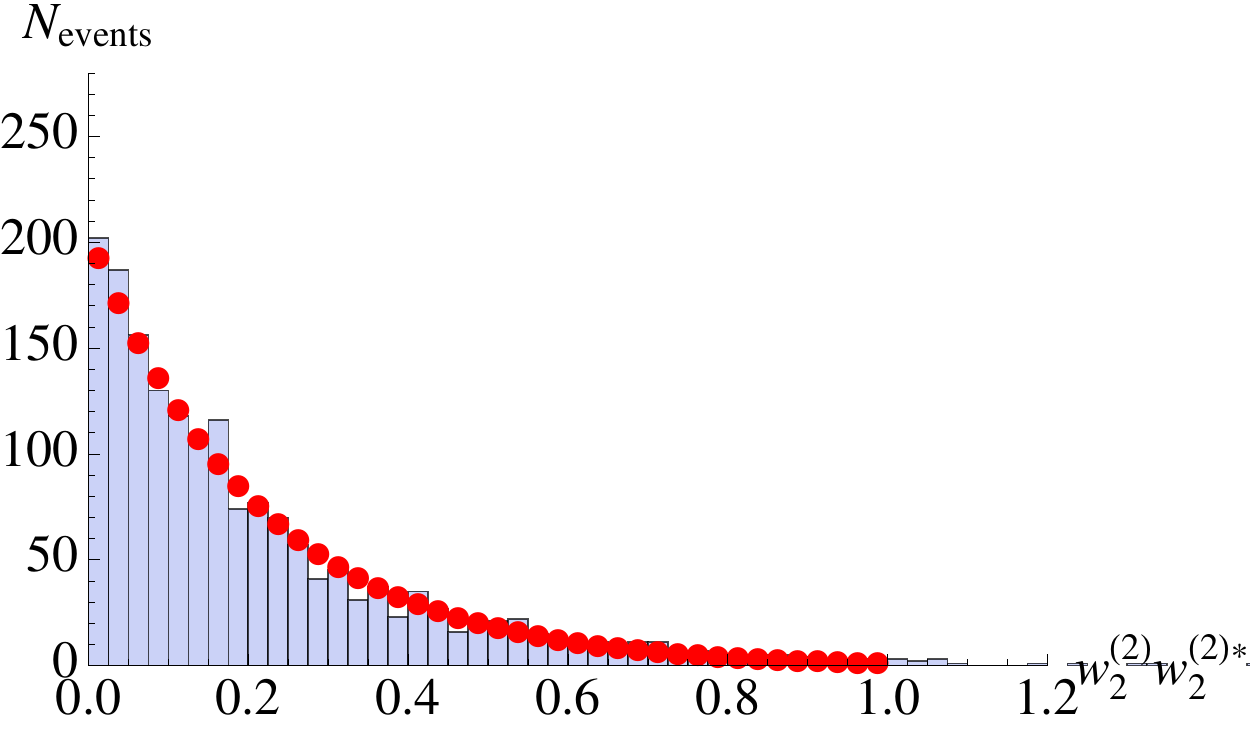}
\includegraphics[width=7.cm]{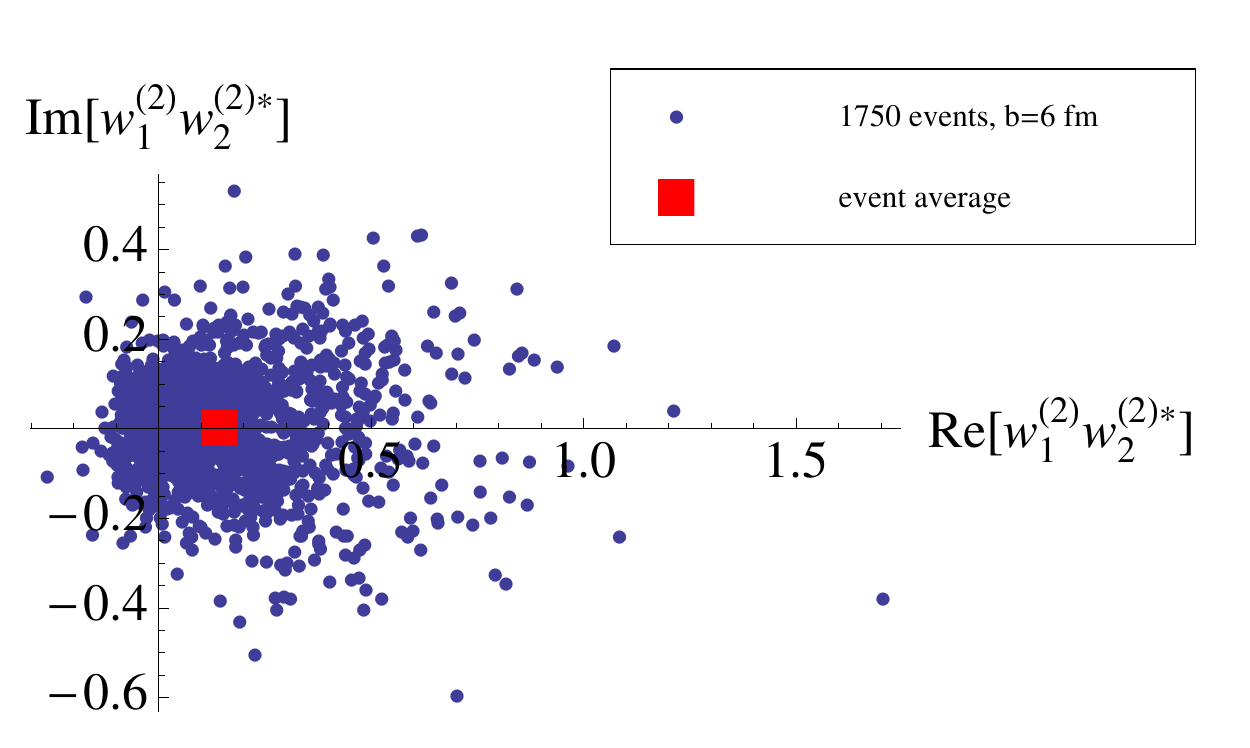}
\includegraphics[width=7.cm]{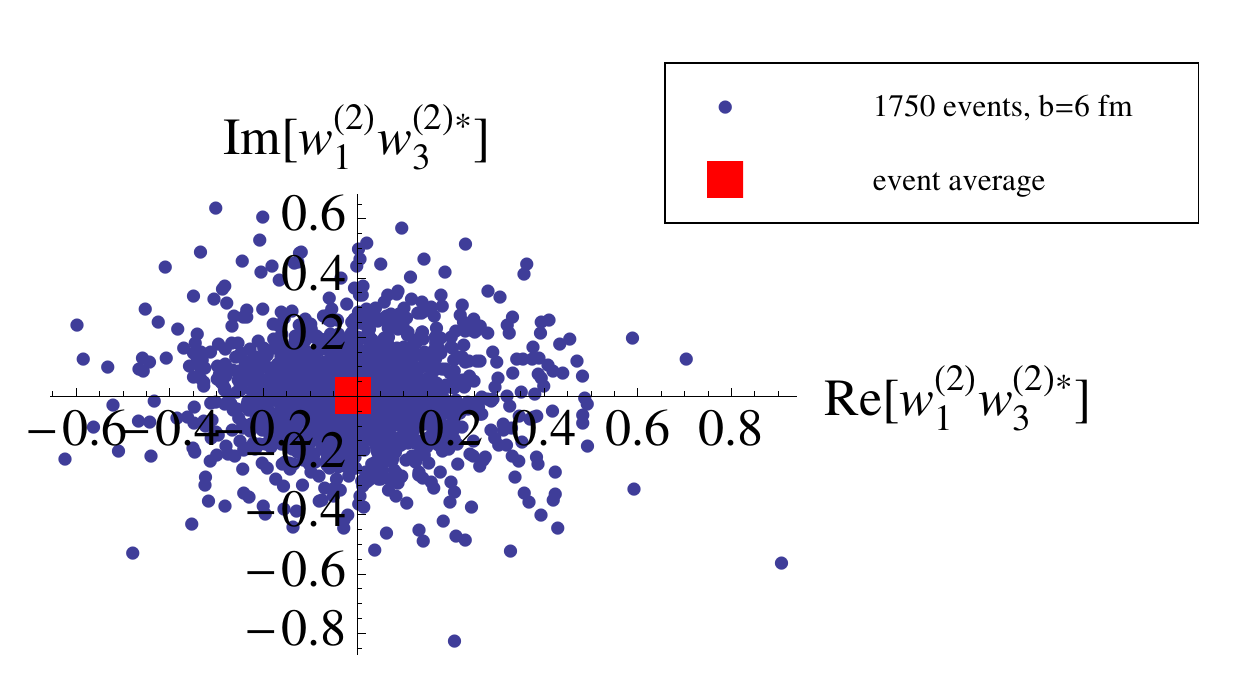}
\end{center}
\vspace{-0.5cm}
\caption{Event distributions of several two-mode correlators.
Same as Fig.~\ref{fig4} but for finite impact parameter $b= 6$ fm.
}\label{fig10}
\end{figure}

Fig.~\ref{fig10} shows event distributions for diagonal and non-diagonal two-mode products of the 
type $\xi_a = w_{l_a}^{(m)}\, w_{l_a}^{(m)*}$ and 
$\chi_{ab} = w_{l_a}^{(m)}\, w_{l_b}^{(m)*}$ respectively. In the upper part of Fig.~\ref{fig10},
we compare the simulated event distribution 
${\cal P}_{\xi}\left( \xi_a \right)$ to the analytical 
expectation (\ref{eq3.16}) for a Gaussian probability distribution (\ref{eq3.8}). 
This comparison is analogous to the one shown in Fig.~\ref{fig3} for vanishing
impact parameter, but it involves now a non-vanishing expectation value 
$\langle w_{l_a}^{(m_1)}\rangle$. Fig.~\ref{fig10} then illustrates that while the
event distributions ${\cal P}_{\xi}$ of different modes can have rather different
shapes, they are all satisfactorily accounted for in a Gaussian probability distribution (\ref{eq3.8}) 
that is determined by a small number of easily accessible
event-averaged values, $\langle w_{l_1}^{(m_1)} \rangle$ and 
$\langle w_{l_1}^{(m_1)}\, w_{l_1}^{(m_1)*}\rangle$.

\begin{figure}[h]
\begin{center}
\includegraphics[width=7.cm]{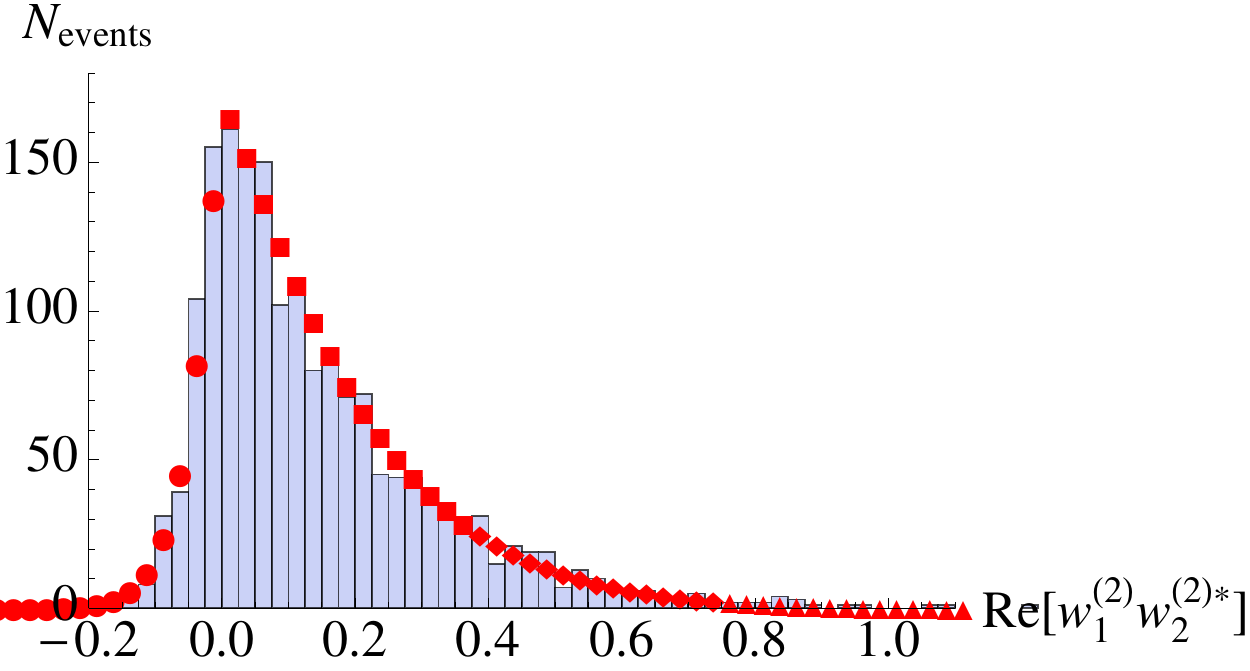}
\includegraphics[width=7.cm]{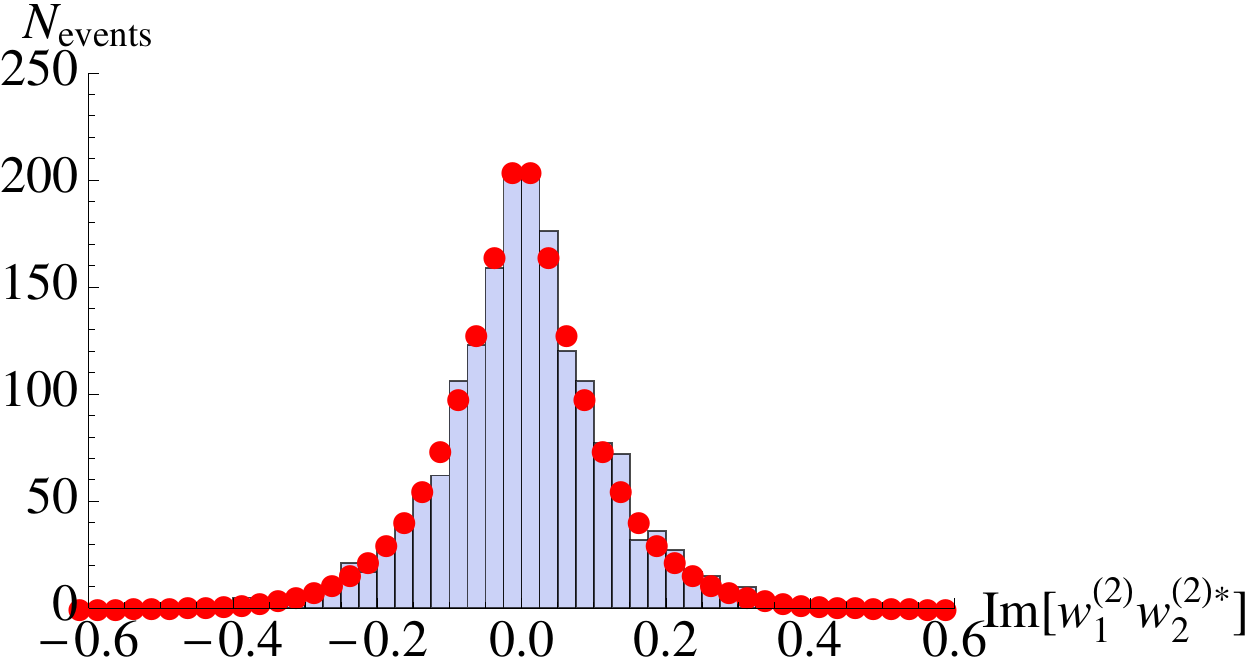}
\includegraphics[width=7.cm]{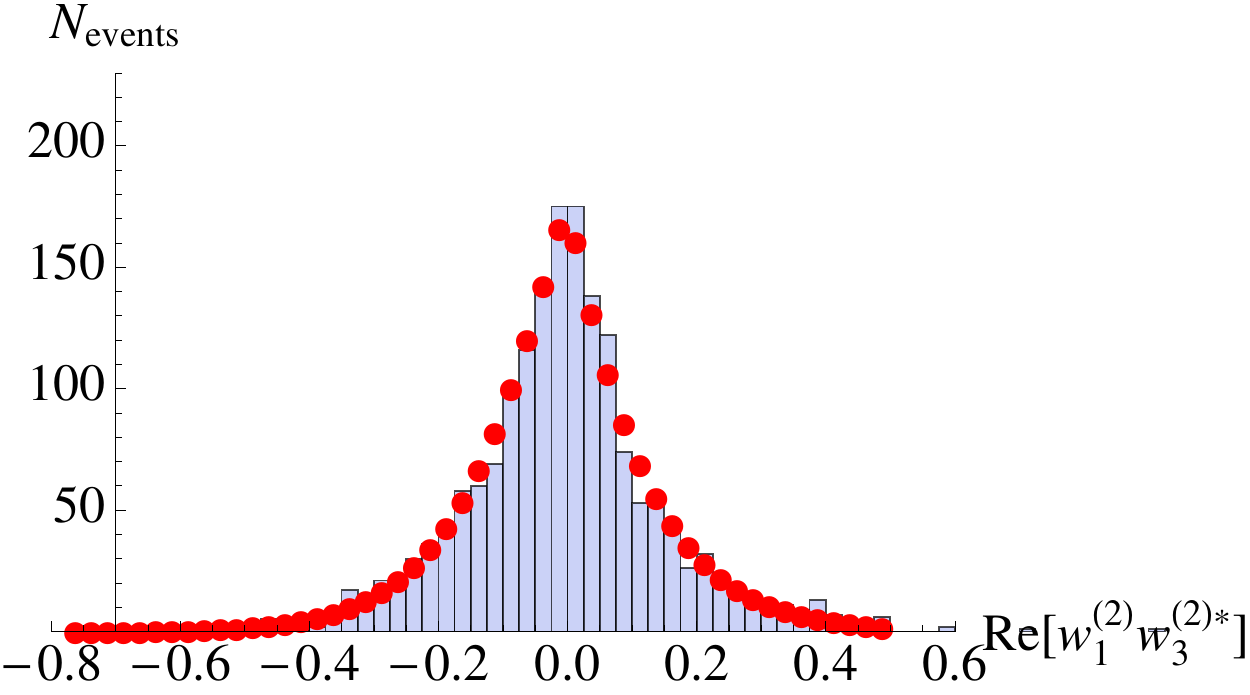}
\includegraphics[width=7.cm]{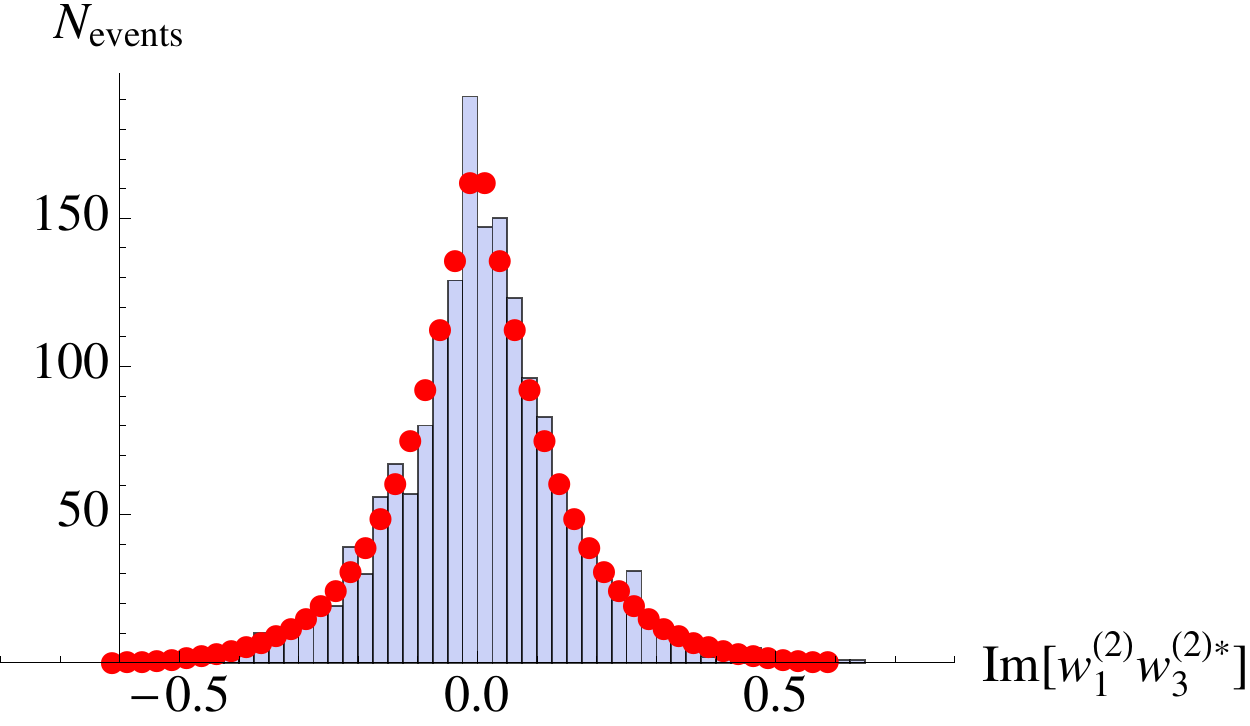}
\end{center}
\vspace{-0.5cm}
\caption{One-dimensional projections of the event distributions shown in Fig.~\ref{fig10}
(blue histograms) compared to the analytical expectation (\ref{eq3.19}). Without any
adjustment of parameters, the Gaussian ansatz (\ref{eq3.8}) fixed in terms of event
averages only, can account for the shapes of all event distributions 
${\cal P}_{\chi}$.
}\label{fig11}
\end{figure}

Fig.~\ref{fig10} also shows examples of event distributions for  
off-diagonal two-mode correlators $\chi_{ab} = w_{l_a}^{(m)}\, w_{l_b}^{(m)*}$. In Fig.~\ref{fig11},
we compare one-dimensional projections of these to analytical expectations for
a Gaussian probability distribution (\ref{eq3.8}). For finite impact, when the expectations values 
$\langle w_{l}^{(m)} \rangle$ do not vanish, one finds from equation (\ref{eq3.15})
\begin{eqnarray}
      &&{\cal P}_{\chi}\left( \chi_{ab}^r,\chi_{ab}^i\right)
      = \frac{{\rm Det}\left[ {\cal T}\right]}{4\pi}
      \int \frac{dx}{x} I_0\left[ \sqrt{ \frac{A^2\, \chi_{ab}\chi_{ab}^*}{x} + 
     B^2\, x + 2 {\rm Re}\chi_{ab}\, A\, B} \right] \nonumber \\
     &&\qquad \qquad
      \times 
      \exp\left[-\frac{1}{2}\left( {\cal T}_{l_al_a}x 
      + {\cal T}_{l_bl_b} \chi_{ab}\chi_{ab}^*/x + 2 {\cal T}_{l_al_b} {\rm Re}\chi_{ab} 
      + \sum_{i,j = l_a,l_b}\langle w_i^{(m)}\rangle {\cal T}_{ij} 
      \langle w_j^{(m)} \rangle \right) \right]\, ,
      \label{eq3.19}
\end{eqnarray}
where
\begin{eqnarray}
	A &\equiv&  {\cal T}_{l_al_b} \langle w_{l_a}^{(m)}\rangle
			+ {\cal T}_{l_bl_b} \langle w_{l_b}^{(m)}\rangle\, , \label{eq3.20}\\
	B &\equiv& {\cal T}_{l_al_a} \langle w_{l_a}^{(m)} \rangle +
			+ {\cal T}_{12} \langle h_2\rangle\, . \label{eq3.21}
\end{eqnarray}
For $\langle w_{l_a}^{(m)}\rangle = \langle w_{l_b}^{(m)}\rangle = 0$, expression (\ref{eq3.19})
reduces to the simple analytical form (\ref{eq3.18}). In general, however,
the event distribution ${\cal P}_{\chi}$
depends on the two event-averages $\langle w_{l_a}^{(m)}\rangle$,
$\langle w_{l_b}^{(m)}\rangle$, and on the
three independent matrix elements ${\cal T}_{ij}$, $i,j = l_a,l_b$. Determining
the latter from (\ref{eq3.10}), we confirm also at finite impact parameter
that the Gaussian approximation (\ref{eq3.8}) accounts very satisfactorily for the shape of event 
distributions of off-diagonal products $\chi_{ab}$. At finite, as well as at 
vanishing impact parameter, a very small set of event-averages (\ref{eq3.9}), (\ref{eq3.10}) is therefore
sufficient to specify fully the shape of all two-mode event distributions. 
 
\section{Normalized density fluctuation}
\label{sec4}

So far, we have discussed in section~\ref{sec2} general properties of the Bessel-Fourier expansion 
(\ref{eq2.6}) for transverse scalar densities, and we have studied in section~\ref{sec3} applications of this
expansion to a simple model of the transverse enthalpy density at vanishing and at finite impact parameter. 
The full enthalpy density $w$ is, of course, positive everywhere and for each event.
However, each fluctuating mode $w^{(m)}_l\, J_m(k^{(m)}_l\, r)$ in the Bessel-Fourier expansion will 
take negative values in some spatial regions. Moreover, at large radial distance $r$, the maximal
amplitudes of the oscillating modes falls off with the root of the radial distance $\propto \sqrt{1/r}$ only, 
while the enthalpy density of each event is expected to fall off exponentially. Therefore, after truncation
at a finite number of modes,  the Bessel-Fourier expansion (\ref{eq2.5}) of $w$ is not guaranteed to
be positive, and it may show locally negative entries in particular at large $r$. 
As we have seen in section~\ref{sec3}, this is not a problem for characterizing the initial conditions. 
However, it becomes an unwanted feature if one wants to propagate single modes $w^{(m)}_l\, J_m(k^{(m)}_l\, r)$ 
fluid dynamically. The propagation of locally negative densities poses certainly problems  in fluid dynamics,
and irrespective of whether one deals with those by 'ad hoc' regularizations of locally negative contributions
or in another way, the effort to ensure that the physical results depend only sufficiently weakly on a chosen
prescription will be an unwanted complication. One way to bypass this problem is to seek a Fourier-Bessel
expansion of the enthalpy density normalized by some conveniently chosen background enthalpy $w_{\rm BG}(r)$, 
such that for sufficiently small fluctuations the truncated expansion remains positive by construction. 
Here, we explore the ansatz
\begin{equation}
w^{(m)}(r) = \delta_{m0}\, w_{\rm BG}(r) + w_{\rm BG}(r) 
\sum_{l=1}^\infty \,  \tilde{w}^{(m)}_l J_m({k}^{(m)}_l\,r) \, ,
\label{eq4.1}
\end{equation}
where $\tilde{w}^{(m)}_l $ are the Bessel coefficients of the normalized density 
$(w^{(m)}(r)-\delta_{m0} w_{\rm BG}(r))/w_{\rm BG}(r)$. We chose the background enthalpy $w_{\rm BG}(r)$ in terms of the 
ensemble average of $w^{(0)}(r)$, 
\begin{equation}
 	w_{\rm BG}(r) \equiv \langle w^{(0)}(r) \rangle \, ,
	\label{eq4.2}
\end{equation}
but other choices may be possible as well. By construction, as long as the coefficients $\tilde{w}^{(m)}_l$ of fluctuations are 
sufficiently small, the density (\ref{eq4.1}) is positive everywhere even if truncated. Mainly for this reason,
we have based a first recent study~\cite{Floerchinger:2013rya} of the fluid dynamic propagation of single modes 
on the expansion (\ref{eq4.1}).  Because of the interesting properties of (\ref{eq4.1}), and 
to fully document the starting point of the dynamical study~\cite{Floerchinger:2013rya}, we discuss this expansion now 
in some detail. 

\begin{figure}[h]
\begin{center}
\includegraphics[width=11.cm]{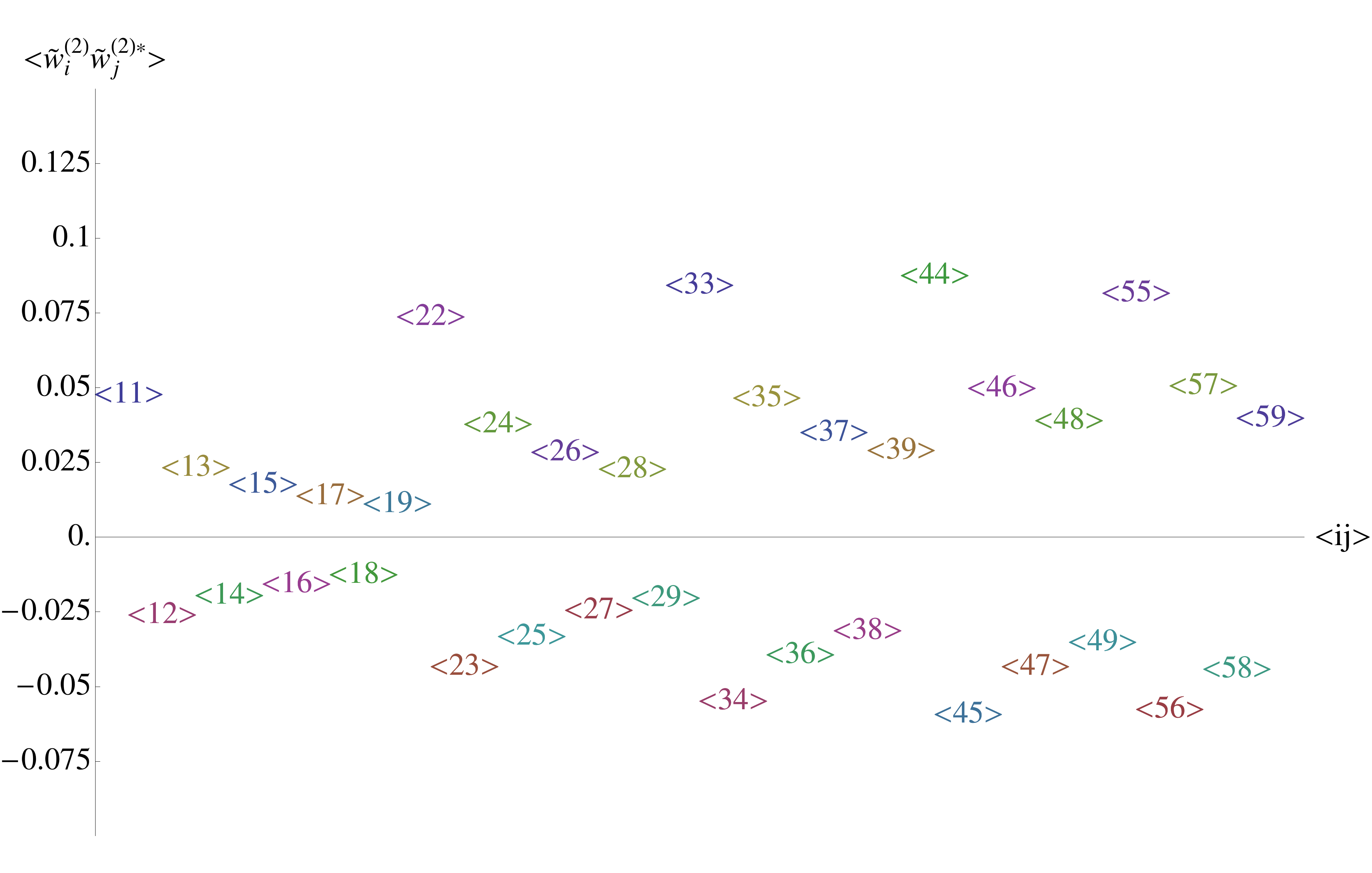}
\includegraphics[width=11.cm]{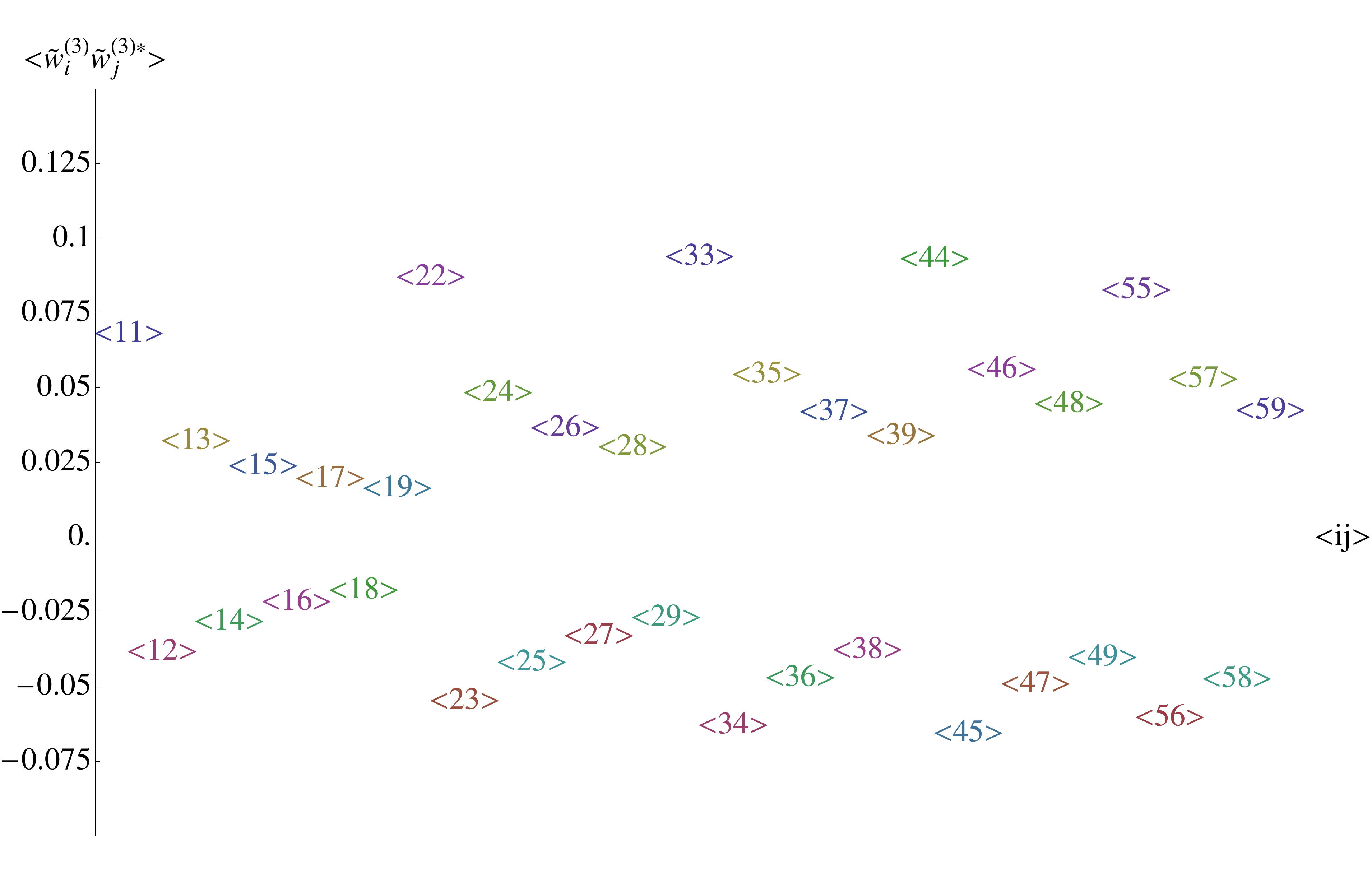}
\end{center}
\vspace{-0.5cm}
\caption{The two-mode correlators 
\protect{$\langle \tilde{w}_{l_1}^{(m)}\, {\tilde{w}_{l_2}^{(m)*}} \rangle$}
for $m=2$ (upper plot) and $m=3$ (lower plot)
of the Bessel-Fourier coefficients entering the normalized enthalpy density distribution (\ref{eq4.1}). 
}\label{fig12}
\end{figure}

We first note that the Bessel-Fourier expansion (\ref{eq2.6}), and the expansion (\ref{eq4.1}) of the normalized enthalpy 
density, share significant commonalities. In particular, we have checked numerically, that both expansions account
with comparable accuracy for a given true density distribution if truncated at the same number of modes (data not shown). 
This indicates that it is not problematic for a good approximation in the physical region $r < R$ that the normalized enthalpy 
density (in contrast to the unnormalized one, see (\ref{eq2.4})) does not vanish at the boundary $r=R$. However, as seen in
Fig.~\ref{fig11}, the two-mode correlators $\langle \tilde{w}_{l_1}^{(m)}\, \tilde{w}_{l_2}^{(m)*}\rangle$ of the normalized 
density show an oscillating structure that is rather different from that seen in Figs.~\ref{fig4} and ~\ref{fig8}.  
Technically, this oscillation arises since for the normalized density $w(r,\phi)/w_{\rm BG}(r)$, fluctuations
at large radius $r$ take much larger values than for the unnormalized case. The Bessel-Fourier expansion tends
to reproduce such structures at large $r$ and the non-vanishing values of $w(r,\phi)/w_{\rm BG}(r)$ at the boundary $r=R$ 
by alternating the sign of neighboring Bessel coefficients, see Fig.~\ref{fig11}.
In general, we find that the structure of $\langle \tilde{w}_{l_1}^{(m)}\, \tilde{w}_{l_2}^{(m)*}\rangle$ in Fig.~\ref{fig11}  
still follows for each $m$ a simple pattern: the sign of the two-mode correlator alternates, and its norm decreases for
fixed $m$ with increasing $l_2 - l_1$, as expected for a density in which radial modes decorrelate with increasing 
difference in wave length. Let us mention as an aside that we have illustrated already in a first dynamical 
study~\cite{Floerchinger:2013rya} how to propagate fluid dynamically event ensembles of small fluctuations 
characterized by the two-point correlators $\langle \tilde{w}_{l_1}^{(m)}\, \tilde{w}_{l_2}^{(m)*}\rangle$, or 
single fluctuating modes of weight $\tilde{w}_{l_1}^{(m)}$, and we have shown how to calculate the contributions of
these fluctuating modes to measured hadron spectra, see also \cite{FWtocome}. 

%
\begin{figure}[h]
\begin{center}
\includegraphics[width=8.cm]{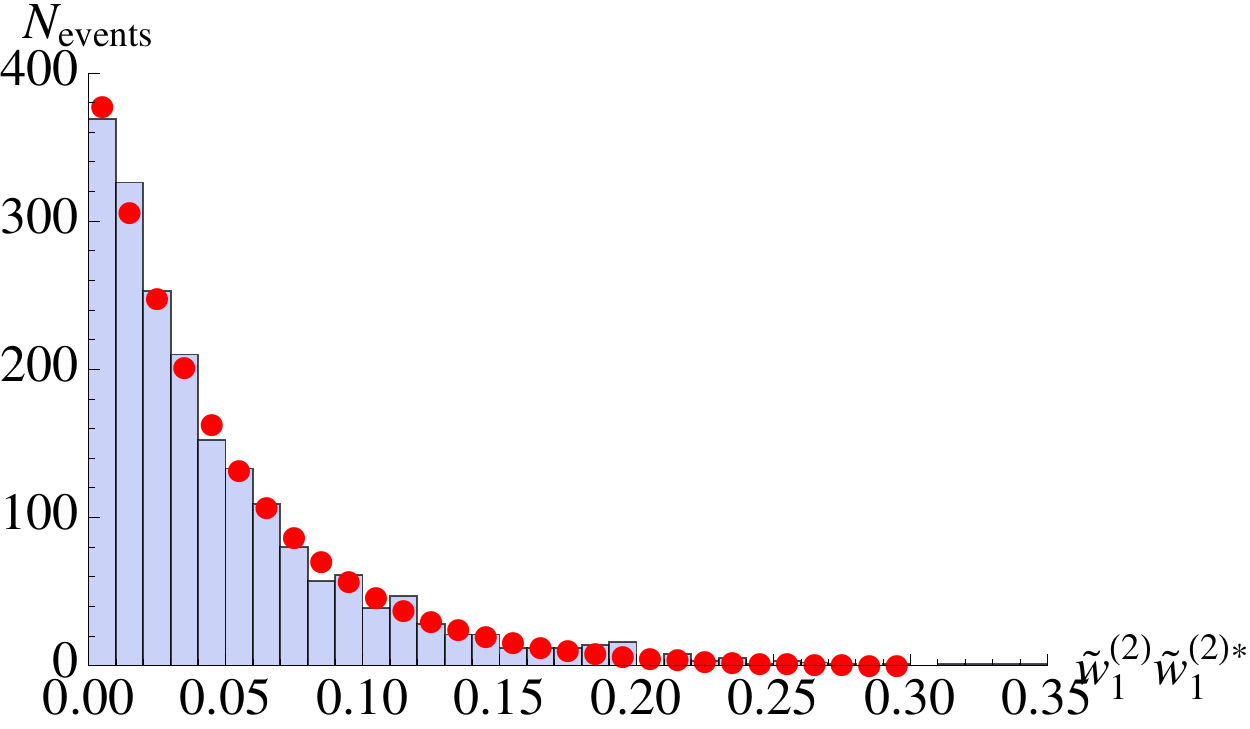}
\includegraphics[width=8.cm]{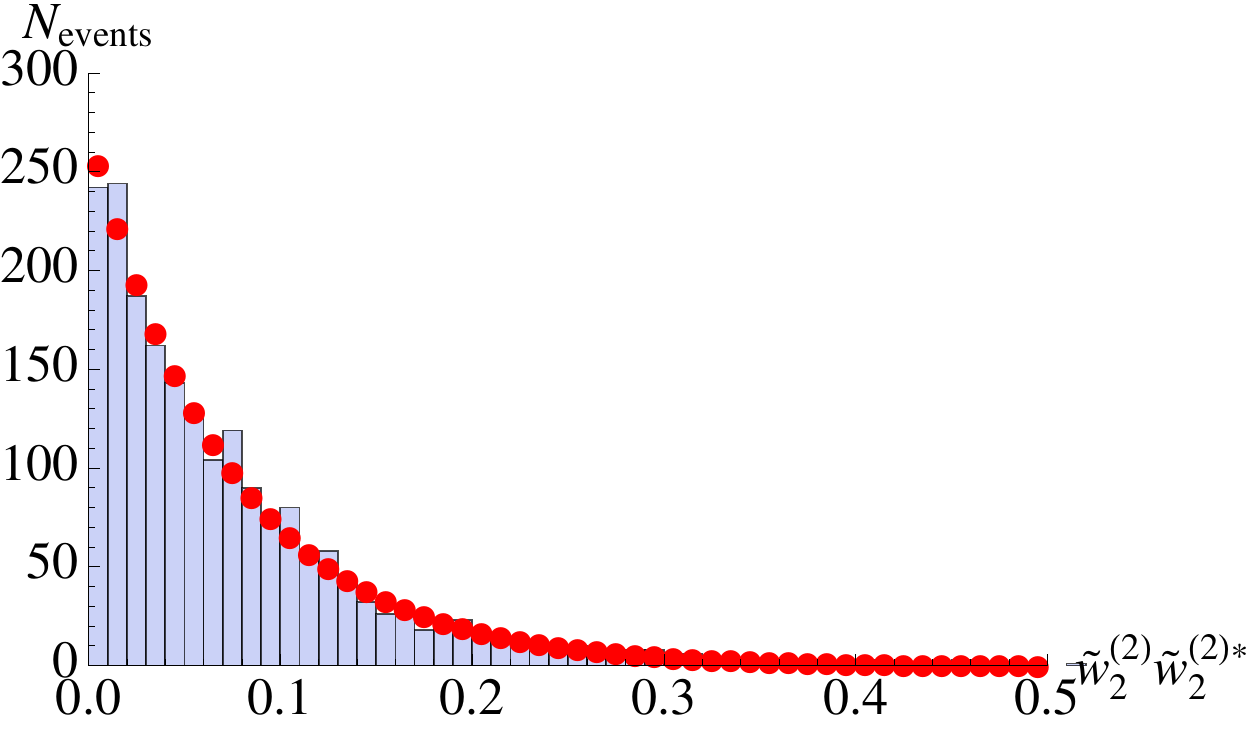}
\includegraphics[width=8.cm]{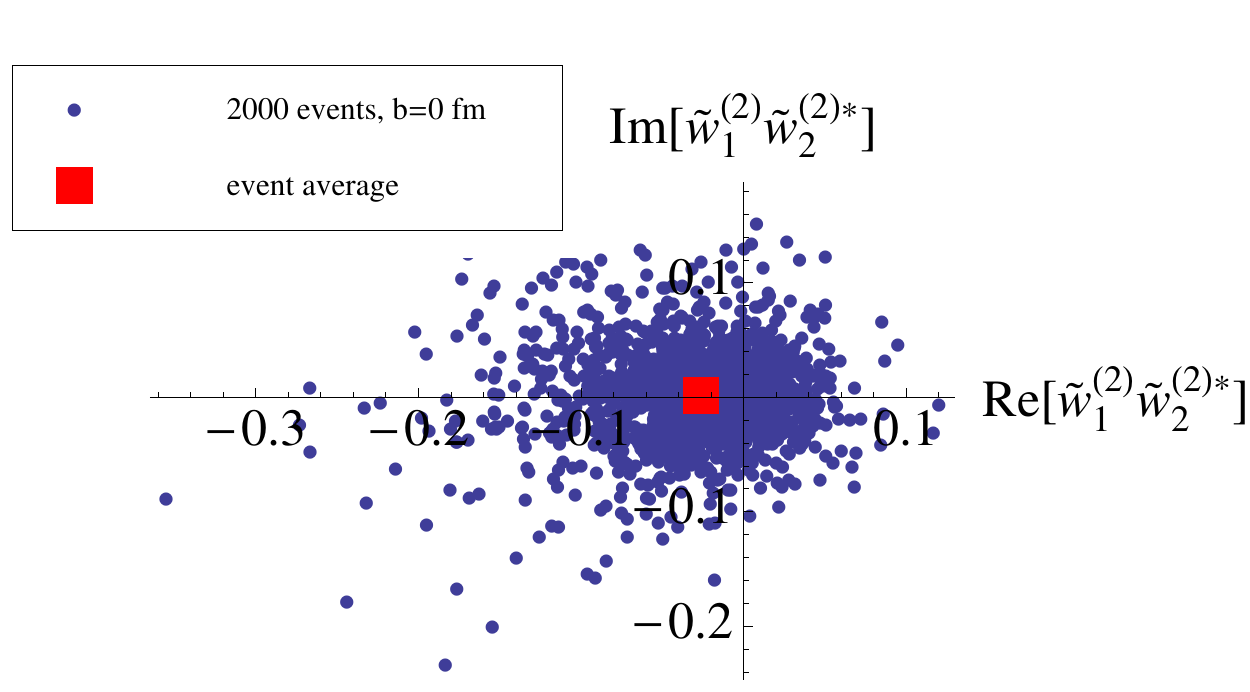}
\includegraphics[width=8.cm]{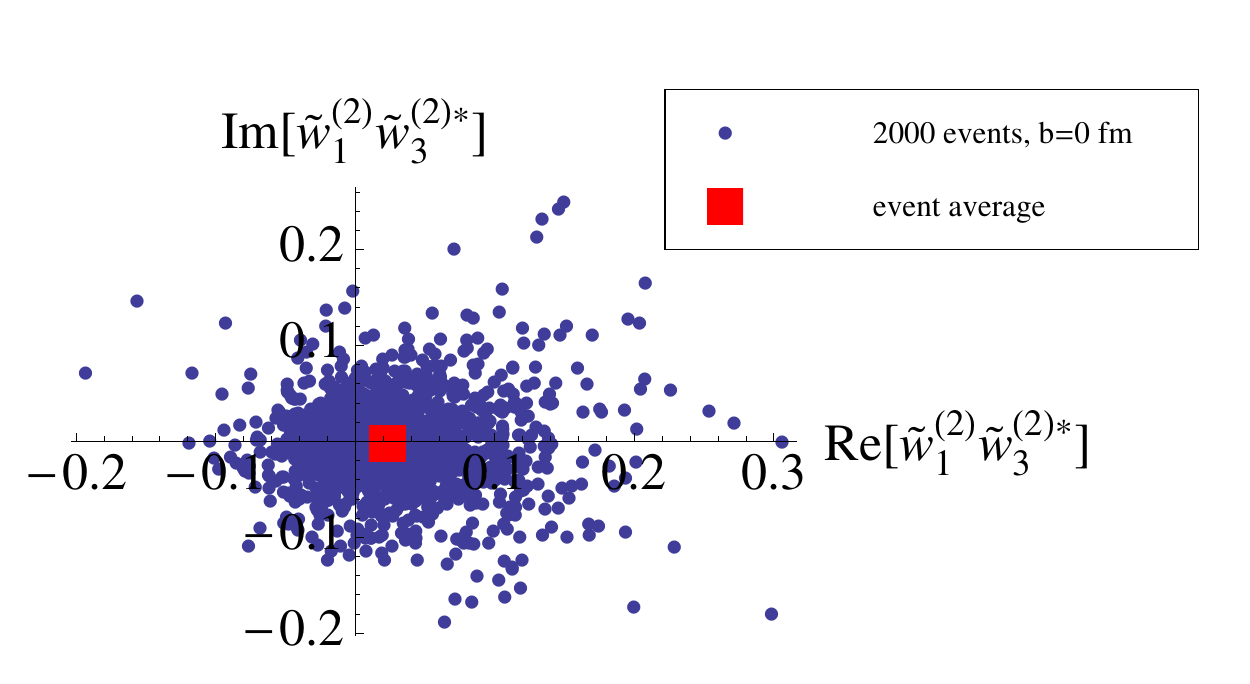}
\end{center}
\vspace{-0.5cm}
\caption{Event distributions of products of two modes $\tilde{w}_{l_1}^{(m)}\, \tilde{w}_{l_2}^{(m)*}$ for $b=0$.
Same as Fig.~\ref{fig3}, but for the normalized enthalpy density $w(r,\phi)/w_{\rm BG}(r)$.
}\label{fig12}
\end{figure}

\begin{figure}[h]
\begin{center}
\includegraphics[width=8.cm]{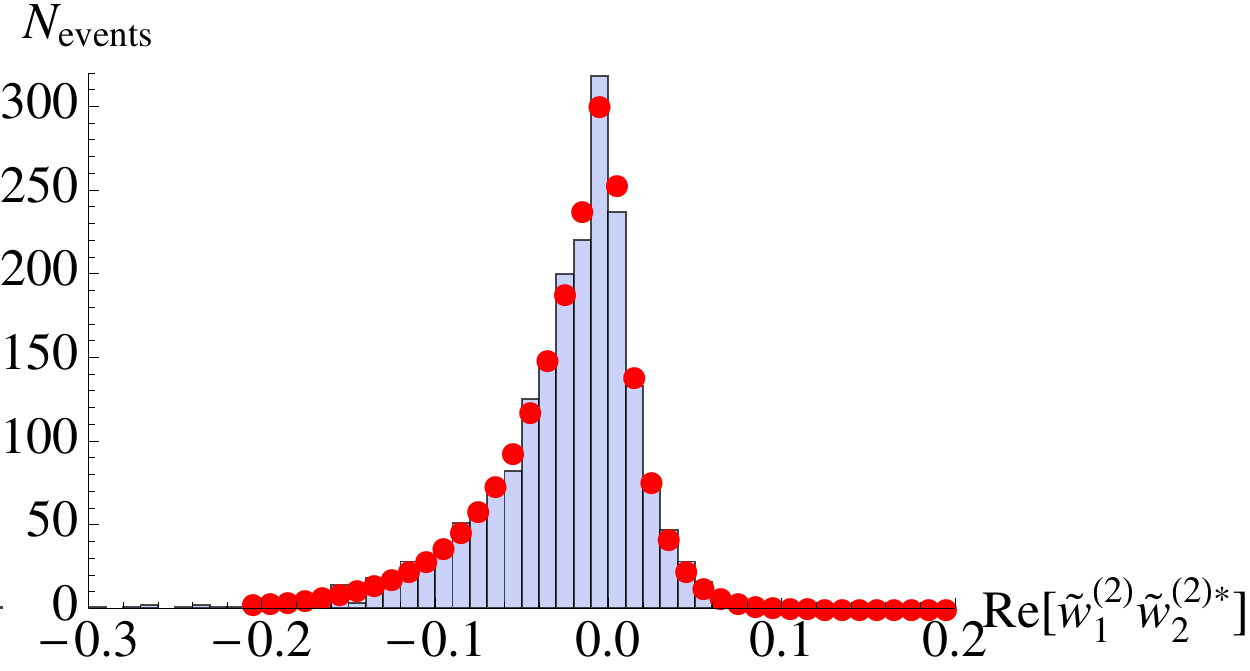}
\includegraphics[width=8.cm]{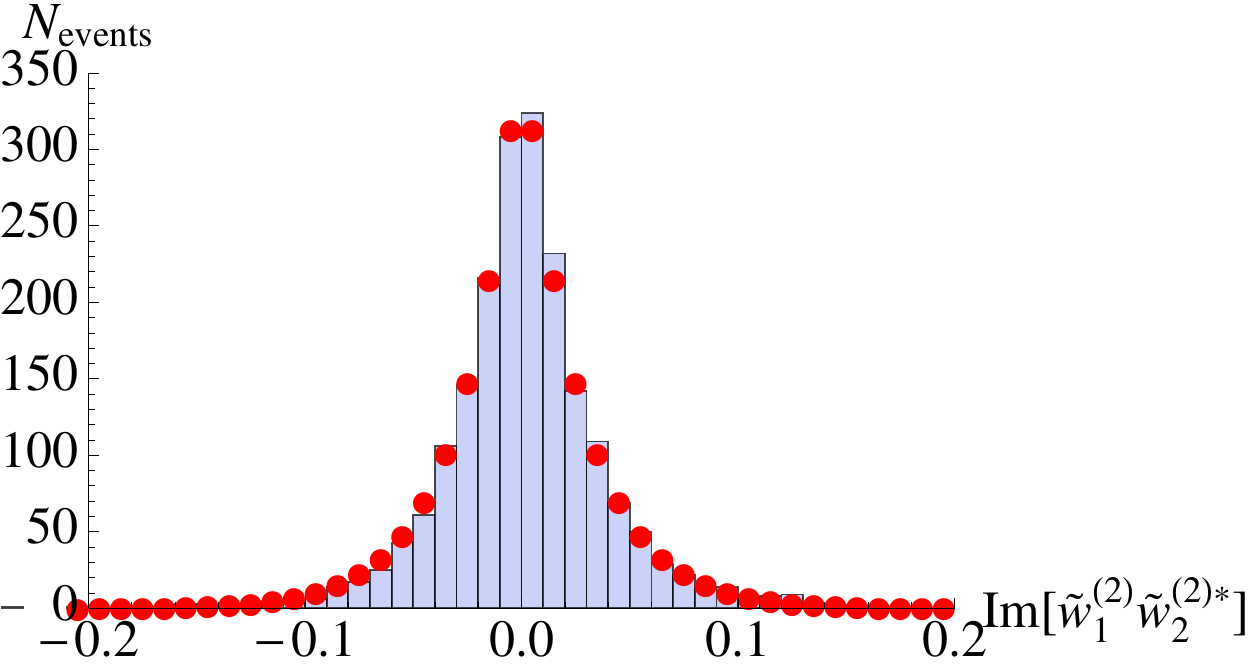}
\includegraphics[width=8.cm]{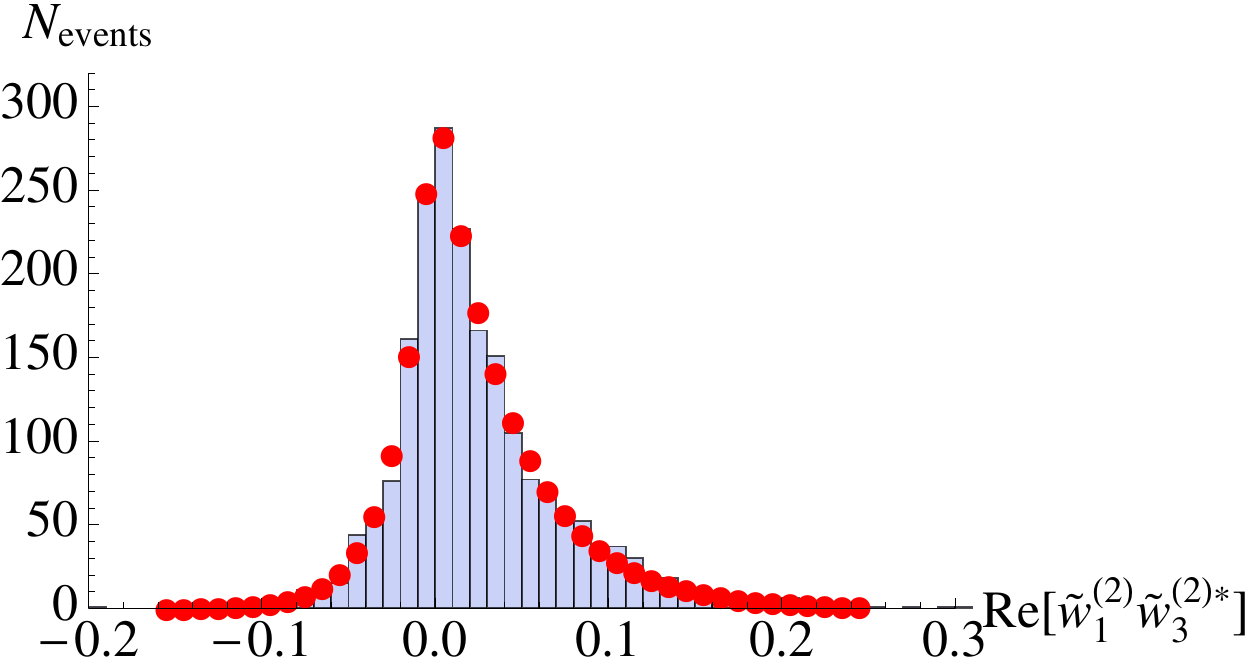}
\includegraphics[width=8.cm]{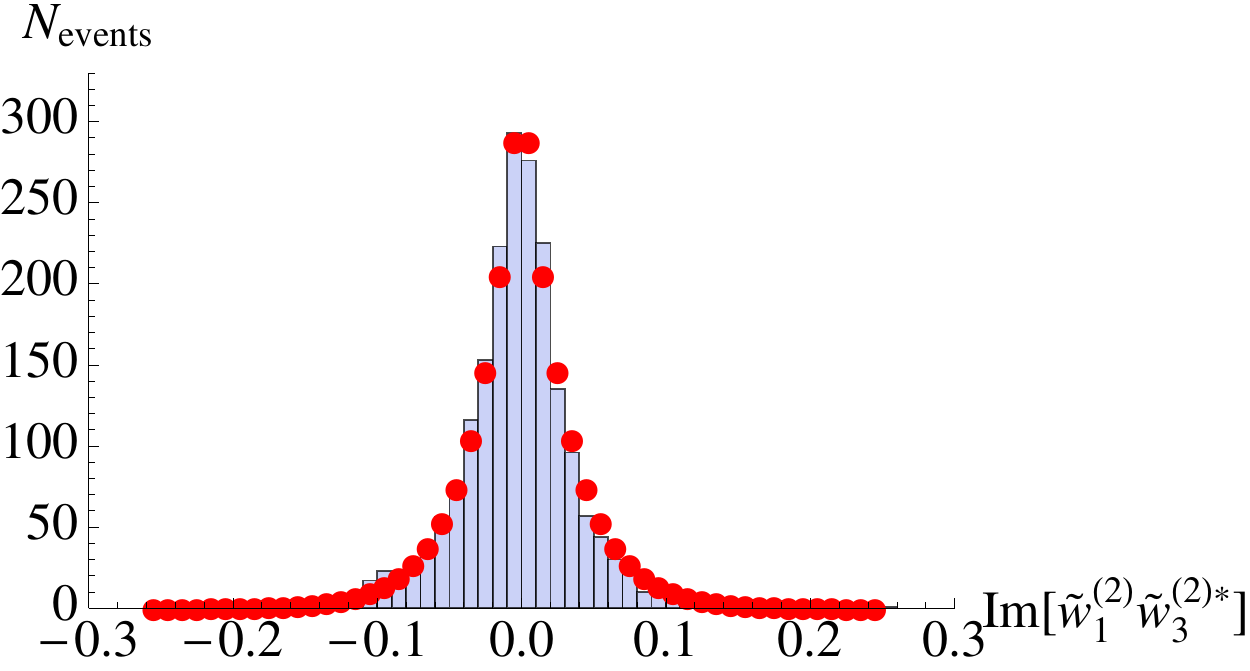}
\end{center}
\vspace{-0.5cm}
\caption{The distribution of real and imaginary values taken by complex-valued two-mode correlators 
$\langle \tilde{w}_{l_1}^{(m)}\, \tilde{w}_{l_2}^{(m)*}\rangle$, $l_1 \not= l_2$. 
Same as Fig.~\ref{fig3}, but for the normalized enthalpy density $w(r)/w_{\rm BG}(r)$.
}\label{fig13}
\end{figure}

As we have seen in section~\ref{sec3}, establishing that the probability distribution of event samples of 
initial conditions is approximately Gaussian provides a significant simplification for the characterization
of event samples. Here, we note that this simplification holds also for the
probability distribution ${\cal P}( \lbrace \tilde{w}_{l}^{(m)}\rbrace )$ of the Bessel coefficients
of the normalized density distribution $w(r,\phi)/w_{\rm BG}(r)$: the information displayed in Fig.~\ref{fig11} provides 
an almost {\it complete} characterization of ${\cal P}( \lbrace \tilde{w}_{l}^{(m)}\rbrace )$. In particular,
as discussed in section ~\ref{sec3.4} already, a Gaussian probability distribution is fully specified  by the two-mode correlators $\langle \tilde{w}_{l_1}^{(m)}\, \tilde{w}_{l_2}^{(m)*}\rangle$, but it makes non-trivial statements 
about the event distributions of $\tilde{w}_{l_1}^{(m)}\, \tilde{w}_{l_2}^{(m)*}$ (and higher order products) around these 
averages. In equations (\ref{eq3.16}) and (\ref{eq3.18}), we have derived explicit expressions for the relevant
event distributions of two-mode products. As we show in Fig.~\ref{fig12} for the distributions of different products of  two modes, and in Fig.~\ref{fig13} for projections of complex-valued products
$\tilde{w}_{l_a}^{(m)} \tilde{w}_{l_b}^{(m)*}$ on the real and imaginary 
axis, these distributions around event-averages indicate that ${\cal P}( \lbrace \tilde{w}_{l}^{(m)}\rbrace )$ is
a close to Gaussian probability distribution, of the form (\ref{eq3.8}).\footnote{Strictly speaking, since the relation between
$ {w}_{l}^{(m)}$ and $\tilde{w}_{l}^{(m)}$ is linear, the later are Gaussian distributed precisely when this holds for the
former. Nevertheless one may expect that possible deviations from Gaussianity are more pronounced in one of the
cases.}

%

In both, the Bessel-Fourier expansion (\ref{eq2.6}) and the scheme (\ref{eq4.1}), the radius $R$ 
must be chosen sufficiently large, so that the area of radius $R$ encompasses the entire physically relevant range.
On the other hand, $R$ should not be chosen too large, since the expansion scheme determines 
the Bessel-Fourier coefficients with respect to the entire radial range $r \in \left[0, R\right]$ without giving 
more weight to the high density region at small $r$ that is physically most relevant. Also, the 
Bessel-Fourier expansion coefficients depend on the choice of $R$. One may wonder whether it
is possible to eliminate this unwanted $R$-dependence and to give in the expansion more weight
to the physically most relevant region of small and intermediate radii that contain most of the enthalpy
density. One idea in this context might be to use a
mapping of the complete range of radii $r \in \left[0, \infty  \right]$ to a compact interval. The outer boundary condition
corresponds then to $r=\infty$, and artificial boundary effects would disappear. A natural mapping of this
kind is induced, for instance, by the background enthalpy density $w_{\rm BG}(r)$ chosen such that it agrees
with an appropriate event average, $w_{\rm BG}(r) = \langle w(r,\phi)\rangle$. Since $w_{\rm BG}(r)$ is
positive, monotonously decreasing with $r$ and integrable in the transverse plane (the total enthalpy is finite),
one can define the transformed radial coordinate such that
\begin{equation}
	\rho(r) = \sqrt{\frac{\int_0^r dr'\, r'\, w_{\rm BG}(r')}{\int_0^\infty dr'\, r'\, w_{\rm BG}(r')}}\, .
\end{equation}
The coordinate $\rho$ is proportional to $r$ for small $r$ and it maps the interval $r \in \left[0, \infty\right]$
onto $\rho \in \left[ 0, 1\right]$.  A reformulation of the expansion (\ref{eq4.1}) in this new radial coordinate
is straightforward, but we shall not further explore this point in the present work.


\section{Vector and tensor fluctuations}
\label{sec5}

In this section we extend the Bessel-Fourier representation of the previous section to hydrodynamical fields that transform as vectors and tensors under rotations. We also make the dependence of these fields on spatial rapidity $\eta$ and time $\tau$ explicit that we have omitted for notational simplicity so far. 
Again we design the expansion such that in a situation where the background field is independent of the coordinates in the transverse plane the evolution equations for different Bessel modes decouple. 
 We will discuss here exemplary the fluctuations in fluid velocity, the extension to other vector fields is then 
 straightforward.

In coordinates $\tau, r, \phi, \eta$ it is sensible to choose the independent components of the fluid velocity as $u^r$, $u^\phi$ and $u^\eta$. The fourth component follows from the normalization condition as $u^\tau=\sqrt{1+(u^r)^2+r^2(u^\phi)^2+\tau^2 (u^\eta)^2}$.
For a background fluid field that satisfies rotational symmetry in the transverse plane as well as Bjorken boost invariance,
the background components $u^\phi$ and $u^\eta$ vanish, but the radial background component $u_{\rm BG}^r$ can be
non-vanishing. We denote the fluctuating part of the velocity fields by a tilde, and we rescale all field components
such that they are dimensionless in units with $c=1$. Expressed with the help of 
transverse cartesian coordinates  $\vec{v}=(v_x,v_y)$, $\vec{s}=(x,y)$, $\vec{s}_\perp=(-y,x)$,
the fluctuating part of the velocity components takes then the form
\begin{eqnarray}
	\tilde{u}^r &=& \vec{s}\cdot\vec{v}/\vert \vec{s}\vert - u_{\rm BG}^r = u^r - u_{\rm BG}^r \, ,\label{eq5.1}\\
	\tilde{u}^{\phi} &=&  \vec{s}_\perp\cdot\vec{v}/\vert \vec{s}_\perp\vert = r \, u^\phi\, , \label{eq5.2}\\
	\tilde{u}^\eta &=& \tau\, u^{\eta}\, .
\end{eqnarray}
%
%
\begin{figure}[t]
\begin{center}
\includegraphics[width=7.cm]{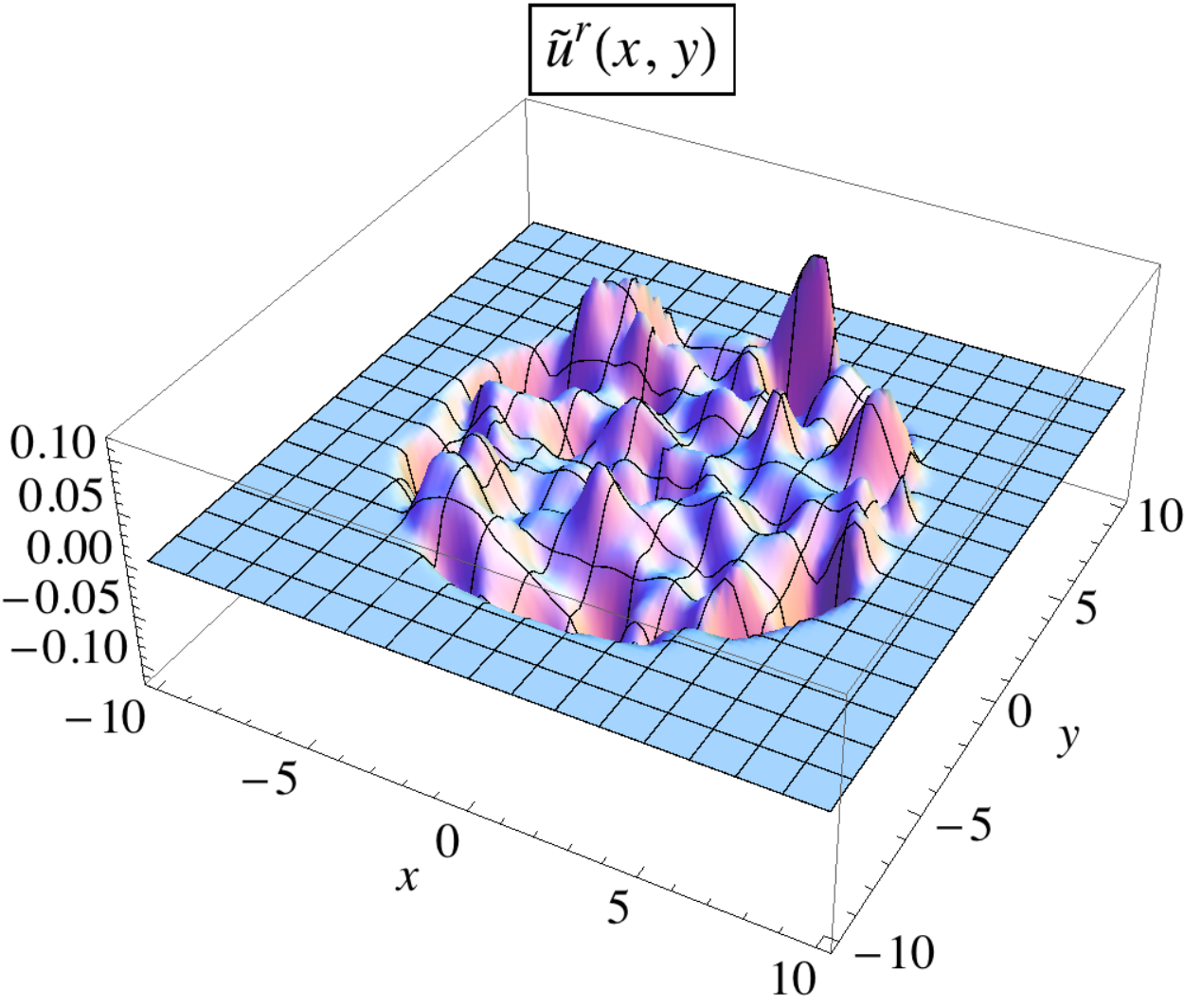}
\includegraphics[width=7.cm]{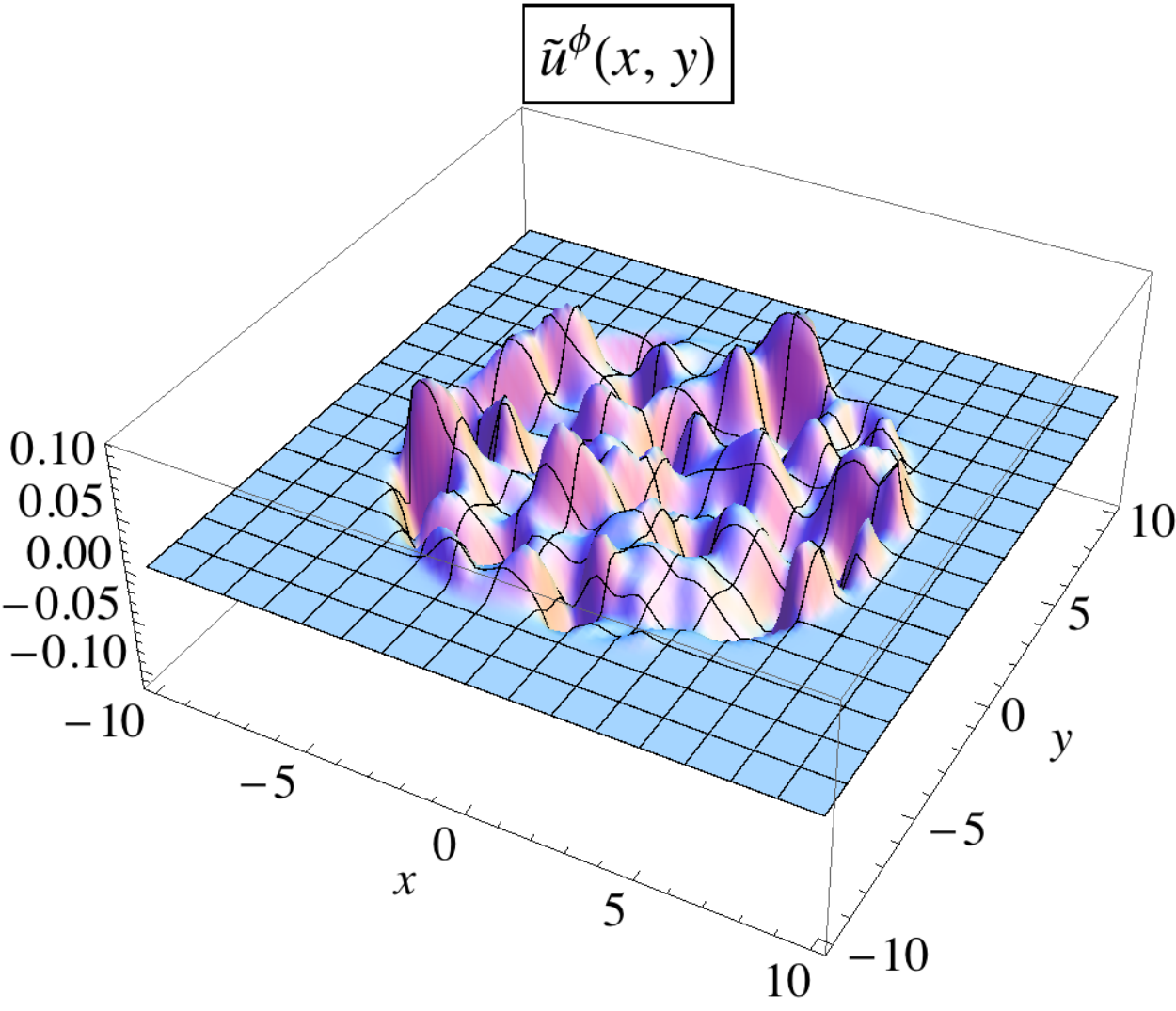}
\end{center}
\vspace{-.5cm}
\caption{Fluctuations in the initial radial (left) and angular (right) velocity fields $u_r$ and $u_\phi$ of a single event.
The event was generated from MC Glauber initial conditions in which the enthalpy density attributed to each participant 
is associated with a small random transverse velocity component, drawn from a Gaussian distribution of width 
$\langle \vert {\bf v}\vert \rangle = 0.1\, c$. 
}\label{figvel1}
\end{figure}

\begin{figure}[t]
\begin{center}
\includegraphics[width=14.cm]{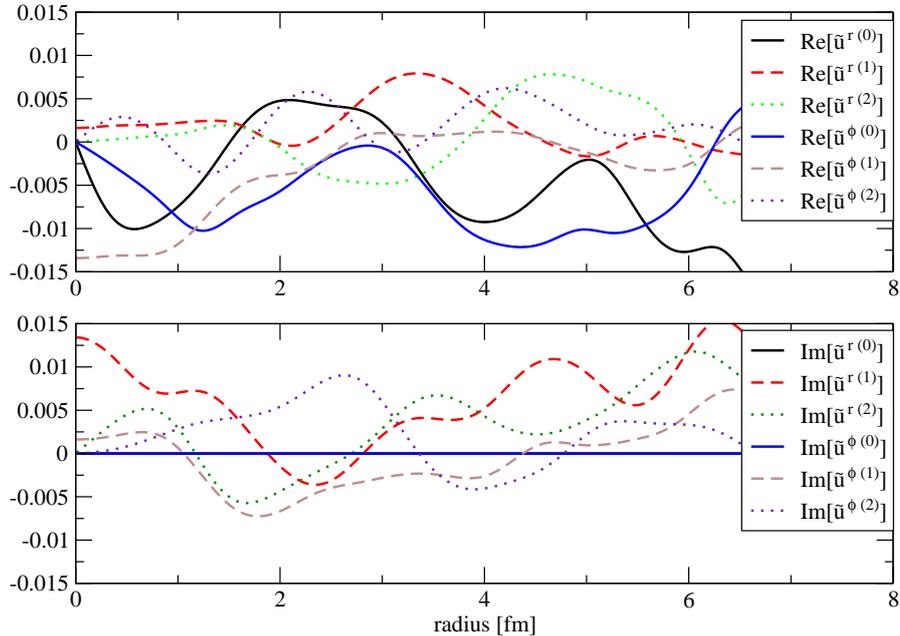}
\end{center}
\vspace{-1.cm}
\caption{The lowest harmonic moments $\tilde{u}^{r\, (m)}$,  $\tilde{u}^{\phi\, (m)}$ of the fields $\tilde{u}^r$ and 
$\tilde{u}^\phi$ of a single
event plotted in Fig.~\ref{figvel1}. The boundary values of these moments for $r\to 0$ are consistent with the choice 
of a Bessel-Fourier expansion in terms of Bessel functions $J_{m+1}$, $J_{m-1}$, see eqs.~(\ref{linearcomb}),  
(\ref{eq:BesselFourieru-u+}) and text.
}\label{figvel2}
\end{figure}

We now discuss how to set up the Bessel-Fourier representation of these fields. Naively one might think that an 
expansion as in \eqref{eq2.5} would do for vector valued quantities as well. However, this
leads to problems as can be seen for example for the $m=0$ modes. At $r=0$ the Bessel functions are non-zero, $J_0(0)=1$ and  an expansion of $\tilde{u}^r$ as in \eqref{eq2.5} would thus contain parts that do not 
vanish for $r\to0$. This is unphysical since the divergence of the fluid velocity would have a $1/r$ singularity; 
the zeroth harmonic moment of $\tilde{u}^r$ must vanish at $r=0$. On the other hand, the first harmonic moments of 
$\tilde{u}^r$ and $\tilde{u}^\phi$ can take finite values at $r=0$, while the expansion \eqref{eq2.5} does not allow for that. 
Instead of \eqref{eq2.5}, one can expand the $m$-th moments of the velocity fields $\tilde{u}^r$ and $\tilde{u}^\phi$ as linear combinations of the Bessel functions $J_{m-1}(k_l^{(m)} r)$ and $J_{m+1}(k_l^{(m)} r)$ that satisfy physical boundary conditions at $r=0$. Here, the wave-numbers $k_l^{(m)}$ are the same as in (\ref{eq2.6}). The functions $J_{m-1}(k_l^{(m)} r)$ and $J_{m+1}(k_l^{(m)} r)$ form an appropriate orthogonal set of functions, see appendix \ref{sec:appA}. 
Closer inspection shows then that physical boundary conditions are realized for the linear 
combinations~\footnote{In general, the $m$-th harmonic moments of $\tilde{u}^r$ and $\tilde{u}^\phi$ vanish for $m\not=1$ at $r=0$, and
for $m=1$ they satisfy ${\rm Re}\left[\tilde{u}_r^{(1)}(r=0) \right] = {\rm Im}\left[\tilde{u}_r^{(1)}(r=0) \right]$ and
${\rm Re}\left[\tilde{u}_\phi^{(1)}(r=0) \right] = -{\rm Im}\left[ \tilde{u}_\phi^{(1)}(r=0) \right]$. 
This can also be seen from Fig.~\ref{figvel2}. 
One checks straightforwardly that these physical boundary conditions are satisfied by the ansatz (\ref{linearcomb}).}
%
%
%
\begin{equation}
\begin{split}
 \tilde{u}^r &= \frac{1}{\sqrt{2}} \left( \tilde{u}^-+ \tilde{u}^+\right)\, ,\\
 \tilde{u}^\phi &= \frac{i}{\sqrt{2}} \left(  \tilde{u}^- -  \tilde{u}^+ \right)\, ,
\end{split}
\label{linearcomb}
\end{equation}
where
\begin{equation}
\begin{split}
\tilde{u}^-(\tau_0,r,\phi,\eta) = & \sum_{l=1}^\infty \sum_{m=-\infty}^\infty \int_{-\infty}^\infty \frac{d k_\eta}{2\pi}  \;\; 
\tilde{u}_l^{-(m)}(\tau_0,k_\eta) \;e^{i(m\phi+k_\eta \eta)} J_{m-1}(k_l^{(m)} r) \, ,   \\
\tilde{u}^+(\tau_0,r,\phi,\eta) = & \sum_{l=1}^\infty \sum_{m=-\infty}^\infty \int_{-\infty}^\infty \frac{d k_\eta}{2\pi} \;\; 
\tilde{u}_l^{+(m)}(\tau_0,k_\eta)\; e^{i(m\phi+k_\eta \eta)} J_{m+1}(k_l^{(m)} r)\, .
\end{split}
\label{eq:BesselFourieru-u+}
\end{equation}
For $\tilde{u}^\eta$ one can use the same expansion as in the scalar case,
\begin{equation}
\tilde{u}^\eta(\tau_0,r,\phi,\eta) = \sum_{l=1}^\infty \sum_{m=-\infty}^\infty \int_{-\infty}^\infty \frac{d k_\eta}{2\pi}  \;\; 
\tilde{u}_l^{\eta(m)}(\tau_0,k_\eta) \;e^{i(m\phi+k_\eta \eta)} J_{m}(k_l^{(m)} r).
\label{eq:BesselFourierueta}
\end{equation}
Note that in \eqref{eq:BesselFourieru-u+} and \eqref{eq:BesselFourierueta} we also expand the dependence on rapidity $\eta$ into an appropriate Fourier transform. In this sense Eqns.\ \eqref{eq:BesselFourieru-u+} and \eqref{eq:BesselFourierueta} provide generalizations of \eqref{eq2.1} and \eqref{eq2.4a}.
%

To shortly illustrate properties of the Bessel-Fourier expansion of vector fields, we have generated initial conditions
with non-vanishing velocity fluctuations according to a model described in Ref.~\cite{Florchinger:2011qf}. This model 
supplements the MC Glauber initial conditions of section~\ref{sec3} with a velocity field by associating a small
random transverse velocity component ${\bf v}$ to each of the participants and their individual enthalpy density  distributions.
For the examples considered here, we draw the random velocity components from a Gaussian distribution of
width $\langle \vert {\bf v}\vert \rangle = 0.1\, c$. Fig.~\ref{figvel1} shows the fluctuations in the radial ($\tilde{u}^r$) and
azimuthal ($\tilde{u}^\phi$) velocity components, generated in such a model for a single event. We mention as an aside 
that the initial velocity fluctuations of this model have divergent and rotational
(a.k.a. vorticity) components of similar size~\cite{Florchinger:2011qf}. It is an open
question whether such initial velocity fluctuations leave characteristic signatures in relativistic heavy ion collisions, but
at least some conceivable scenarios are being explored~\cite{Csernai:2013bqa, Pang:2012he}. 
However, even if velocity fluctuations should turn out to be negligible at initial time $\tau_0$, they will
be generated at $\tau > \tau_0$ in response to fluctuations in the enthalpy density. Understanding how the Bessel-Fourier
expansion extends to vector and tensor fields is therefore relevant for studying how single density modes propagate.
In particular, our first exploratory study of the dynamical evolution of single modes of the enthalpy 
density\cite{Floerchinger:2013rya} was based on a Bessel-Fourier expansion of all vector and tensor fields at times 
$\tau>\tau_0$. 

Here we do not further discuss the physics of initial velocity fluctuation, but we limit our discussion to
the properties of characterizing vector fields with the ansatz (\ref{linearcomb}), (\ref{eq:BesselFourieru-u+}). It is a general
feature of vector fields that in the limit $r\to 0$ their first harmonic moments can take non-vanishing values, while all
other harmonic moments vanish. Fig.~\ref{figvel2} shows the harmonic moments $\tilde{u}^{r\, (m)}$, 
$\tilde{u}^{\phi\, (m)}$ for the velocity fluctuations of Fig.~\ref{figvel1} and illustrates this point.

To determine the Bessel-Fourier coefficients of the expansion (\ref{eq:BesselFourieru-u+}) of vector fields, one
can apply again Lemoine's method of discrete Bessel transformation. The only difference to the scalar case is now
that $J_m(k_l^{(m)}r)$  in (\ref{eq2.9}) gets replaced by  $J_{m-1}(k_l^{(m)}r)$,  $J_{m+1}(k_l^{(m)}r)$, respectively, so that 
\begin{equation}
	\tilde{u}_{l}^{\pm\, (m)} = \sum_{\alpha=1}^{N_l} 
	{\cal M}^{\pm\, (m)}_{l \alpha} \tilde{u}^{\pm(m)}(r^{(m)}_\alpha)\, ,
	\label{eq2.8r}
\end{equation}
where the matrix ${\cal M}^{\pm (m)}_{l \alpha}$ is independent of the properties of $\tilde u^{\pm\, (m)}(r)$
and reads
\begin{equation}
	 {\cal M}^{\pm\, (m)}_{l \alpha} = \frac{4\, J_{m\pm 1}\left(k_l^{(m)} r^{(m)}_\alpha\right)}{
	 (z_{N_l}^{(m)})^2\,  
	 J_{m+1}^2(z_l^{(m)})\, J_{m+1}^2(z_\alpha^{(m)})}\, .
	 \label{eq2.9r}
\end{equation}

Let us now turn to tensor valued fields. The prime example for this is the shear stress tensor $\pi^{\mu\nu}$. 
If the event-averaged background of this shear tensor has rotational symmetry in the transverse plane and 
Bjorken boost invariance , then it depends only on $r$ and the only non-zero components are
$\pi_{\rm BG}^{\tau\tau}$, $\pi_{\rm BG}^{\tau r}$, $\pi_{\rm BG}^{r\tau}$, $\pi_{\rm BG}^{r r}$, $\pi_{\rm BG}^{\phi\phi}$ 
and $\pi_{\rm BG}^{\eta\eta}$. We note that only two of these components are independent, the other are
constrained by
\begin{equation}
\pi^{\mu\nu}=\pi^{\nu\mu}, \quad \pi^\mu_{\;\mu}=0,\quad u_\mu \pi^{\mu\nu} = 0.
\label{eq:constraintsonpi}
\end{equation}
Here, the last constraint is non-linear in the fluid dynamical fields. It is therefore not necessarily true for expectation values. 
It may still be reasonable, however, to assume that the relation \eqref{eq:constraintsonpi} holds when 
$u_\mu$ and $\pi^{\mu\nu}$ are replaced by their background values $u_{\rm BG}^\mu$ and $\pi_{\rm BG}^{\mu\nu}$. 

We now rescale again the components of the shear viscous tensor so that they are dimensionless, 
and we denote their fluctuating parts by a tilde 
\begin{equation}
\begin{split}
& \tilde \pi^{rr} = \frac{1}{w_{\rm BG}} (\pi^{rr}-\pi_{\rm BG}^{rr}), 
\quad \tilde \pi^{r\phi} = \tilde \pi^{\phi r} = \frac{r}{w_{\rm BG}} (\pi^{r \phi}-\pi^{r \phi}_{\rm BG}),
\quad \tilde\pi^{r\eta} = \tilde \pi^{\eta r} = \frac{\tau}{w_{\rm BG}} \pi^{r\eta},\\
& \tilde \pi^{\phi\phi} = \frac{r^2}{w_{\rm BG}} (\pi^{\phi\phi}-\pi_{\rm BG}^{\phi\phi}), \quad \tilde\pi^{\phi\eta} = \tilde\pi^{\eta\phi} = \frac{r\tau}{w_{\rm BG}} \pi^{\phi\eta}, \quad \tilde\pi^{\eta\eta} = \frac{\tau^2}{w_{\rm BG}}(\pi^{\eta\eta}-\pi_{\rm BG}^{\eta\eta}).
\end{split}
\label{eq:tildepicomponents}
\end{equation}
The components involving the temporal direction $\tau$ can be inferred from these using $u_\mu\pi^{\mu\nu}=0$. We note also that one of the components in \eqref{eq:tildepicomponents} can be expressed in terms of the others due to the traceless constraint $\pi^{\mu}_{\;\mu}=0$. 

For the Bessel-Fourier expansion it is furthermore useful to make the following change of variables
\begin{equation}
\begin{split}
\tilde \pi^{r\eta} = & \frac{1}{\sqrt{2}} \left( \tilde \pi^{-\eta} + \tilde \pi^{+\eta} \right), \quad \tilde\pi^{\phi\eta} = \frac{i}{\sqrt{2}} \left( \tilde \pi^{-\eta} -\tilde\pi^{+\eta} \right),\\
\tilde\pi^{r\phi} = & \frac{1}{\sqrt{2}} \left( \tilde\pi^{--} + \tilde\pi^{++} \right), \quad \tilde\pi^{\phi\phi}+\frac{1}{2}\tilde\pi^{\eta\eta} = \frac{i}{\sqrt{2}} \left( \tilde\pi^{--}-\tilde\pi^{++} \right).
\end{split}
\end{equation}
As a 5th independent component we take $\tilde\pi^{\eta\eta}$. One can then use the expansion scheme
\begin{equation}
\begin{split}
\tilde \pi^{\eta\eta}(\tau_0,r,\phi,\eta) = & \sum_{l=1}^\infty \sum_{m=-\infty}^\infty \int_{-\infty}^\infty \frac{d k_\eta}{2\pi}  \;\; 
\tilde \pi_l^{\eta\eta\, (m)}(\tau_0,k_\eta) \;e^{i(m\phi+k_\eta \eta)} J_{m}(k_l^{(m)} r),\\
\tilde \pi^{-\eta}(\tau_0,r,\phi,\eta) = & \sum_{l=1}^\infty \sum_{m=-\infty}^\infty \int_{-\infty}^\infty \frac{d k_\eta}{2\pi}  \;\; 
\tilde \pi_l^{-\eta\, (m)}(\tau_0,k_\eta) \;e^{i(m\phi+k_\eta \eta)} J_{m-1}(k_l^{(m)} r),\\
\tilde \pi^{+\eta}(\tau_0,r,\phi,\eta) = & \sum_{l=1}^\infty \sum_{m=-\infty}^\infty \int_{-\infty}^\infty \frac{d k_\eta}{2\pi} \;\; 
\tilde \pi_l^{+\eta\, (m)}(\tau_0,k_\eta)\; e^{i(m\phi+k_\eta \eta)} J_{m+1}(k_l^{(m)} r),\\
\tilde \pi^{--}(\tau_0,r,\phi,\eta) = & \sum_{l=1}^\infty \sum_{m=-\infty}^\infty \int_{-\infty}^\infty \frac{d k_\eta}{2\pi} \;\; 
\tilde \pi_l^{--\, (m)}(\tau_0,k_\eta)\; e^{i(m\phi+k_\eta \eta)} J_{m-2}(k_l^{(m)} r),\\
\tilde \pi^{++}(\tau_0,r,\phi,\eta) = & \sum_{l=1}^\infty \sum_{m=-\infty}^\infty \int_{-\infty}^\infty \frac{d k_\eta}{2\pi} \;\; 
\tilde \pi_l^{++\, (m)}(\tau_0,k_\eta)\; e^{i(m\phi+k_\eta \eta)} J_{m+2}(k_l^{(m)} r).
\end{split}
\label{eq:BesselFourierPi}
\end{equation}
The inverse relations for $\tilde\pi^{\eta\eta}$, $\tilde\pi^{-\eta}$ and $\tilde\pi^{+\eta}$ are analogous to the vector case \eqref{eq2.9r}. For the components $\tilde\pi^{--}$ and $\tilde\pi^{++}$ one has additional boundary terms,
\begin{equation}
\begin{split}
\tilde \pi^{--}(\tau_0,l,m,k_\eta) = & \frac{1}{2\pi} \frac{2}{R^2(J_{m+1}(z_l^{(m)}))^2} \int_0^R r dr \int_0^{2\pi} d\phi \int_{-\infty}^\infty d\eta \; \tilde \pi^{--}(\tau_0,r,\phi,\eta)\; e^{-i(m\phi+k_\eta \eta)} J_{m-2}(k_l^{(m)} r)\\
& + \frac{1}{2\pi} \frac{2}{z^{(m)}_l J_{m+1}(z_l^{(m)})} \int_0^{2\pi}d\phi \int_{-\infty}^\infty d\eta \; \pi^{--}(\tau_0,R,\phi,\eta)\; e^{-i(m\phi+k_\eta \eta)},\\
\tilde \pi^{++}(\tau_0,l,m,k_\eta) = & \frac{1}{2\pi} \frac{2}{R^2(J_{m+1}(z_l^{(m)}))^2} \int_0^R r dr \int_0^{2\pi} d\phi \int_{-\infty}^\infty d\eta \; \tilde \pi^{++}(\tau_0,r,\phi,\eta)\; e^{-i(m\phi+k_\eta \eta)} J_{m+2}(k_l^{(m)} r)\\
& + \frac{1}{2\pi} \frac{2}{z^{(m)}_l J_{m+1}(z_l^{(m)})} \int_0^{2\pi}d\phi \int_{-\infty}^\infty d\eta \; \pi^{++}(\tau_0,R,\phi,\eta)\; e^{-i(m\phi+k_\eta \eta)}.
\end{split}
\label{eq:inverseBesselFourierpi--}
\end{equation}
The reason is the modified orthogonality relation for the functions $J_{m-2}$ and $J_{m+2}$ in \eqref{eq:A4}.
Again, these relations can be inverted with Lemoine's method. We emphasize that the expressions given in this section 
are of practical use. In particular, the calculation of the fluid dynamical propagation of single fluctuating modes presented 
in Ref.~\cite{Floerchinger:2013rya} involves a Bessel-Fourier decomposition of all scalar, vector and tensor 
fluid dynamic fields at each time step of the simulation.

\section{Summary and Outlook}

In summary, we have shown in the present work that a Bessel-Fourier expansion provides a convenient
orthonormal basis for the characterization of fluctuating initial conditions in all fluid dynamic fields. The
form of the Bessel-Fourier expansion explored in section~\ref{sec3} was proposed for scalar fields
already in Ref.~\cite{ColemanSmith:2012ka}, where in particular results closely related to Figs.~\ref{fig1}
and ~\ref{fig2} of the present work were presented. Here, we have extended these studies to the 
characterization of vector and tensor fields, we have extended it to the characterization 
of correlations between fluctuating modes, and we have explained how the weights of these modes can be determined
in practice in a CPU-inexpensive way based on Lemoine's method. 
 Moreover, in section~\ref{sec4}, we have introduced a variant of the Bessel-Fourier expansion
for normalized densities that remains by construction positive definite if truncated after a finite number of
modes. As we have argued here on general grounds, and as we have demonstrated in a first  
fluid dynamical study of fluctuations recently~\cite{Floerchinger:2013rya}, this property allows one to
propagate single modes fluid dynamically. The Bessel-Fourier expansion, in the form given in 
section~\ref{sec4} is therefore a suitable starting point for the program of mode-by-mode hydrodynamics
that we plan to pursue in future work.

We have also shown that the orthonormal Bessel-Fourier expansion provides for a simple and efficient
characterization of the functional event-by-event probability distribution ${\cal P}$. To illustrate this point, we have 
characterized ${\cal P}$ in sections~\ref{sec3} and~\ref{sec4} for the MC Glauber model of fluctuating initial conditions.
We have shown for this model in particular that event distributions of single modes and distributions of products of two
modes are described by a Gaussian ansatz for ${\cal P}$ with high accuracy. This is important since it allows for the
discussion of event distributions in terms of simple analytic expressions  that depend on a finite number of 
event-averaged quantities only. 

For a general classification of the initial conditions of ultra-relativistic nucleus-nucleus
collisions, it would be interesting to understand in the future to what extent the event probability 
distributions ${\cal P}$ that characterize other models of fluctuating initial conditions are also well-approximated 
by a Gaussian ansatz. We note in this context that the framework presented in sections~\ref{sec3} and~\ref{sec4} 
need not be limited to the analysis of event-distributions of single modes and of products of two modes. In 
close analogy to our discussion of equations (\ref{eq3.14}), (\ref{eq3.15}), one can also compare  
for event distributions of three or more modes the true model distributions to the results of a Gaussian ansatz.
And one can test, of course, whether higher-mode correlators of the form (\ref{eq3.6}) factorize into products
of two-point correlators, as expected for a Gaussian distribution. 
This would establish to what extent non-Gaussianities arise in different models of fluctuating initial conditions,
and it could thus contribute to a general classification of these initial conditions.  

\medskip
\section*{Acknowledgments}
\medskip
We thank U. Heinz, M. Luzum, J.Y. Ollitrault and H. Petersen for useful discussions at various stages of this work. 
\begin{appendix}
\section{Bessel functions and Bessel transformation}
\label{sec:appA}

In this appendix we gather some properties of Bessel functions that we found useful in manipulating the representation of fluctuating initial conditions proposed in the main text.

Denoting by $z_l^{(m)}$ the $l$-th zero of the Bessel function of the first kind $J_m(z)$ one can write the standard orthogonality property as
\begin{equation}
\int_0^R  dr\; r\; J_m\left(z_l^{(m)} \frac{r}{R}\right) J_m\left(z^{(m)}_{l^\prime} \frac{r}{R}\right) = \frac{R^2}{2} \left[J_{m+1}(z^{(m)}_l)\right]^2 \delta_{ll^\prime}.
\label{eq:A1}
\end{equation}
In essence this relation states that for given $m$ one can use the functions $f_l=J_m\left(z_l^{(m)} r/R\right)$ with $l=1,\dots,\infty$ as an orthogonal set of functions on the interval $0\dots R$ when the integration measure is $r \;dr$.

Particularly useful are also the following expressions for derivatives and for multiplying with $m/r$
\begin{equation}
\begin{split}
J^\prime\left( z^{(m)}_l \frac{r}{R} \right) = & \frac{1}{2} \left[ J_{m-1}\left(z_l^{(m)}\frac{r}{R}\right) - J_{m+1}\left( z_l^{(m)} \frac{r}{R} \right)\right],\\
\frac{m R}{z^{(m)}_l r} J\left( z^{(m)}_l \frac{r}{R} \right) = & \frac{1}{2} \left[ J_{m-1}\left(z_l^{(m)}\frac{r}{R}\right) + J_{m+1}\left( z_l^{(m)} \frac{r}{R} \right)\right].
\end{split}
\label{eq:A2}
\end{equation}
From the Bessel differential equation one can also derive the following relations
\begin{equation}
\begin{split}
\int_0^R  dr\; r\; J_{m-1}\left(z_l^{(m)} \frac{r}{R}\right) J_{m-1}\left(z^{(m)}_{l^\prime} \frac{r}{R}\right) = &\frac{R^2}{2} \left[J_{m+1}(z^{(m)}_l)\right]^2 \delta_{ll^\prime},\\
\int_0^R  dr\; r\; J_{m+1}\left(z_l^{(m)} \frac{r}{R}\right) J_{m+1}\left(z^{(m)}_{l^\prime} \frac{r}{R}\right) = &\frac{R^2}{2} \left[J_{m+1}(z^{(m)}_l)\right]^2 \delta_{ll^\prime}.\\
\end{split}
\label{eq:A3}
\end{equation}
Note that $z^{(m)}_l$ is here still the $l$-th zero of $J_m(x)$. The significance of \eqref{eq:A3} is that in addition to $J_m(z_l^{(m)}r/R)$ also the set of functions $J_{m-1}(z_l^{(m)}r/R)$ or $J_{m+1}(z_l^{(m)}r/R)$ for $l=1,\dots,\infty$ constitute orthogonal sets of functions on the interval $0\dots R$. This feature is important for the Bessel expansion of vector valued functions such as the fluctuations in the fluid velocity.

Finally, we note a related property for the sets of functions $J_{m-2}(z_l^{(m)}r/R)$ and $J_{m+2}(z_l^{(m)}r/R)$. The orthogonality relations are now slightly more complicated,
\begin{equation}
\begin{split}
\int_0^R  dr\; r\; J_{m-2}\left(z_l^{(m)} \frac{r}{R}\right) J_{m-2}\left(z^{(m)}_{l^\prime} \frac{r}{R}\right) - \frac{R^2}{z_l^{(m)}} J_{m-1}\left(z_l^{(m)}\right) J_{m-2}\left(z_{l^\prime}^{(m)}\right)= &\frac{R^2}{2} \left[J_{m+1}(z^{(m)}_l)\right]^2 \delta_{ll^\prime},\\
\int_0^R  dr\; r\; J_{m+2}\left(z_l^{(m)} \frac{r}{R}\right) J_{m+2}\left(z^{(m)}_{l^\prime} \frac{r}{R}\right)+ \frac{R^2}{z_l^{(m)}} J_{m+1}\left(z_l^{(m)}\right) J_{m+2}\left(z_{l^\prime}^{(m)}\right) = &\frac{R^2}{2} \left[J_{m+1}(z^{(m)}_l)\right]^2 \delta_{ll^\prime}.
\end{split}
\label{eq:A4}
\end{equation}
These relations can still be used for an expansion of tensor  valued fluctuations in terms of the set of functions $J_{m-2}(z_l^{(m)}r/R)$ or $J_{m+2}(z_l^{(m)}r/R)$, $l=1,\dots,\infty$ although some expressions contain additional boundary terms, as discussed in section~\ref{sec5}.
\section{Discrete Bessel transformation}
\label{sec:appB}

In this appendix we discuss an efficient numerical scheme due to Lemoine \cite{Lemoine} to do Bessel transformations by converting integrals into finite numerical sums. In this scheme $r$-dependent functions $h$ are represented in position space by their value on $N$ discretization points
\begin{equation}
h(r_\alpha^{(m)}),\quad \alpha=1,\dots,N
\label{eq:B1}
\end{equation}
where 
\begin{equation}
r^{(m)}_\alpha=\frac{z_\alpha^{(m)}}{z_{N}^{(m)}}R
\label{eq:B2}
\end{equation}
and $z_\alpha^{(m)}$ is the $\alpha$th zero crossing of the Bessel function $J_m(z)$. Note that $r_{N}^{(m)}=R$ is on the boundary where one assumes $h(R)=0$.

Consider now the Bessel function expansion
\begin{equation}
h(r) = \sum_{l=1}^\infty h_l J_m(k_l^{(m)}r)
\end{equation}
with 
\begin{equation}
k_l^{(m)}=z_l^{(m)}\frac{1}{R}.
\end{equation}
We truncate this expansion at $l=N$ or for $k_l^{(m)} = k_{N}^{(m)}=z^{(m)}_{N}/R$. Restricting also to the points $r_\alpha^{(m)}$,
\begin{equation}
h(r_\alpha^{(m)})\approx\sum_{l=1}^{N}  \; h_l\;  J_m\left(k_l^{(m)}r_\alpha^{(m)}\right),
\label{eq:B5}
\end{equation}
thereby possibly cutting off the very fine structures of the function $h(r)$. The virtue of the spatial discretization \eqref{eq:B1}, \eqref{eq:B2} is now that one can efficiently approximate the inverse relation
\begin{equation}
h_l = \frac{2}{R^2 [J_{m+1}(k_l^{(m)} R)]^2} \int_0^R dr\; r\; h(r) \;J_m\left(k_l^{(m)}r\right)
\label{eq:B6}
\end{equation}
by a finite sum
\begin{equation}
h_l \approx \sum_{\alpha=1}^{N} \frac{4}{\left[z^{(m)}_{N}\right]^2 \left[ J_{m+1}(z_l^{(m)}) \right]^2 \left[ J_{m+1}(z_\alpha^{(m)}) \right]^2} \; h\left(r_\alpha^{(m)}\right) \; J_m\left(k_l^{(m)} r_\alpha^{(m)}\right).
\label{eq:B7}
\end{equation}
Note that the last term with $\alpha=N$ vanishes so that the sum goes effectively over the range $\alpha=1,\dots,N-1$. Note that \eqref{eq:B5} and \eqref{eq:B7} constitute matrix relations between the two representations of the function $h$ in position space \eqref{eq:B1} and in Bessel space represented by $h_l$, $l=1,\dots,N$.

We emphasize at this point that even in a situation where $h_l=0$ for $l>N$ so that \eqref{eq:B5} is exact, this is not necessarily the case for discrete version of the inverse relation \eqref{eq:B7}. This is in contrast to other relations of similar kind such as the discrete Fourier transforms. Equation \eqref{eq:B6} is exact and \eqref{eq:B7} is getting better and better as $N\to\infty$.
\section{Probability distribution of enthalpy densities}
\label{sec:appC}

In this appendix we discuss some general properties of the event-by-event probability distribution of enthalpy density in the transverse plane at time $\tau_0$ where hydrodynamics is initialized. For notational simplicity we neglect the dependence on the longitudinal rapidity coordinate. 

Since the initial transverse enthalpy density $w(r,\phi)$ is a function of radius $r$ and azimuthal angle $\phi$, the probability distribution that describes an ensemble of events, is a functional 
\begin{equation}
{\cal P}[w].
\label{eq:appc0}
\end{equation}
It can be characterized in different ways, for example by the expectation value
\begin{equation}
\langle w(r,\phi) \rangle\, ,
\label{eq:appc1}
\end{equation}
and the set of $n$-point correlation functions
\begin{equation}
\langle w(r_1,\phi_1)\ldots w(r_n,\phi_n) \rangle\, .
\label{eq:appc2}
\end{equation}
Note that the enthalpy density is real and positive definite which therefore has to be the case for the expectation values and correlation functions, as well.

If the event ensemble in question consists of events with arbitrary orientation in the transverse plane, azimuthal rotation invariance $\phi\to \phi+\Delta \phi$ and invariance under reflections $\phi \to - \phi$ are realized as statistical symmetries. This means that the transverse enthalpy distribution of a single event is not invariant under these transformations but appropriate event averages are. In particular, the expectation value in \eqref{eq:appc1} is then independent of $\phi$ and the correlation functions in \eqref{eq:appc2} depend only on differences between the azimuthal angles. The statistical symmetry must also be realized for the probability distribution \eqref{eq:appc0}.

Let us now discuss the particularly simple and important case of a functional probability distribution of Gaussian form (${\cal N}$ is an appropriate normalization factor),
\begin{equation}
\begin{split}
{\cal P}[w] = & {\cal N} \exp {\Bigg [} -\frac{1}{2} \int dr_1 dr_2 d\phi_1 d\phi_2 \, r_1 \, r_2 \\
& \times \left( w(r_1,\phi_1) - \langle w(r_1,\phi_1) \rangle \right) M(r_1,r_2,\phi_1,\phi_2) \left( w(r_2,\phi_2) - \langle w(r_2,\phi_2) \rangle \right) {\Bigg ]}.
\end{split}
\label{eq:appc3}
\end{equation}
The statistical azimuthal rotation and reflection symmetry would imply that $M\in \mathbb{R}$ depends on $\phi_1$ and $\phi_2$ only via the difference $|\phi_1-\phi_2|$. 
We expand now $w(r,\phi)$ in terms of the Bessel-Fourier decomposition proposed in section \ref{sec2},
\begin{equation}
w(r,\phi) = \sum_{m=-N_m}^{N_m} \sum_{l=1}^{N_l} w^{(m)}_l \, e^{i m \phi} \, J_m\left(k^{(m)}_l r\right).
\end{equation}
The coefficients $w^{(m)}_l$ are complex but fulfill $w^{(m)*}_l = w^{(-m)}_l$ since $w(r,\phi) \in \mathbb{R}$. The Gaussian distribution \eqref{eq:appc3} becomes in this basis
\begin{equation}
\begin{split}
{\cal P}[w] = & {\cal N} \exp{\Bigg [} -\frac{1}{2} \sum_{m_1,m_2=-N_m}^{N_m} \sum_{l_1,l_2=1}^{N_l} \\
& \times \left(w^{(m_1)}_{l_1} - \langle w^{(m_1)}_{l_1} \rangle\right)^* T^{(m_1)(m_2)}_{l_1 l_2} \left(w^{(m_2)}_{l_2} - \langle w^{(m_2)}_{l_2} \rangle\right) {\Bigg ]},
\end{split}
\label{eq:appc5}
\end{equation}
with
\begin{equation}
\begin{split}
T^{(m_1)(m_2)}_{l_1 l_2} = & \int dr_1 dr_2 d\phi_1 d\phi_2 \, r_1 r_2\, e^{-i m_1 \phi_1} e^{im_2 \phi_2}\\
& \times J_{m_1}\left( k^{(m_1)}_{l_1} r_1 \right) J_{m_2}\left( k^{(m_2)}_{l_2} r_2 \right) \, M(r_1,r_2,\phi_1,\phi_2).
\end{split}
\end{equation}
The matrix $T^{(m_1)(m_2)}_{l_1 l_2}$ is hermitean since ${\cal P}$ is real. It also fulfills
\begin{equation}
T^{(m_1)(m_2)*}_{l_1 l_2} = T^{(-m_1)(-m_2)}_{l_1 l_2}.
\end{equation}
For statistical rotation symmetry it is diagonal with respect to the indices $m_1$ and $m_2$,
\begin{equation}
T^{(m_1)(m_2)}_{l_1 l_2} = \delta_{m_1 m_2} T^{(m_1)}_{l_1 l_2}.
\end{equation}
Statistical azimuthal reflection symmetry implies
\begin{equation}
T^{(m_1)(m_2)}_{l_1 l_2} = T^{(-m_1)(-m_2)}_{l_1 l_2} \in \mathbb{R}.
\end{equation}
We finally note that general properties of Gaussian distributions imply
\begin{equation}
(T^{-1})^{(m_1)(m_2)}_{l_1 l_2} = \langle  w^{(m_1)}_{l_1} w^{(m_2)*}_{l_2} \rangle - \langle  w^{(m_1)}_{l_1}\rangle \langle w^{(m_2)*}_{l_2} \rangle .
\end{equation}
The functional probability distribution \eqref{eq:appc5} is therefore completely determined by the expectation values and two-mode correlators.

\end{appendix}

\end{document}